\documentclass[traditabstract]{aa}
\usepackage{graphicx}
\usepackage{txfonts}
\usepackage{xcolor}

\begin{document}

\title{The rotation curves shapes of late-type dwarf galaxies}

\author{R. A. Swaters\inst{1,2}\fnmsep
  \thanks{Present address: Department of Astronomy, University of
    Mary\-land, College Park, MD 20742-2421. Email:
    swaters@astro.umd.edu}\and R. Sancisi\inst{3,4}\and T. S. van
  Albada\inst{3}\and J. M.  van der Hulst\inst{3}}

\institute{Department of Physics and Astronomy, Johns Hopkins
  University, 3400 N. Charles Str., Baltimore, MD 21218
\and 
  Space Telescope Science Institute, 3700 San Martin Dr.,
  Baltimore, MD 21218
\and
  Kapteyn Astronomical Institute, PO Box 800, 9700 AV
  Groningen, The Netherlands
\and
  INAF - Osservatorio Astronomico di Bologna, via Ranzani 1, 40127
  Bologna, Italy
}

\date{Received 2008; accepted 2008}

\def\HI{\ion{H}{i}}
\def\Halpha{H$\alpha$}
\def\sk{\kern4pt}
\def\kms{km s$^{-1}$}
\def\figHI{\HI}
\def\notes#1{\noindent{\sl UGC~#1\/}}

\newcommand{\tskip}{\noalign{\vspace{2pt}}}
\newcommand{\crdr}[3]{\hbox{\hbox to14pt{\hfil#1$^h$}\hbox to14pt{\hfil#2$^m$}\hbox to19pt{\hfil#3$^s$}}}
\newcommand{\crdd}[3]{\hbox{\hbox to14pt{\hfil#1$^\circ$}\hbox to12pt{\hfil#2$'$}\hbox to13pt{\hfil#3$''$}}}

\abstract{
  We present rotation curves derived from \HI\ observations for a
  sample of 62 galaxies that have been observed as part of the
  Westerbork \HI\ Survey of Spiral and Irregular Galaxies (WHISP)
  project. These rotation curves have been derived by interactively
  fitting model data cubes to the observed cubes. This procedure takes
  the rotation curve shape, the \HI\ distribution, the inclination,
  and the size of the beam into account, and makes it possible to
  correct for the effects of beam smearing.  A comparison with higher
  spatial resolution H$\alpha$ rotation curves available in the
  literature shows that there is general agreement between the
  two. The late-type dwarf galaxies in our sample have rotation-curve
  shapes that are similar to those of late-type spiral galaxies, in
  the sense that their rotation curves, when expressed in units of
  disk scale lengths, rise as steeply in the inner parts and start to
  flatten at two disk scale lengths. None of the galaxies in our
  sample have solid-body rotation curves that extend beyond three disk
  scale lengths. The logarithmic slopes between two disk scale lengths
  and the last measured point on the rotation curve is similar between
  late-type dwarf and spiral galaxies. Thus, whether the flat part of
  the rotation curve is reached or not seems to depend more on the
  extent of the rotation curve than on its amplitude. We also find
  that the outer rotation curve shape does not strongly depend on
  luminosity, at least for galaxies fainter than $M_R\sim-19$.  We
  find that in spiral galaxies and even in the central regions of
  late-type dwarf galaxies, the shape of the central distribution of
  light and the inner rise of the rotation curve are related. This
  implies that galaxies with stronger central concentrations of light
  also have higher central mass densities, and it suggests that the
  luminous mass dominates the gravitational potential in the central
  regions, even in low surface brightness dwarf galaxies.
}


\maketitle

\section{Introduction}
\label{secintro}

The shape and amplitude of rotation curves are directly related to the
gravitational potential in the midplane of the galaxy.  Hence, a
relation is expected between the main mass components in a galaxy and
the shape of the rotation curve. If the luminous mass density is
dynamically significant, the rotation curve shape will be related to
the distribution of light. On the other hand, dark matter may weaken
such a relation, in particular if the dark matter dominates the
potential.  Therefore, in principle, a comparison between rotation
curves shapes and the distribution of luminous mass can provide
insight in the amount and distribution of dark matter in galaxies, and
it can indicate how the structural properties of the dark and luminous
components are linked.  Moreover, because this link is probably the
result of the process of galaxy formation, studying this link between
dark and luminous matter is relevant for theories of galaxy formation.

The relation between the luminous properties of galaxies and their
rotation curves has been a topic of study for many years (e.g., Rubin
et al.\ 1985; Kent 1986, 1987; Corradi \& Capaccioli 1990; Persic et
al.\ 1996; Rhee 1996; Verheijen 1997; Catinella et al.\ 2006).  These
studies, however, are mainly based on high surface brightness spiral
galaxies with absolute magnitudes brighter than $M_B\sim-18$. To
better understand the relation between luminous and dark mass and its
implications for the structure and formation of galaxies it is
necessary to include galaxies with lower luminosities and lower
surface brightnesses.

In recent years, considerable effort has been invested in studies of
rotation curve shapes of dwarf and low surface brightness galaxies
(e.g., de Blok et al.\ 1996; de Blok \& McGaugh 1997; Stil
  1999; Swaters 1999; Swaters et al.\ 2000; van den Bosch et
al.\ 2001; van den Bosch \& Swaters 2001; de Blok et al.\ 2001;
Marchesini et al.\ 2002; de Blok \& Bosma 2002, herafter dBB; Swaters
et al.\ 2003a (hereafter SMvdBB); Rhee et al.\ 2004; Simon et
al.\ 2005; Spekkens et al.\ 2005; Kuzio de Naray et al.\ 2006;
Chengalur et al.\ 2008). Most of these studies focused on the
shape of the rotation curve in the central regions of these galaxies,
because these inner rotation curve shapes provide a powerful test of
models of galaxy formation in a cold dark matter dominated universe
(e.g., Navarro et al.\ 1996, 1997; Bullock et al.\ 2001; Power et
al.\ 2003). Because high spatial resolution is necessary to determine
the inner rotation curve shapes, these studies of dwarf and low
surface brightness galaxies have mostly relied on optical spectroscopy
of the H$\alpha$ emission line. Unfortunately, the H$\alpha$ emission
is usually only detected in the optical disk. Hence, most of these
rotation curves do not extend to large enough radii to measure the
rotation curve shape outside the optical disk. Instead, HI rotation
curves are needed for that purpose.

To date, relatively little work has been done to investigate the
relations between luminous properties and the \HI\ rotation curve
shape in the dwarf galaxy regime. The studies by Tully et al.\ (1978),
Broeils (1992a) and C\^ot\'e et al. (2000), which each comprised a
handful of dwarf galaxies, led the picture that late-type dwarf
galaxies have low amplitude, slowly rising rotation curves that keep
rising to the last measured point.  Thus, the rotation curves shapes
of dwarf galaxies seemed different from those of spiral galaxies,
which, in general, have a steep rise in the inner regions, followed by
relatively flat outer parts (e.g., Bosma 1978, 1981a,b; Begeman 1987,
1989). This was confirmed and refined by Casertano \& van Gorkom
(1991, hereafter CvG) and Broeils (1992a). They investigated the
correlations between the outer slopes of the rotation curves and the
maximum rotation velocities, and found that the rotation curves, when
expressed in units of the optical radius, rise more slowly towards
lower rotation amplitudes.

To further our understanding of the relation between the distribution
of light and the shape of the rotation curves for dwarf galaxies, a
large sample with well-defined rotation curves and optical properties
is needed.  Such a sample is presented in Swaters et al.\ (2002,
hereafter Paper I); it contains 73 late-type dwarf galaxies.  For
these galaxies optical properties have been presented in Swaters \&
Balcells (2002, hereafter Paper~II). This sample forms the basis of
the study of the rotation curves of late-type dwarf galaxies presented
here (see also Swaters 1999), and of studies of their dark matter
properties published earlier (Swaters 1999; van den Bosch \& Swaters
2001; SMvdBB).

The structure of this paper is as follows. In Section~\ref{secsample}
we describe the sample selection.  Section~\ref{secrc} describes in
detail the procedure used to derive the rotation curves.  We give a
brief description of a number of individual galaxies in
Section~\ref{secnotes}, and in Section~\ref{seccomp} we compare our
\HI\ rotation curves with long-slit \Halpha\ rotation curves available
in the literature.  The relations between the rotation curve shapes
and the luminous properties are discussed in
Section~\ref{secrcshapes}. In Section~\ref{secdisc} we discuss the
results and present our conclusions.  In Appendix~\ref{appfigures} we
present the data in graphical form.

\section{The sample}
\label{secsample}

The galaxies in this sample have been observed as part of the WHISP
project (Westerbork \HI\ Survey of Spiral and Irregular Galaxies)
which aimed at mapping several hundred nearby spiral and irregular
galaxies in \HI\ with the WSRT. The WHISP sample was selected from the
Uppsala General Catalogue of Galaxies (UGC, Nilson 1973), taking all
galaxies with declinations north of $20^\circ$, blue major axis
diameters larger than $1.5'$ and measured \HI\ flux densities larger
than 100 mJy. For the present purpose, we defined as dwarf galaxies
all galaxies with Hubble types later than Sd, and spiral galaxies of
earlier Hubble types for which the absolute $B$-band magnitudes are
fainter than $-17$.  For a detailed description of the selection
criteria and the general goals of the WHISP project, see Paper~I.

The \HI\ observations on which the rotation curves presented here are
based, are discussed in detail in Paper~I. The typical full resolution
of the observations is $14''\times 14''/ \sin\delta$. In general the
signal-to-noise ratios at this resolution are low, and therefore \HI\ 
maps and velocity fields are given at $30''$ resolution (see
Appendix~\ref{appfigures}).

The sample of late-type dwarf galaxies presented here is identical to
the sample of Paper~I, except for four galaxies that have
been omitted because they are interacting with a close companion:
UGC~1249, UGC~5935, UGC~6944 and UGC~7592. The remaining sample,
listed in Table~\ref{tabsample}, consists of 69 late-type dwarf
galaxies. For 62 of those rotation curves have been derived.

In addition to the \HI\ data, optical $R$-band data have also been
obtained for these galaxies. Details of these observations and the
data reduction are presented in Swaters \& Balcells (2002). The
absolute magnitudes, scale lengths and surface brightnesses as derived
in Paper~II are listed in Table~\ref{tabsample}.

\section{The rotation curves}
\label{secrc}

\subsection{Derivation of the rotation curves}

\begin{figure}
\resizebox{\hsize}{!}{\includegraphics{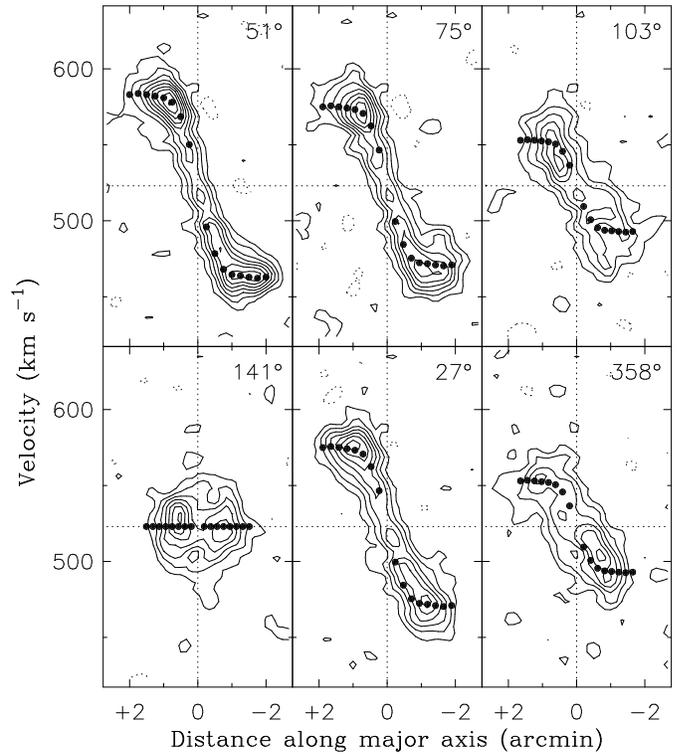}}
\caption{%
  Example of an interactively determined rotation curve (solid points)
  fit to six position-velocity diagrams (contours), which were taken
  along the position angels indicated in the top right of each panel.
  The galaxy shown here is UGC~4325.  To determine the rotation curve
  the position angle, the inclination and the rotation velocity were
  changed until the projected points on the position-velocity diagrams
  were found to best match the data.}
\label{figderrc}
\end{figure}

Usually, rotation curves are derived by fitting tilted-ring models to
the observed velocity fields (e.g., Begeman 1987). In principle, this
method can be applied to the galaxies in this sample as well, provided
that the requirements for fitting a tilted-ring model are met (see
Begeman 1987). However, because the galaxies in our sample are often
small and the HI distribution is usually irregular and clumpy, and
because of the limited angular resolution and sensitivity of the
present observations, the derivation of the velocity fields and of the
rotation curves is not straightforward. In particular the possible
effects of beam smearing must be understood and corrected for (see
also Bosma 1978; Begeman 1989; Rubin et al.\ 1989; Broeils 1992b; de
Blok \& McGaugh 1997; Swaters 1999; Swaters et al.\ 2000; van den
Bosch et al.\ 2000; van den Bosch \& Swaters 2001; McGaugh et
al.\ 2001).

Due to the finite beam size, the \HI\ emission will be smeared out,
leading to apparently larger \HI\ disks and to filling of holes in the
\HI\ distribution.  Moreover, as a result of beam smearing the
gradients in the velocity fields, as determined for example from
Gaussian fits to the line profiles, may become shallower. Hence,
rotation curves derived from these velocity fields have shallower
gradients as well.  The magnitude of the effect of beam smearing thus
depends on the combination of the size of the beam, the distribution
of the \HI, the inclination angle of the galaxy and the intrinsic
velocity gradients.  Despite the effects of beam smearing, information
on the true rotation curve is still contained in the data, and may be
retrieved if an appropriate method to derive the rotation curve is
used.

To take the effects of beam smearing into account, we have adopted a
two-step procedure. First, we obtained the orientation parameters and
an initial estimate of the rotation curve, if possible from a tilted
ring fit, and using an interactive procedure if not.  In the second
step, this estimate of the rotation curve was refined iteratively by
constructing model data cubes, using the input rotation curve and
adjusting this rotation curve to match the model cubes to the observed
ones.

Because for most galaxies the signal-to-noise ratios of the full
resolution data were too low to derive reliable rotation curves, we
have derived the rotation curves from the $30''$ data.

\subsubsection{Initial estimate of the rotation curves}

Before deriving the initial estimate of the rotation curve for a
galaxy, we determined the position of its center. For most of the
galaxies it was not possible to derive the center from the velocity
field because of the often close to linear inner rise of the rotation
curves.  Therefore, we have used the optical centers from Paper~II to
define the centers, except when noted otherwise in
Table~\ref{tabcenters}.

Next, the systemic velocity was determined from a tilted-ring fit to
the Gaussian-fit velocity field, derived from the $30''$ velocity
field (see Paper~I), with the center fixed, by taking the average over
all rings, except the outer rings which usually have large
uncertainties. While determining the systemic velocity, the
  position angle was kept fixed at the initial estimate derived from
  the velocity field, and the inclination was kept fixed at the value
  derived from the ellipticity of the optical isophotes (see
  Paper~II). We then made a new tilted ring fit with center and
systemic velocity fixed to obtain the initial estimates for the
rotation curve and the run of inclination and position angle with
radius. If there were no appreciable variations of the inclination or
position angle with radius, a fixed value was used (the average over
all rings, excluding the rings with large uncertainties).  If it was
not possible to estimate the inclination from the velocity field, the
inclination was determined from the ellipticity of the optical
isophotes (see Paper~II) or \HI\ contours (see Paper~I), assuming an
intrinsic disk thickness of 0.2.

Because many of the galaxies in our sample are small and affected by
beam smearing, we found that in many cases it was difficult to obtain
a good initial estimate of the rotation curve and orientation
parameters from the tilted ring fits, especially in the central parts
of the galaxies (see also de Blok, McGaugh, \& van der Hulst 1996). In
those cases, we interactively fit the rotation curve and the
orientation parameters to a set of six position-velocity
diagrams\footnote{This interactive fitting was done with the task {\sc
    inspector}, implemented in the Groningen Image Processing System
  (GIPSY, http://www.astro.rug.nl/$\sim$gipsy)} as illustrated in
Fig.~\ref{figderrc} to improve the initial estimates (one along the
major axis, two at a $30^\circ$ angle away from the major axis as
measured in the plane of the galaxy, two $60^\circ$ degrees from the
major axis, and one along the minor axis of the galaxy). To assure
ourselves we are able to accurately retrieve the rotation curves and
orientation parameters, we created several model galaxies with
different and radially changing rotation velocities and orientation
parameters. Except for pure solid-body rotators, we were able to
retrieve the input parameters to within the uncertainties expected
based on the noise added to the models.

In this interactive fitting, the rotation velocity, inclination and
position angle at each radius were adjusted until the projected point
was deemed to best match the six position-velocity diagrams
simultaneously. Most weight was given to the major axis, and least to
the minor axis. For determining the run of inclination and position
angle with radius, equal weight was given to all position-velocity
diagrams. Inclinations of around $30^\circ$ and lower are uncertain.
We applied this method to the Hanning smoothed, $30''$ resolution
data.

\begin{table}
\caption{Galaxy centers}
\label{tabcenters}
\centering
\begin{tabular}{lrrl}
\hline\hline
\tskip
UGC & \multicolumn{2}{c}{~R.A.~~~(2000)~~~Dec.} & Comments\\
\hline
\tskip
731  & \crdr{ 1}{10}{43.0} & \crdd{49}{36}{ 7} & Kinematic center \\
2455 & \crdr{ 2}{59}{42.6} & \crdd{25}{14}{34} & Center of \HI\ distribution \\
4173 & \crdr{ 8}{ 7}{ 8.7} & \crdd{80}{ 7}{38} & New optical center \\
4325 & \crdr{ 8}{19}{20.3} & \crdd{50}{ 0}{36} & Kinematic center \\
5414 & \crdr{10}{ 3}{57.2} & \crdd{40}{45}{22} & New optical center \\
5721 & \crdr{10}{32}{16.6} & \crdd{27}{40}{ 8} & Kinematic center \\
8188 & \crdr{13}{ 5}{51.4} & \crdd{37}{35}{56} & Center of \HI\ distribution \\
8490 & \crdr{13}{29}{36.4} & \crdd{58}{25}{10} & Kinematic center \\
8837 & \crdr{13}{54}{43.2} & \crdd{53}{53}{42} & Center of \HI\ distribution \\
\hline
\end{tabular}
\end{table}

The inclinations and position angles as derived from the tilted rings
fits or the interactive fits are listed in Table~\ref{tabsample}.  For
warped galaxies the orientation of the inner, unwarped part of the
disks are given.  For galaxies with clear differences between the
approaching and receding sides, rotation curve were determined from
both sides independently, using the same orientation parameters. The
rotation curves have been sampled every 15 arcseconds, and the
individual points in the rotation curves are therefore not
independent.

For seven galaxies (UGC~4274, UGC~6817, UGC~7199, UGC~7408, UGC~8201,
UGC~8331 and UGC~9128) we have not derived a rotation curve, because of a
combination of small \HI\ extent ($R_\mathrm{\HI}\la 1'$), clumpy \HI\
distribution and lack of ordered rotation.

\subsubsection{Final rotation curves}

A potential shortcoming of our method to derive the initial estimates
of the rotation curves is that it is subjective. To obtain a more
objective rotation curve and to investigate the effects of beam
smearing further, we refined the initial estimate of the rotation
curves by detailed modeling of the observations.

First, for each galaxy, an axisymmetric velocity field was
constructed, based on the rotation curve determined as described
above.  Next, a three-dimensional model datacube was constructed from
this velocity field and the observed \HI\ distribution at the full
resolution, assuming a constant Gaussian velocity dispersion of 8
\kms. The model was subsequently convolved to a resolution of $30''$.
The datacube obtained in this way was compared to the observations by
calculating the $\chi^2$ value of the difference between model and
observations over a small sub-cube that contained all HI emission, and
also by visually comparing six model position-velocity diagrams (like
those shown in Fig.~\ref{figderrc}) with those of the observations.
Although we used the $\chi^2$ values to find better matching models,
we have not used an algorithm to minimize $\chi^2$ because of the
prohibitively large amounts of computing time needed.  If the match
between model and observations was not satisfactory, as determined
from the visual inspection, the input rotation curve was adjusted
interactively and the procedure was repeated until the $\chi^2$ had
reached an approximate minimum. For most galaxies only the rotation
velocities in the inner $60''$ needed to be adjusted. Position angles
and inclinations have not been adjusted using this method. For
galaxies with differences between the approaching and receding sides,
the modeling was done for both sides independently.

To verify that our method works, we constructed model galaxies with
known rotation curves, orientation parameters, and HI distributions,
and derived the rotation curves with the method described here. As
long as there was detectable HI emission in the center and the model
galaxies were not close to edge-on, we were able to reproduce the
rotation velocities accurate to a few km/s (depending on the noise).
Independent verification that the method works well is given in
Section~\ref{seccomp}, where we compare our \HI\ rotation curves with
\Halpha\ rotation curves available in the literature.

\begin{figure}
\resizebox{\hsize}{!}{\includegraphics{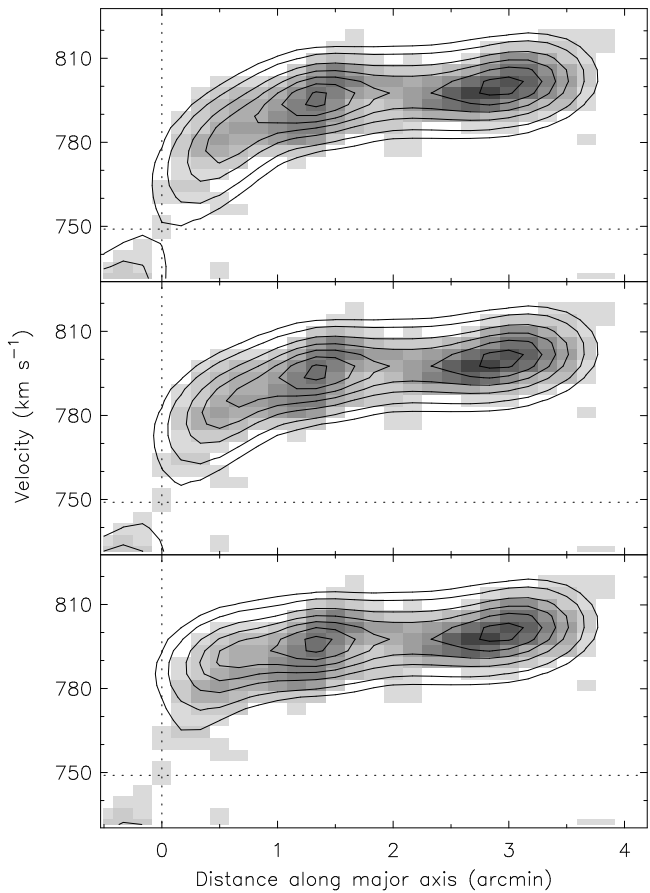}}
\caption{%
  Comparison between the data for the receding side of UGC~12732,
  shown in grayscale, and models of the observations (contours), based
  on an input rotation curve and the observed \figHI\ distribution at
  the full resolution. The input rotation curve is adjusted until the
  best match to the data is obtained (shown in the middle panel). The
  top panel shows the model based on the rotation curve derived by
  visual inspection, which underestimates the rotation velocities in
  the inner parts. The bottom panel shows a model based on a rotation
  curve that overestimates the rotation velocities between the center
  and 1$'$.}
\label{figmodgal}
\end{figure}

An example of the modeling procedure is given in Fig.~\ref{figmodgal}.
The data for the receding side of UGC~12732 are shown in grayscale,
and different models are overlayed in contours. The top panel shows
the model based on the rotation curve determined interactively.  Close
inspection reveals that between the center and 1 arcmin the model
fails to reproduce the observations.  The model systematically
underestimates the velocities in the central regions.  In the middle
panel, we show the model that we considered the best match, as found
from the iterative modeling procedure described above. In the bottom
panel, a model is shown based on a rotation curve for which the
rotation velocities in the inner parts have been overestimated on
purpose. This is also reflected in the model, which systematically
overestimates the velocities between the center and $1'$. Note that
the comparison of the model and the observation was not done based on
one position-velocity diagram, like the one shown in
Fig.~\ref{figmodgal}, but using the entire data cube for calculating
$\chi^2$, and using six position-velocity diagrams.

The final rotation curves are shown in Appendix~\ref{appfigures}. In
the cases where the rotation curves have been derived for both sides
independently, the average of the two is shown as well. The
determination of the uncertainties in the derived rotation velocities is
described below. In Appendix~\ref{appfigures} the rotation curves are
also shown overlayed on the position-velocity diagrams along the major
axes. Where applicable, the rotation curves for the approaching and
receding sides are overlayed rather than the average rotation curve.

\subsection{Rotation curves for edge-on galaxies}

In principle, it would have been possible to derive the rotation
curves for the edge-on galaxies in our sample using a procedure
similar to the modeling of NGC~1560 (Broeils 1992b) or NGC~891
(Swaters et al.\ 1997).  However, such a procedure
is extremely time-consuming, and therefore we did not use it.
Instead, we chose to follow the approach used by Sancisi \& Allen
(1979) to derive the rotation curve of NGC~891. At each
position along the major axis the rotation velocity was determined
from the edge of the line profile, correcting for the effects of
instrumental broadening and random motions.  The random motions of the
gas were assumed to be 8 \kms. In this method, it is assumed that
there is gas everywhere along the line of nodes. Consequently, if a
galaxy has a ring-like distribution of \HI\ or a central depression in
the \HI\ distribution, an incorrect rotation curve may be derived.
\footnote{After we derived the rotation curves for the galaxies in our
  sample, alternative methods to derive the rotation curves for
  edge-on galaxies have been published that are more objective (e.g.,
  Kregel \& van der Kruit 2004, Garc{\'{\i}}a-Ruiz et
  al.\ 2002). Because of the intrinsic uncertainties in the derivation
  of the rotation curves of edge-on galaxies, and given that we do not
  use the rotation curve for the edge-on galaxies in our analysis, we
  have not rederived the rotation curves.}

The rotation curves for edge-on galaxies derived in this way for the
approaching and receding sides separately are shown overlayed on the
position-velocity diagrams along the major axis in
Appendix~\ref{appfigures}. The average rotation curves are shown in
Appendix~\ref{appfigures} as well.

\subsection{Uncertainties of the rotation velocities}

The method outlined above to derive the rotation curves does not
provide an estimate of the uncertainties of the rotation velocities.
Therefore, we have estimated the uncertainties resulting from the two
main sources of error. The first is the accuracy with which the
position of a point on the position-velocity diagrams can be measured.
This measurement error depends on the width and the signal-to-noise
ratio of the profile. Based on trial and error for a range of
profiles, we have found that the $1\sigma$ error with which the radial
velocity can be determined from a profile is about 2 \kms.  The
$1\sigma$ errors for the points on the rotation curve is therefore
$2/\sin i$ \kms.

The second source of error stems from non-circular motions and the asymmetry of
the galaxy, which can be estimated by the difference between the
approaching and the receding sides.  We made the {\it ad hoc} assumption
that the difference in rotation velocity between the mean rotation
curve and the rotation curve measured on either the approaching or
receding side represents a $2\sigma$ difference.  We have added the
asymmetry error and the measurement error quadratically to obtain an
estimate of the $1\sigma$ errors in the rotation velocities.

For the edge-on galaxies we have used the same procedure to estimate
the errors. This estimate does not include possible systematic effects
as a result of lack of gas on the line of nodes.  As a result, some of
the rotation curves for the edge-on galaxies may be more uncertain
than indicated by the errors.  This systematic uncertainty was taken
into account in determining the rotation curve quality, as
described below.

\subsection{Rotation curve quality}

The quality of the derived rotation curves differs from galaxy to
galaxy.  Some galaxies are well resolved, others are only a few beams
across. Some galaxies have strong \HI\ emission, others are weak in
\HI.  Some galaxies have distorted kinematics, others show ordered
rotation. Hence, for some galaxies we could derive reliable rotation
curves, whereas for others only lower quality rotation curves could be
derived.  We have tried to quantify this by tentatively dividing the
derived rotation curves into four different categories of quality
$q$. The quality depends on the signal-to-noise ratio, on the presence
of non-circular motions and on deviations from axial symmetry.

The quality estimates (tabulated in Table~\ref{tabsample}) range in
value from 1 to 4, where 1 indicates a rotation curve deemed to be
reliable, 2 indicates an uncertain rotation curve, 3 indicates a
highly uncertain rotation curve, and 4 means that no rotation curve
could be derived, either because the \HI\ distribution or kinematics
were too irregular, or because there was no significant rotation.

\begin{table*}[ht]
\setlength{\tabcolsep}{1.75mm}
\caption{Optical and rotation curve properties}
\label{tabsample}
\centering
\begin{tabular}{rrrcrrrrrrrrrrrrrrr}
\noalign{\vspace{-0.2cm}}\hline\noalign{\smallskip}
 UGC &$D_\mathrm{a}$ &$M_R$ & Type& $h$ &$\mu_0^R$ &$q$ &$i$ &P.A. &$\varv_\mathrm{sys}$ &$\varv_1$ &$\varv_2$ &$\varv_3$ &$\varv_4$ &$\varv_\mathrm{last}$ &$r_\mathrm{last}$ &$S_\mathrm{(1,2)}$ &$S_\mathrm{(2,3)}$ &$S_\mathrm{(2,L)}$ \\
& Mpc & mag & & kpc & mag${\prime\prime}^{-2}$ & & $^\circ$ & $^\circ$ & \multicolumn{6}{c}{-- \hfil -- \hfil -- \hfil -- \hfil \kms \hfil -- \hfil -- \hfil -- \hfil --} & kpc &&&\\
\noalign{\smallskip}
 (1) & (2) & (3) & (4) & (5) & (6) & (7) & (8) & (9) & (10) & (11) & (12) & (13) & (14) & (15) & (16) & (17) & (18) & (19) \\
\noalign{\smallskip}\hline\noalign{\smallskip}
731 & 8.0 &-16.6 &.IB.9\$.&1.65 &23.0 &1 &57 &257 &638 &50 &63 &73 &74 &74 &6.98 &0.35 &0.34 &0.20 \\
\noalign{\vspace{1.5pt}}
1281 & 5.5 &-16.2 &.S..8..&1.66 &22.7 &2 &90 &220 &157 &32 &53 &57 &-- &57 &5.20 &0.73 &0.15 &0.14 \\
\noalign{\vspace{1.5pt}}
2023 & 10.1 &-17.2 &.I..9$\ast$.&1.22 &21.8 &2 &19 &315 &603 &24 &41 &59 &-- &59 &3.67 &0.77 &0.90 &0.90 \\
\noalign{\vspace{1.5pt}}
2034 & 10.1 &-17.4 &.I..9..&1.29 &21.6 &1 &19 &342 &578 &29 &37 &40 &45 &47 &5.88 &0.39 &0.18 &0.29 \\
\noalign{\vspace{1.5pt}}
2053 & 11.8 &-16.0 &.I..9..&1.09 &22.5 &3 &40 &323 &1025 &62 &85 &97 &-- &99 &3.43 &0.45 &0.32 &0.33 \\
\noalign{\vspace{1.5pt}}
2455 & 7.8 &-18.5 &.IBS9..&1.06 &19.8 &2 &51 &263 &371 &23 &32 &42 &57 &61 &4.54 &0.45 &0.68 &0.87 \\
\noalign{\vspace{1.5pt}}
3137 & 18.4 &-18.7 &.S?....&1.95 &24.2 &2 &90 &74 &993 &31 &66 &94 &103 &100 &30.8 &1.08 &0.88 &0.20 \\
\noalign{\vspace{1.5pt}}
3371 & 12.8 &-17.7 &.I..9$\ast$.&3.09 &23.3 &1 &49 &133 &818 &51 &75 &81 &-- &86 &10.2 &0.55 &0.18 &0.26 \\
\noalign{\vspace{1.5pt}}
3698 & 8.5 &-15.4 &.I..9$\ast$.&0.41 &21.2 &3 &40 &334 &420 &51 &75 &-- &-- &27 &1.24 &0.55 &-- &-- \\
\noalign{\vspace{1.5pt}}
3711 & 8.6 &-17.8 &.IB.9..&0.96 &20.9 &2 &60 &281 &433 &79 &92 &93 &-- &95 &3.75 &0.21 &0.03 &0.05 \\
\noalign{\vspace{1.5pt}}
3817 & 8.7 &-15.1 &.I..9$\ast$.&0.71 &22.5 &1 &30 &0 &436 &16 &29 &38 &-- &45 &2.53 &0.82 &0.70 &0.78 \\
\noalign{\vspace{1.5pt}}
3851 & 3.4 &-16.9 &.IBS9..&1.31 &22.6 &2 &59 &42 &104 &31 &50 &55 &54 &60 &5.85 &0.70 &0.21 &0.22 \\
\noalign{\vspace{1.5pt}}
3966 & 6.0 &-14.9 &.I..9..&0.58 &22.2 &2 &41 &270 &364 &38 &46 &47 &-- &50 &2.18 &0.28 &0.07 &0.12 \\
\noalign{\vspace{1.5pt}}
4173 & 16.8 &-17.8 &.I..9$\ast$.&4.46 &24.3 &2 &40 &168 &865 &36 &49 &-- &-- &57 &12.2 &0.46 &-- &0.44 \\
\noalign{\vspace{1.5pt}}
4274 & 6.6 &-17.9 &.SBS9P.&0.71 &20.7 &4 &00 &0 &447 &-- &-- &-- &-- &-- &-- &-- &-- &-- \\
\noalign{\vspace{1.5pt}}
4278 & 10.5 &-17.7 &.SBS7$\ast$\/&1.42 &22.5 &2 &90 &353 &559 &29 &53 &65 &77 &86 &7.64 &0.89 &0.51 &0.48 \\
\noalign{\vspace{1.5pt}}
4305 & 3.4 &-16.8 &.I..9..&1.04 &21.7 &2 &40 &172 &156 &24 &37 &33 &33 &33 &5.44 &0.62 &-0.26 &-0.11 \\
\noalign{\vspace{1.5pt}}
4325 & 10.1 &-18.1 &.SAS9\$.&1.63 &21.6 &1 &41 &51 &523 &73 &89 &92 &-- &92 &5.88 &0.29 &0.08 &0.04 \\
\noalign{\vspace{1.5pt}}
4499 & 13.0 &-17.8 &.SX.8..&1.49 &21.5 &1 &50 &140 &691 &38 &58 &66 &71 &74 &8.51 &0.63 &0.30 &0.24 \\
\noalign{\vspace{1.5pt}}
4543 & 30.3 &-19.2 &.SA.8..&4.00 &22.0 &2 &46 &331 &1956 &59 &61 &64 &66 &67 &17.6 &0.03 &0.12 &0.13 \\
\noalign{\vspace{1.5pt}}
5272 & 6.1 &-15.1 &.I..9..&0.60 &22.4 &2 &59 &97 &525 &19 &33 &-- &-- &45 &1.78 &0.81 &-- &0.83 \\
\noalign{\vspace{1.5pt}}
5414 & 10.0 &-17.6 &.IXS9..&1.49 &21.8 &1 &55 &220 &607 &35 &52 &-- &-- &61 &4.36 &0.57 &-- &0.42 \\
\noalign{\vspace{1.5pt}}
5721 & 6.7 &-16.6 &.SX.7?.&0.45 &20.2 &2 &61 &279 &542 &39 &54 &68 &78 &79 &7.31 &0.46 &0.59 &0.18 \\
\noalign{\vspace{1.5pt}}
5829 & 9.0 &-17.3 &.I..9..&1.94 &22.4 &2 &34 &197 &627 &34 &48 &60 &-- &69 &7.20 &0.49 &0.59 &0.59 \\
\noalign{\vspace{1.5pt}}
5846 & 13.2 &-16.1 &.I..9..&1.18 &22.9 &2 &30 &305 &1019 &30 &46 &51 &-- &51 &3.84 &0.63 &0.24 &0.19 \\
\noalign{\vspace{1.5pt}}
5918 & 7.7 &-15.4 &.I..9$\ast$.&1.27 &24.2 &2 &46 &239 &337 &30 &38 &42 &-- &45 &4.48 &0.38 &0.24 &0.26 \\
\noalign{\vspace{1.5pt}}
5986 & 8.7 &-18.6 &.SBS9.\/&2.18 &21.4 &2 &90 &219 &624 &76 &112 &116 &110 &125 &12.0 &0.56 &0.10 &0.11 \\
\noalign{\vspace{1.5pt}}
6446 & 12.0 &-18.4 &.SA.7..&1.87 &21.4 &1 &52 &188 &647 &58 &70 &75 &78 &80 &9.60 &0.27 &0.17 &0.15 \\
\noalign{\vspace{1.5pt}}
6628 & 15.3 &-18.9 &.SA.9..&2.70 &21.8 &2 &20 &204 &851 &40 &42 &-- &-- &42 &7.79 &0.08 &-- &0.01 \\
\noalign{\vspace{1.5pt}}
6817 & 4.02 &-15.2 &.I..9..&0.97 &23.1 &4 &60 &140 &243 &-- &-- &-- &-- &-- &-- &-- &-- &-- \\
\noalign{\vspace{1.5pt}}
6956 & 15.7 &-17.2 &.SBS9..&2.37 &23.4 &3 &30 &118 &917 &34 &49 &-- &-- &58 &5.71 &0.52 &-- &0.82 \\
\noalign{\vspace{1.5pt}}
7047 & 3.5 &-15.2 &.IA.9..&0.48 &21.6 &2 &46 &34 &208 &16 &26 &36 &-- &38 &1.53 &0.70 &0.86 &0.85 \\
\noalign{\vspace{1.5pt}}
7125 & 19.5 &-18.3 &.S..9..&1.73 &22.8 &2 &90 &84 &1078 &30 &46 &51 &56 &70 &21.3 &0.61 &0.23 &0.23 \\
\noalign{\vspace{1.5pt}}
7151 & 3.5 &-15.7 &.SXS6\$\/&0.54 &22.3 &2 &90 &282 &267 &44 &64 &68 &72 &76 &2.80 &0.53 &0.12 &0.18 \\
\noalign{\vspace{1.5pt}}
7199 & 3.5 &-15.1 &.IA.9..&0.37 &21.4 &4 &50 &192 &166 &-- &-- &-- &-- &-- &-- &-- &-- &-- \\
\noalign{\vspace{1.5pt}}
7232 & 3.5 &-15.3 &.I..9P.&0.33 &20.2 &2 &59 &0 &230 &19 &32 &43 &-- &44 &1.02 &0.71 &0.76 &0.76 \\
\noalign{\vspace{1.5pt}}
7261 & 9.1 &-17.7 &.SBS8..&1.69 &21.9 &2 &30 &262 &856 &66 &74 &-- &-- &76 &4.63 &0.16 &-- &0.11 \\
\noalign{\vspace{1.5pt}}
\noalign{\smallskip}\hline
\end{tabular}
\end{table*}

\addtocounter{table}{-1}
\begin{table*}[ht]
\setlength{\tabcolsep}{1.75mm}
\caption[]{-- Continued}
{\centering
\begin{tabular}{rrrcrrrrrrrrrrrrrrr}
\noalign{\vspace{-0.2cm}}\hline\noalign{\smallskip}
 UGC &$D_\mathrm{D}$ &$M_R$ &Type&$h$ &$\mu_0^R$ &$q$ &$i$ &P.A. &$\varv_\mathrm{sys}$ &$\varv_1$ &$\varv_2$ &$\varv_3$ &$\varv_4$ &$\varv_\mathrm{last}$ &$r_\mathrm{last}$ &$S_\mathrm{(1,2)}$ &$S_\mathrm{(2,3)}$ &$S_\mathrm{(2,L)}$ \\
& Mpc & mag & & kpc & mag${\prime\prime}^{-2}$ & & $^\circ$ & $^\circ$ & \multicolumn{6}{c}{-- \hfil -- \hfil -- \hfil -- \hfil \kms \hfil -- \hfil -- \hfil -- \hfil --} & kpc &&&\\
\noalign{\smallskip}
 (1) & (2) & (3) & (4) & (5) & (6) & (7) & (8) & (9) & (10) & (11) & (12) & (13) & (14) & (15) & (16) & (17) & (18) & (19) \\
\noalign{\smallskip}\hline\noalign{\smallskip}
7278 & 3.5 &-18.3 &.IXS9..&0.93 &20.2 &2 &30 &74 &292 &47 &68 &75 &79 &81 &5.80 &0.51 &0.26 &0.15 \\
\noalign{\vspace{1.5pt}}
7323 & 8.1 &-18.9 &.SXS8..&2.20 &21.2 &1 &47 &38 &518 &49 &78 &-- &-- &86 &5.89 &0.66 &-- &0.33 \\
\noalign{\vspace{1.5pt}}
7399 & 8.4 &-17.1 &.SBS8..&0.79 &20.7 &1 &55 &320 &535 &55 &79 &89 &92 &109 &11.0 &0.54 &0.27 &0.16 \\
\noalign{\vspace{1.5pt}}
7408 & 8.4 &-16.6 &.IA.9..&0.99 &21.9 &4 &45 &275 &462 &-- &-- &-- &-- &-- &-- &-- &-- &-- \\
\noalign{\vspace{1.5pt}}
7490 & 8.5 &-17.4 &.SA.9..&1.19 &21.3 &3 &20 &113 &464 &57 &71 &-- &-- &79 &3.09 &0.32 &-- &0.42 \\
\noalign{\vspace{1.5pt}}
7524 & 3.5 &-18.1 &.SAS9$\ast$.&2.58 &22.2 &1 &46 &327 &320 &58 &75 &83 &-- &79 &7.89 &0.38 &0.24 &0.12 \\
\noalign{\vspace{1.5pt}}
7559 & 3.2 &-13.7 &.IB.9..&0.67 &23.8 &1 &61 &137 &216 &21 &31 &33 &-- &33 &2.10 &0.57 &0.15 &0.15 \\
\noalign{\vspace{1.5pt}}
7577 & 3.5 &-15.6 &.I..9..&0.84 &22.5 &1 &63 &128 &196 &8 &13 &-- &-- &18 &2.29 &0.78 &-- &0.92 \\
\noalign{\vspace{1.5pt}}
7603 & 6.8 &-16.9 &.SBS7?\/&0.90 &20.8 &1 &78 &197 &644 &30 &47 &59 &60 &64 &5.94 &0.65 &0.54 &0.25 \\
\noalign{\vspace{1.5pt}}
7608 & 8.4 &-16.4 &.I..9..&1.24 &22.6 &1 &25 &257 &535 &32 &51 &62 &-- &69 &4.89 &0.67 &0.48 &0.46 \\
\noalign{\vspace{1.5pt}}
7690 & 7.9 &-17.0 &.I..9$\ast$.&0.54 &19.9 &2 &41 &41 &536 &44 &59 &61 &59 &56 &4.02 &0.41 &0.10 &-0.04 \\
\noalign{\vspace{1.5pt}}
7866 & 4.8 &-15.2 &.IXS9..&0.57 &22.1 &2 &44 &338 &354 &17 &25 &28 &32 &33 &2.44 &0.54 &0.32 &0.39 \\
\noalign{\vspace{1.5pt}}
7916 & 8.4 &-14.9 &.I..9..&1.81 &24.4 &2 &74 &0 &603 &21 &33 &-- &-- &36 &4.28 &0.62 &-- &0.57 \\
\noalign{\vspace{1.5pt}}
7971 & 8.4 &-17.1 &.S..9$\ast$.&0.99 &21.3 &2 &38 &35 &470 &22 &37 &45 &-- &45 &3.06 &0.73 &0.51 &0.50 \\
\noalign{\vspace{1.5pt}}
8188 & 4.7 &-17.4 &.SAS9..&1.16 &21.3 &3 &20 &0 &318 &28 &-- &-- &-- &45 &2.05 &-- &-- &-- \\
\noalign{\vspace{1.5pt}}
8201 & 4.9 &-15.8 &.I..9..&0.80 &21.9 &4 &63 &90 &30 &-- &-- &-- &-- &-- &-- &-- &-- &-- \\
\noalign{\vspace{1.5pt}}
8286 & 4.8 &-17.2 &.S..6$\ast$\/&1.23 &20.9 &2 &90 &28 &407 &66 &76 &82 &83 &84 &5.94 &0.20 &0.18 &0.12 \\
\noalign{\vspace{1.5pt}}
8331 & 5.9 &-15.1 &.IA.9..&0.70 &22.9 &4 &90 &139 &262 &4 &7 &11 &13 &21 &4.29 &0.96 &0.89 &0.92 \\
\noalign{\vspace{1.5pt}}
8490 & 4.9 &-17.3 &.SAS9..&0.66 &20.5 &1 &50 &175 &201 &48 &66 &74 &77 &78 &10.7 &0.45 &0.29 &0.08 \\
\noalign{\vspace{1.5pt}}
8550 & 5.3 &-15.6 &.SBS7?\/&0.67 &22.0 &2 &90 &166 &358 &39 &49 &51 &53 &58 &4.24 &0.33 &0.10 &0.14 \\
\noalign{\vspace{1.5pt}}
8683 & 12.6 &-16.7 &.I..9..&1.37 &22.5 &3 &28 &349 &659 &17 &31 &-- &-- &31 &2.75 &0.87 &-- &0.84 \\
\noalign{\vspace{1.5pt}}
8837 & 5.1 &-15.7 &.IBS9.\/&1.63 &23.2 &2 &80 &22 &135 &31 &52 &-- &-- &54 &3.71 &0.77 &-- &0.22 \\
\noalign{\vspace{1.5pt}}
9128 & 4.4 &-14.3 &.I..9..&0.36 &21.9 &4 &40 &135 &155 &-- &-- &-- &-- &-- &-- &-- &-- &-- \\
\noalign{\vspace{1.5pt}}
9211 & 12.6 &-16.2 &.I..9$\ast$.&1.32 &22.6 &1 &44 &287 &686 &35 &53 &63 &66 &65 &8.25 &0.59 &0.41 &0.17 \\
\noalign{\vspace{1.5pt}}
9992 & 10.4 &-15.9 &.I..9..&0.75 &22.2 &2 &30 &35 &428 &28 &31 &33 &33 &34 &3.78 &0.15 &0.13 &0.10 \\
\noalign{\vspace{1.5pt}}
10310 & 15.6 &-17.9 &.SBS9..&1.66 &22.0 &1 &34 &199 &718 &44 &65 &70 &72 &74 &9.08 &0.57 &0.17 &0.13 \\
\noalign{\vspace{1.5pt}}
11557 & 23.8 &-19.7 &.SXS8..&3.10 &21.0 &2 &30 &274 &1386 &53 &76 &83 &-- &85 &10.4 &0.52 &0.20 &0.20 \\
\noalign{\vspace{1.5pt}}
11707 & 15.9 &-18.6 &.SA.8..&4.30 &23.1 &1 &68 &57 &904 &68 &89 &94 &-- &100 &15.0 &0.39 &0.14 &0.20 \\
\noalign{\vspace{1.5pt}}
11861 & 25.1 &-20.8 &.SX.8..&6.06 &21.4 &1 &50 &219 &1481 &111 &147 &-- &-- &153 &16.4 &0.41 &-- &0.12 \\
\noalign{\vspace{1.5pt}}
12060 & 15.7 &-17.9 &.IB.9..&1.76 &21.6 &1 &40 &183 &883 &61 &72 &74 &75 &74 &10.3 &0.24 &0.07 &0.03 \\
\noalign{\vspace{1.5pt}}
12632 & 6.9 &-17.1 &.S..9$\ast$.&2.57 &23.5 &1 &46 &36 &421 &58 &69 &74 &-- &76 &8.53 &0.27 &0.16 &0.17 \\
\noalign{\vspace{1.5pt}}
12732 & 13.2 &-18.0 &.S..9$\ast$.&2.21 &22.4 &1 &39 &14 &749 &53 &68 &76 &81 &98 &15.4 &0.35 &0.27 &0.29 \\
\noalign{\vspace{1.5pt}}
\noalign{\smallskip}\hline
\noalign{\vspace{4pt}}
\end{tabular}
} 
\hbox to\hsize{\hfil\vbox{\hsize=\hsize
(1) UGC number;
(2) the adopted distance, from Paper~II;
(3) the absolute $R$-band as determined in paper~II;
(4) the morphological type according to the RC3 catalog (de Vaucouleurs et al.\ 1991), using the same coding;
(5) and (6) the disk scale length and the $R$-band central disk surface
brightness as determined in Paper~II;
(7) the rotation curve quality $q$ as
defined in Section~\ref{secrc};
(8) and (9) the adopted inclination and
position angles;
(10) the systemic velocity 
derived from the velocity field;
(11) to (14) the rotation velocities at
one, two, three and four disk scale lengths;
(15) and (16) the rotation velocity at the
last measured point and the radius of the last measured point;
(17), (18) and (19) the logarithmic shape
of the rotation curve between one and two disk scale lengths, between
two and three disk scale lengths, and between two disk scale lengths
and the last measured point.}\hfil}
\end{table*}

\subsection{Asymmetric drift}

The stability of a galaxy against gravitational collapse is provided
by rotation and pressure gradients. To derive the true circular
velocity that reflects the underlying mass distribution, the observed
rotation curve should in principle be corrected for the effects of
pressure. The difference between the observed and the true circular
velocity can be derived from the Jeans equations (Binney \& Tremaine
1987, eq.~4-33) and is given by:
\begin{equation}
\varv_\varphi^2-\varv_c^2={R\over{\rho}}{\partial P\over{\partial R}},
\label{eqasdrift}
\end{equation}
where $\varv_\varphi$ is the observed rotation velocity, $\varv_c$ the
true circular velocity, $R$ the radius, $\rho$ the density at that
radius, and $P$ the pressure. The pressure can also be written as
$\rho\sigma^2$, where $\sigma$ is the one-dimensional random velocity
of the gas. Inserting this in Eq.~\ref{eqasdrift}, one obtains:
\begin{equation}
\varv_c^2=\varv_\varphi^2-\sigma^2\left({\partial\ln\rho\over{\partial
\ln R}}+{\partial\ln\sigma^2\over{\partial\ln R}}\right).
\label{eqadcorr}
\end{equation}
We found that the velocity dispersions for the late-type dwarf
galaxies in this sample were almost constant with radius, hence the
second term within parentheses in Eq.~\ref{eqadcorr} can be ignored.
If we assume that the scale height is approximately constant with
radius as well (as is suggested by measurements of the thickness of
the \HI\ layer in the edge-on dwarf galaxy NGC~5023, Bottema et
al.\ 1986), and if we assume that the \HI\ is the dominant gaseous
component, the volume density is proportional to the measured
\HI\ column density.

With these assumptions, the correction for asymmetric drift is smaller
than 3 \kms\ at all radii, for all but three galaxies. For most
galaxies the maximum correction occurs at the last measured point,
where the amplitude of the rotation curve is highest. In the inner
parts the correction is negligible, i.e., smaller than 1 \kms, for
90\% of the galaxies, and smaller than 2 \kms\ for all except three
galaxies.  These are the same galaxies for which the maximum
correction is larger than 3 \kms: UGC~7278, UGC~7577 and UGC~8331.

Because the corrections for asymmetric drift as calculated above are
small for the vast majority of the galaxies in our sample, and because
of the uncertainties in the corrections due to ignoring the
contribution of other gas components, we have not corrected the
rotation curves for asymmetric drift, and we assumed that our derived
rotation curves corrected for inclination represent the circular
velocity.

\subsection{Parametrized rotation curve shapes}

The radial scale of the rotation curves derived above is given in
arcseconds. In order to facilitate the comparison of the rotation
curves between galaxies and the comparison between the rotation curves
and the light profiles, the rotation curves have been resampled in
units of disk scale lengths.  Table~\ref{tabsample} lists the rotation
velocities at one, two, three, and four disk scale lengths, as
$\varv_1$ to $\varv_4$. In addition, it lists the velocity at the last
measured point of the rotation curve $\varv_\mathrm{last}$ and the
corresponding radius $r_\mathrm{last}$.

We have also calculated the logarithmic slope of the rotation
curves. The logarithmic slope is defined as a
straight line fit to the rotation curve in logarithmic coordinates
between two radii:
\begin{equation}
S_\mathrm{(a,b)} = {\Delta\log\varv\over{\Delta\log R}} =
{\log(\varv_b/\varv_a)\over{\log(R_b/R_a)}}
\end{equation}
For a flat rotation curve, $S_\mathrm{(a,b)}$ equals zero, and for a
solid body rotation curve $S_\mathrm{(a,b)}$ equals unity.  In
Table~\ref{tabsample}, we list $S_\mathrm{(1,2)}^{\,h}$,
$S_\mathrm{(2,3)}^{\,h}$ and $S_\mathrm{(2,L)}^{\,h}$, where the superscript
$h$ refers to the fact that the logarithmic slopes have been
calculated from the rotation curves expressed in units of disk scale
length. The subscript numbers between parentheses refer to the number
of scale lengths, and the `L' refers to the radius of the last
measured point.

\section{Notes on individual galaxy rotation curves}
\label{secnotes}

Because of the relatively large size of the sample, we restrict
ourselves here to a discussion of properties relevant to the
derivation of the rotation curves.

\notes{2034} has a small kinematic warp. The position angle changes by
about $35^\circ$. Because the galaxy is almost face-on, the variation
of inclination angle with radius is difficult to determine.

\notes{2053} has its kinematic major axis and its morphological major
axis almost at right angles. Its kinematics are reminiscent
of that expected for a bar potential. The bright optical image appears
to have a bar-like shape, but no faint disk is detected, neither in
the optical nor in the \HI.

\notes{3711} has twisted isovelocity contours. Because this galaxy is
poorly resolved, it is difficult to determine the cause; the twisted
contours may indicate a kinematic warp, but may also be related to the
bar-shaped morphology.

\notes{3851} has an irregular kinematic structure. On the north side,
the isovelocity contours curve strongly, suggesting a nearly flat
rotation curve. On the south side, the isovelocity contours are almost
straight, and not symmetric with respect to the major axis, but
instead run more or less north to south.

\notes{4173} has a faint optical disk surrounding a central bar-like
structure. The kinematical major axis is aligned with the faint
disk. The bar is not clearly reflected in the velocity field.

\notes{4274} is asymmetric and \HI\ is only detected in one clump on
the receding side. No rotation curve has been derived.

\notes{4305} has previously been found to have a violent interstellar
medium (e.g., Puche et al.\ 1992). The observations presented here
also show signs of holes, shells, and small-scale kinematic
irregularities, making an accurate determination of the rotation curve
difficult.

\notes{5272} displays a misalignment between the optical and the
kinematic major axes.

\notes{5721} has the most extended rotation curve of this sample when
the radius is expressed in optical disk scale lengths. The last
measured point of the rotation curve is at $16.3h$. The inner points
of this rotation curve are uncertain, because of insufficient angular
resolution.

\notes{5986} is warped and has a strong asymmetry between the
approaching and the receding side, both in kinematics and in density.
A possible cause is an interaction with a small companion to the south
west.

\notes{6817} has a clumpy and irregular \HI\ distribution and
kinematics. No rotation curve could be derived.

\notes{7199} has little \HI\ and is poorly resolved by the $30''$
beam. No rotation curve could be derived.

\notes{7261} is strongly barred. The S-shape distortion in the
kinematics that is associated with the bar is clearly visible in the
velocity field.

\notes{7278} has a strong kinematic warp, with the inclination
changing from about $30^\circ$ in the central parts to about
$10^\circ$ in the outer parts, and the position angle changing from
$74^\circ$ to $90^\circ$. Because of the low inclination, the actual
values of the inclination, and hence the amplitude of the rotation
curve, are uncertain. The \HI\ in the central parts has a high
velocity dispersion and seems to counterrotate with respect to the
outer parts.

\notes{7399} has an extended \HI\ rotation curve, yet the inner rise
of the rotation curve is poorly resolved because of the small size of
the optical galaxy.

\notes{7408} has little \HI. No rotation curve could be derived.

\notes{7577} has the lowest rotation amplitude of the galaxies in this
sample, yet its velocity field is still regular in appearance.

\notes{7524} is a prototype of a kinematically lopsided galaxy. The
rotation curve on the receding side continues to rise, whereas on the
approaching side the rotation curve quickly reaches a flat part
(see also Swaters et al.\ 1999).

\notes{7603} has a high inclination of $78^\circ$. As a result, the
inner parts are effectively seen edge-on by the $30''$ beam, making
the inner rotation curve more uncertain.

\notes{7690} appears to have a declining rotation curve. However, this
galaxy also has a modest kinematic warp, with an inclination changing
from about $40^\circ$ in the central parts to $30^\circ$ in the outer
parts. Because the inclination as a function of radius is uncertain,
so is the shape of the rotation curve.

\notes{8201} has most of its \HI\ located in two clumps that are
roughly located on the major axis, but that show clear difference in
velocity. No rotation curve could be derived.

\notes{8331} has \HI\ in one clump in the center and two clumps at
larger radii, making it impossible to measure the rotation curve.

\notes{8490} is strongly warped. The position angle changes from
$160^\circ$ to $225^\circ$, the inclination changes from $50^\circ$ to
$30^\circ$, and then rises again to about $50^\circ$. This system was
studied in detail by Sicotte et al.\ (1996) and Sicotte \& Carignan
(1997). The outer half of the rotation curve may be more uncertain
than indicated by the error bars, due to uncertainties in the
orientation angles.

\notes{9128} is hardly resolved by the $30''$ beam, and hence no
rotation curve could be determined.

\section{Comparison to \Halpha\ rotation curves}
\label{seccomp}

\begin{figure*}[p]
\resizebox{\hsize}{!}{\includegraphics{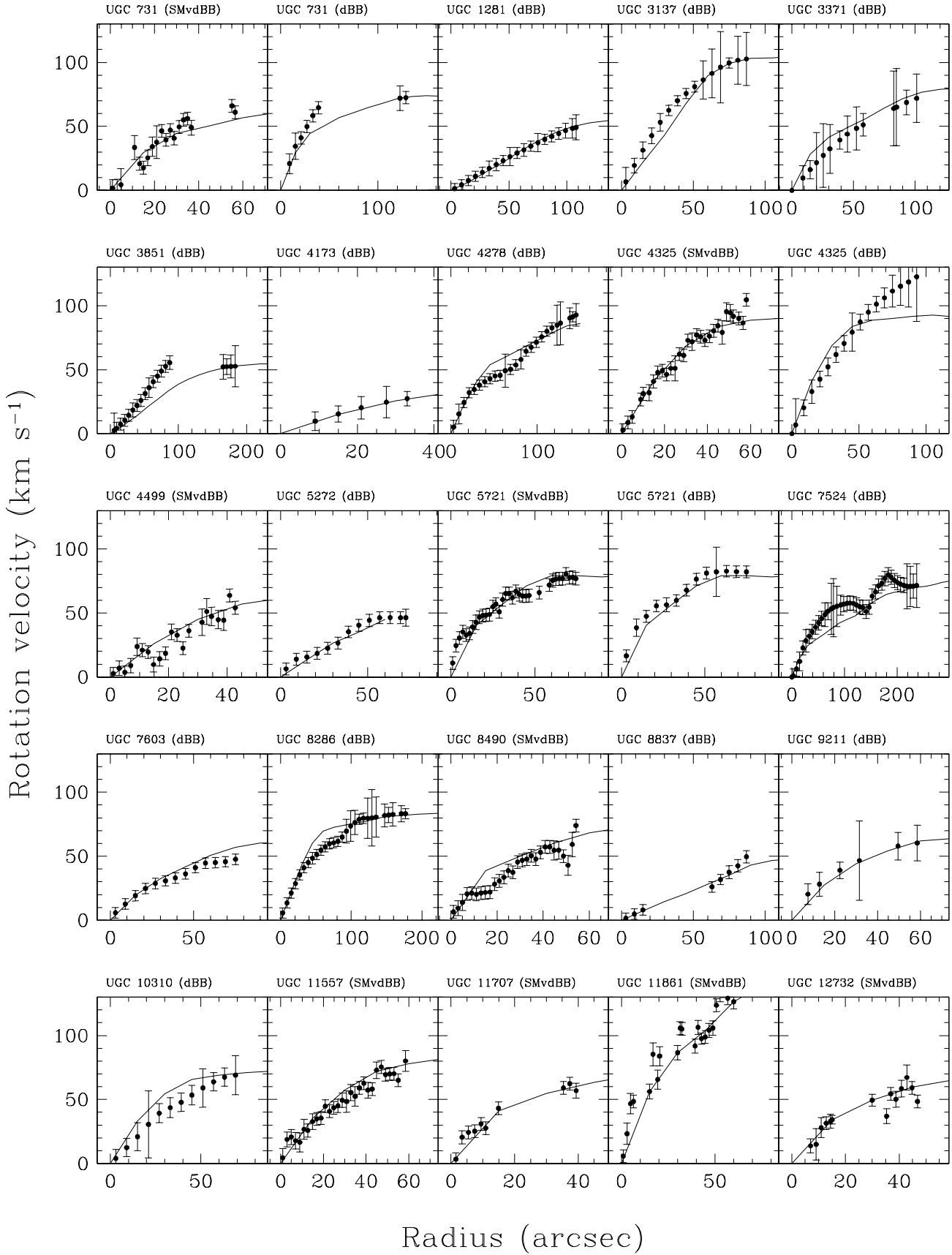}}
\caption{Comparison of the \Halpha\ rotation curves (dots with error
bars) with \HI\ rotation curves (solid lines).}
\label{figcomphiha}
\end{figure*}

In Fig.~\ref{figcomphiha} we compare the \HI\ rotation curves
presented here to high-resolution rotation curves based on long-slit
H$\alpha$ spectroscopy from the studies of dBB and SMvdBB. There is a
significant overlap between the samples of those two studies and the
sample presented here because they are both based on the sample of
Swaters (1999).

dBB have derived their rotation curves by combining their
\Halpha\ data with the S99 \HI\ data, and then fitting a spline to the
combined data to obtain a smooth curve. The dBB data plotted in
Fig.~\ref{figcomphiha} are the points from their spline-interpolated
rotation curves over the range in radii where they have \Halpha\ data.

The comparison between the \HI\ and H$\alpha$ rotation curves shows
that they generally agree fairly well, although there are significant
differences in individual galaxies. Below we discuss the comparison of
the \HI\ and \Halpha\ rotation curves in detail.

\notes{731}. Observed in \Halpha\ both by SMvdBB and dBB. The
agreement between the \HI\ data and the SMVdBB data is good, whereas
the \Halpha\ rotation curve by de dBB rises more steeply than the
\HI\ rotation curve. The difference between these two \Halpha\ curves
can be explained by the lopsided kinematics of this galaxy (see
Swaters et al.\ 1999). Its approaching side has a more steeply rising
rotation curve than the receding side. The \Halpha\ data in SMvdBB
samples both sides, whereas in the dBB data, which is less deep,
\Halpha\ emission is biased to the approaching side.

\notes{1281}. The \HI\ and \Halpha\ rotation curves for this edge-on
galaxy are in excellent agreement.

\notes{3137}. The \Halpha\ rotation curve rises more steeply than the \HI\
one. This galaxy is edge-on, and has a rotation curve that rises
steeply with respect to the size of the \HI\ beam. This, in
combination with its inclination, makes the \HI\ rotation curve hard to
determine.

\notes{3371}. The \HI\ rotation curve rises more steeply than the
\Halpha\ rotation curve. The reason for this difference is
unclear. Swaters et al.\ (2003b) obtained an H$\alpha$ rotation curve
derived from integral field spectroscopy and find a good match between
their \Halpha\ rotation curve and our \HI\ rotation curve.

\notes{3851}. As noted above, this galaxy has an irregular kinematic
structure (see also Hunter et al. 2002). The difference between the
\HI\ and \Halpha\ curves could be caused by the fact that the \HI\
rotation curve is derived from the entire velocity field, whereas the
\Halpha\ rotation curve is derived from a single, $1''$ wide slice
along the major axis.

\notes{4173}. The \HI\ and \Halpha\ rotation curves are in excellent
agreement. However, the \Halpha\ data by dBB is of low quality, and
their interpolated rotation curve is dominated by the \HI\ data.

\notes{4278}. The \HI\ rotation curve of this edge-on galaxy rises
more steeply than the \Halpha\ one.

\begin{figure*}[ht]
\resizebox{\hsize}{!}{\includegraphics{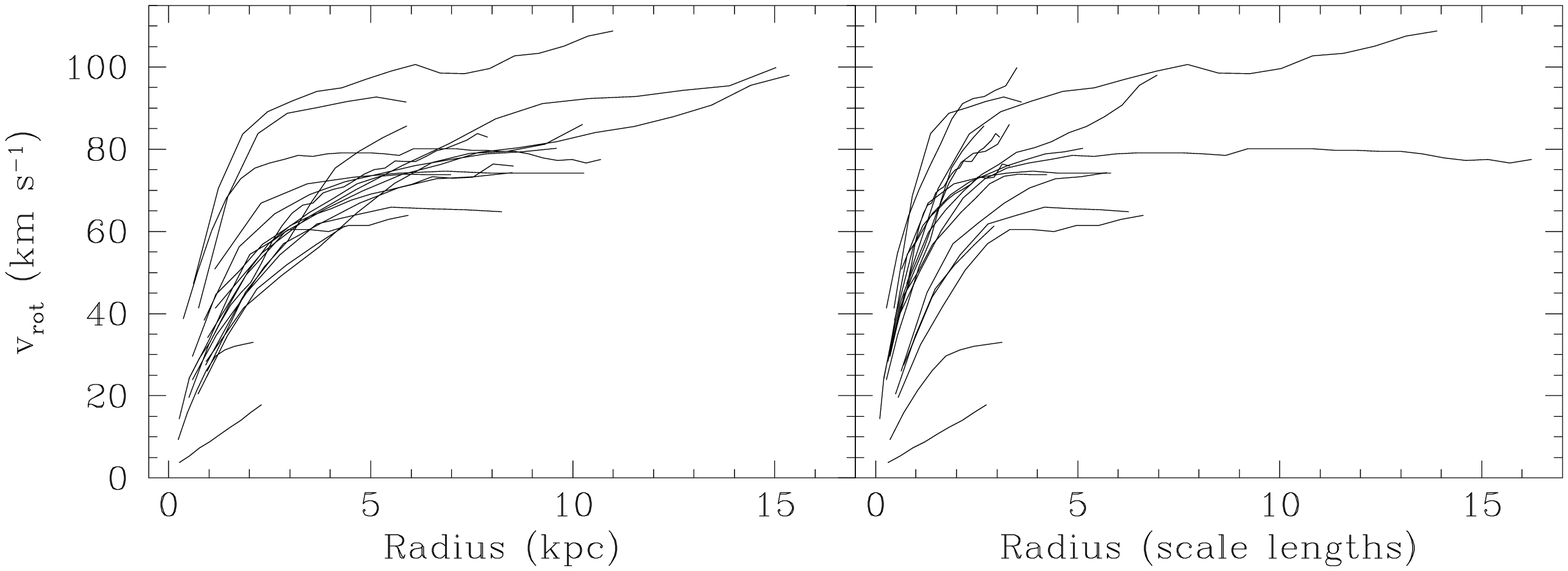}}
\caption{Compilation of the rotation curves of the galaxies in the high
  quality sample. The left panel shows the rotation curves with radii
  expressed in kpc, in the right panel the radii are
  expressed in units of disk scale lengths. }
\label{figallrc}
\end{figure*}

\notes{4325}. This galaxy has been observed by both SMvdBB and
dBB. The SMvdBB rotation curve is in good agreement with the \HI\
curve, whereas the dBB curve rises more slowly and reaches a higher
amplitude. The latter is difficult to understand, as the amplitude of
the rotation curve can be determined accurately from the
two-dimensional velocity field. A possible explanation is that dBB
picked up a localized disturbance in the kinematics.

\notes{4499}. The \Halpha\ rotation curve has considerable scatter,
but there is good general agreement between the \HI\ and \Halpha\
curves.

\notes{5272}. The \HI\ and \Halpha\ are in excellent agreement.

\notes{5721}. This galaxy has been observed by both SMvdBB and
dBB. Both \Halpha\ curves agree well with each other, and both show a
steep rise in the inner $10''$ that is not seen in the \HI\ data.

\notes{7524}. This is the best resolved galaxy in our sample, yet the
\HI\ and \Halpha\ rotation curves show significant differences. The
velocity field of this galaxy shows a great deal of structure,
including kinematic lopsidedness (see Swaters et al.\ 1999). The
structure seen in the dBB curve is also seen along the major axis of
the \HI\ data.

\notes{7603}. The \HI\ and \Halpha\ data are in good agreement,
although the outer points on the \Halpha\ curve fall below the \HI\
curve. As this galaxy has a fairly high inclination ($76^\circ$), and
a steeply rising rotation curve that is not well resolved by the \HI\
beam, it is likely that the difference is due to residual affects of
beam smearing.

\notes{8286}. The \HI\ and \Halpha\ rotation curves for this edge-on
galaxy agree except in the region where the rotation curve turns over.

\notes{8490}. The \Halpha\ rotation curve falls well below the \HI\
rotation curve near a radius of about $15''$. As this corresponds to
the first point on a steeply rising rotation curve, it is possible
that this difference is the result of an incorrect correction for beam
smearing.

\notes{8837}. The \HI\ and the poorly sampled \Halpha\ rotation curves
are in good agreement.

\notes{9211}. The \HI\ and \Halpha\ rotation curves are in good agreement.

\notes{10310}. The \Halpha\ rotation curve falls below the \HI\ rotation
curve. The reason for this difference is unclear.

\notes{11557}. The \HI\ and \Halpha\ rotation curves agree well, but
in the outer parts the \Halpha\ curve falls slightly below the \HI\
curve.

\notes{11707}. Although the \Halpha\ rotation curve is poorly sampled,
the inner rise in the \Halpha\ rotation curve seems somewhat steeper
than that of the \HI\ curve.

\notes{11861}. The \Halpha\ rotation curve has significant scatter,
but seems to rise more steeply in the central regions. In the outer
parts both curves agree well.

\notes{12732}. The \Halpha\ and \HI\ rotation curves are in agreement.

The detailed comparison of \HI\ and \Halpha\ rotation curves shows a
fairly good general agreement between the two sets, indicating that
the method used to derive beam smearing corrected rotation curves is
successful to a large degree. There are, however, several systematic
differences. For example, the \HI\ data are not sensitive to features
in the central rotation curve if they are smaller than the size of the
\HI\ beam (UGC~5721, UGC~11707, UGC~11861). In addition, if the
velocity gradient over the beam is large, the derived \HI\ rotation
curve becomes more uncertain (UGC~3137, UGC~7603, UGC~8490). On the
other hand, the \Halpha\ data may be sensitive to small scale
structure in the velocity field (UGC~731, UGC~3851, UGC~4325,
UGC~7524), because it is derived from a $1''$ wide slice through the
velocity field, whereas the \HI\ rotation curve is derived from the
velocity field as a whole. Systematic differences between the \HI\ and
the \Halpha\ rotation curves are common in edge-on galaxies (UGC~3137,
UGC~4278, UGC~8286). For these galaxies, derivation of the rotation
curves is difficult (see Section~\ref{secrc}), especially for the
\Halpha\ data, for which the typical instrumental velocity
resolution of around 75 \kms\ and the possibility of extinction along
the line of sight make it difficult to determine the extreme velocity
accurately (see e.g. SMvdBB).

For two galaxies, UGC~3371 and UGC~10310 there were no obvious reasons
for the difference between the \HI\ and \Halpha\ rotation curves. The
effects discussed above may all have contributed. Interestingly,
Swaters et al. (2003b) have reobserved UGC~3371. Based on a
high-resolution, two-dimensional \Halpha\ velocity field, they derive
a rotation curve that agrees well with the \HI\ rotation curve. They
suggest that the differences between their rotation curve and the dBB
rotation curve may be due to that fact that they used two-dimensional
data, and the fact that dBB spline-interpolated their data.

Blais-Ouellette et al.\ (2001) and Garrido et al. (2002)
compared rotation curves derived from Fabry-Perot imaging spectroscopy
to those derived from \HI\ synthesis observations, and found good
general agreement between the \Halpha\ and \HI\ rotation curves for
galaxies that are well-resolved. For galaxies that are less
well-resolved the relation between the \HI\ and \Halpha\ shows larger
scatter, leading Garrido et al. (2002) to conclude that corrections
for beam smearing have introduced uncertainties.  Although corrections
for beam smearing may explain part of this scatter, uncertainties in
the \Halpha\ Fabry-Perot data will also play a role e.g., because
Fabry-Perot are less sensitive to the faint \Halpha\ emission, and
hence are biased towards regions of strong star formation, which may
be associated with shocks, spiral arms, and bars.

Although we argue that the effects of beam smearing on \HI\ data can
usually be corrected for the type of galaxies presented here, and that
the \HI\ rotation curves are therefore reliable, it is obvious that
higher angular resolution observations are needed to sample the
detailed rotation curve shapes in the central $~30''$.

\section{Rotation curve shapes}
\label{secrcshapes}

\subsection{The rotation curves}

For the study of the properties of late-type dwarf galaxy rotation
curves, our sample has been divided into two subsamples.  The high
quality sample contains 19 galaxies with inclinations $39^\circ\le
i\le 80^\circ$ and rotation curves with quality $q=1$. A lower limit
was applied because at low inclinations the rotation curve amplitude
and shape are uncertain due to higher inclination uncertainties. The
particular value of $39^\circ$ was chosen so that UGC~12732 is
included in the sample. UGC~11861 was excluded from the analysis below
because of its high luminosity, resulting in a high quality sample of
18 galaxies.  The lower quality sample contains 16 galaxies with
inclinations $39^\circ\le i\le 80^\circ$ and rotation curves with
$q=2$. UGC~4543 was excluded because of its high luminosity. The lower
quality sample thus selected contains 15 dwarf galaxies. 

In Fig.~\ref{figallrc} the rotation curves of the galaxies in the high
quality sample are shown. The left panel shows the rotation curves
with the radii expressed in kpc, in the right panel the radii are
expressed in units of optical disk scale lengths.  A striking feature
of Fig.~\ref{figallrc} is that the rotation curves of the dwarf
galaxies in our sample look similar to those of spiral galaxies, in
particular when the radii are expressed in units of disk scale
lengths: most of the rotation curves rise rapidly in the inner parts
and start to flatten after about two disk scale lengths.  Although
about half of the galaxies in our high-quality sample do not extend
significantly beyond three disk scale lengths, those that do have
rotation curves that are fairly flat, including amplitudes as low as
60 \kms.  Thus, it seems that whether the flat part of a rotation
curve is reached or not depends more on the radial extent of the
rotation curve than on its amplitude. None of the galaxies in our
sample have declining rotation curves, even though some of the
rotation curves extend to beyond 10 disk scale lengths.

To compare the properties of the rotation curves of the late-type
dwarf galaxies in our sample to the rotation curve properties of other
galaxies, we have constructed a comparison sample of extended \HI\
rotation curves. We used the \HI\ rotation curves of Ursa Major
galaxies by Verheijen (1997) and Verheijen \& Sancisi (2001). This
sample spans a range of galaxy types, but has relatively few
early-type galaxies and no galaxies with $M_R>-17$.  We only included
galaxies with at least 5 measured points in the rotation curve to
ensure sufficient resolution. We also included the sample presented in
Broeils (1992a), which consists of a collection of rotation curves
presented in the literature. It includes a large range of galaxy types
and luminosities, spanning from $M_R=-23$ to $M_R=-15$. Most of the
photometric data presented in Broeils (1992a) are in the $B$-band.
These were converted to $R$-band assuming $B-R=1.0$.  A third source
is the study by Spekkens \& Giovanelli (2006), who have studied a
sample of massive late-type spiral galaxies (their average Hubble type
is Sbc). We assumed $R-I=0.5$ to convert their $I$-band photometry to
the $R$-band.  For each of these literature samples we only used
galaxies with inclinations $39^\circ\le i\le 80^\circ$. The combined
literature sample will be used as comparison in the following
discussion.

\begin{table*}[h!t]
\caption{Optical and rotation curve properties of the comparison sample}
\label{tablitsample}
{\centering
\begin{tabular}{lcrrcrrrrrrrrrrrrrr}
\noalign{\vspace{-0.2cm}}\hline\noalign{\smallskip}
Name & Ref. &$D_\mathrm{a}$ &$M_R$ & Type &$h$ &$\mu_0^R$ &$\varv_1$ &$\varv_2$ &$\varv_3$ &$\varv_4$ &$\varv_\mathrm{last}$ &$r_\mathrm{last}$ &$S_\mathrm{(1,2)}$ &$S_\mathrm{(2,3)}$ &$S_\mathrm{(2,L)}$ \\
&& Mpc & mag && kpc & mag${\prime\prime}^{-2}$ & \multicolumn{5}{c}{-- \hfil -- \hfil -- \hfil -- \hfil \kms \hfil -- \hfil -- \hfil -- \hfil --} & kpc &&&\\
\noalign{\smallskip}
 (1) & (2) & (3) & (4) & (5) & (6) & (7) & (8) & (9) & (10) & (11) & (12) & (13) & (14) & (15) & (16) \\
\noalign{\smallskip}\hline\noalign{\smallskip}
NGC 55   &B   & 1.6  &-19.6  &.SBS9$\ast$\/&1.6  &20.5  &   47 &   73 &   85 &   86 &   87 & 7.7 & 0.63 & 0.40 & 0.20 \\
\noalign{\vspace{1.40pt}}
NGC 247  &B   & 2.5  &-19.0  &.SXS7..&2.9  &22.4  &   77 &  104 &   -- &   -- &  108 & 7.4 & 0.43 &   -- &   -- \\
\noalign{\vspace{1.40pt}}
NGC 300  &B   & 1.8  &-18.8  &.SAS7..&2.1  &21.2  &   72 &   89 &   97 &   -- &   93 & 8.0 & 0.30 & 0.22 & 0.08 \\
\noalign{\vspace{1.40pt}}
NGC 801  &B   &79.2  &-22.7  &.S..5..& 12  &20.9  &  223 &  222 &  214 &   -- &  216 &  44 &-0.01 &-0.09 &-0.05 \\
\noalign{\vspace{1.40pt}}
UGC 2259 &B   & 9.8  &-18.0  &.SBS8..&1.3  &21.3  &   70 &   83 &   85 &   88 &   90 & 5.7 & 0.25 & 0.06 & 0.11 \\
\noalign{\vspace{1.40pt}}
NGC 1324 &SG  &  79  &-23.1  &.S..3$\ast$.&5.6  &18.9  &  258 &  282 &  300 &  299 &  264 &  35 & 0.13 & 0.15 &-0.06 \\
\noalign{\vspace{1.40pt}}
UGC 2849 &SG  & 116  &-23.0  &.S..6$\ast$.&6.2  &20.1  &  245 &  277 &  280 &  277 &  267 &  34 & 0.18 & 0.03 &-0.04 \\
\noalign{\vspace{1.40pt}}
UGC 2885 &B   &78.7  &-23.8  &.SAT5..& 13  &21.0  &  275 &  280 &  296 &  298 &  298 &  54 & 0.03 & 0.14 & 0.08 \\
\noalign{\vspace{1.40pt}}
NGC 1560 &B   &   3  &-16.9  &.SAS7.\/&1.3  &22.2  &   42 &   59 &   65 &   74 &   77 & 6.2 & 0.49 & 0.24 & 0.29 \\
\noalign{\vspace{1.40pt}}
UGC 3371 &dBB & 12.8 &-17.74 &.I..9$\ast$.&3.09 &23.3  &   47 &   71 &   -- &   -- &   72 & 6.3 & 0.60 &   -- &   -- \\
\noalign{\vspace{1.40pt}}
NGC 2403 &B   & 3.3  &-20.3  &.SXS6..&2.1  &20.4  &  102 &  127 &  128 &  131 &  134 &14.9 & 0.31 & 0.02 & 0.04 \\
\noalign{\vspace{1.40pt}}
NGC 2841 &B   &  18  &-22.7  &.SAR3$\ast$.&4.6  &20.1  &  322 &  327 &  319 &  298 &  294 &  61 & 0.02 &-0.06 &-0.06 \\
\noalign{\vspace{1.40pt}}
NGC 2903 &B   & 6.4  &-21.0  &.SXT4..&2.0  &19.5  &  209 &  211 &  202 &  198 &  180 &18.2 & 0.01 &-0.11 &-0.10 \\
\noalign{\vspace{1.40pt}}
NGC 2955 &SG  & 105  &-22.8  &PSAR3..&4.9  &19.4  &  244 &  244 &  262 &  241 &  208 &  47 & 0.00 & 0.18 &-0.10 \\
\noalign{\vspace{1.40pt}}
NGC 2998 &B   &67.4  &-22.9  &.SXT5..&5.4  &19.3  &  206 &  215 &  213 &  214 &  198 &  35 & 0.06 &-0.02 &-0.07 \\
\noalign{\vspace{1.40pt}}
UGC 5272 &dBB & 6.1  &-15.11 &.I..9..&0.60 &22.4  &   22 &   40 &   47 &   -- &   46 & 2.1 & 0.87 & 0.37 &0.21 \\
\noalign{\vspace{1.40pt}}
NGC 3109 &B   & 1.7  &-17.8  &.SBS9.\/&1.6  &22.2  &   31 &   50 &   61 &   -- &   66 & 6.2 & 0.67 & 0.51 & 0.42 \\
\noalign{\vspace{1.40pt}}
NGC 3198 &B   & 9.4  &-20.4  &.SBT5..&2.6  &20.6  &  135 &  153 &  155 &  154 &  149 &22.6 & 0.18 & 0.03 &-0.02 \\
\noalign{\vspace{1.40pt}}
UGC 5721 &SMBB& 6.7  &-16.58 &.SX.7?.&0.45 &20.2  &   41 &   57 &   64 &   68 &   77 &2.42 & 0.47 & 0.29 &0.33 \\
\noalign{\vspace{1.40pt}}
UGC 6399 &VS  &15.5  &-18.3  &.S..9$\ast$.&1.8  &22.2  &   53 &   76 &   84 &   -- &   88 & 6.8 & 0.52 & 0.25 & 0.22 \\
\noalign{\vspace{1.40pt}}
UGC 6446 &VS  &15.5  &-18.5  &.SA.7..&2.6  &22.2  &   62 &   74 &   81 &   85 &   80 &13.2 & 0.24 & 0.23 & 0.09 \\
\noalign{\vspace{1.40pt}}
NGC 3726 &VS  &15.5  &-21.3  &.SXR5..&4.4  &20.1  &  124 &  163 &  150 &  150 &  167 &  28 & 0.40 &-0.21 & 0.02 \\
\noalign{\vspace{1.40pt}}
NGC 3877 &VS  &15.5  &-21.2  &.SAS5$\ast$.&2.5  &19.1  &  131 &  161 &  171 &   -- &  169 & 9.8 & 0.31 & 0.14 & 0.07 \\
\noalign{\vspace{1.40pt}}
NGC 3917 &VS  &15.5  &-20.4  &.SA.6$\ast$.&2.4  &20.9  &   80 &  119 &  136 &  137 &  137 &12.8 & 0.57 & 0.33 & 0.15 \\
\noalign{\vspace{1.40pt}}
NGC 3949 &VS  &15.5  &-20.6  &.SAS4$\ast$.&1.4  &19.2  &  118 &  151 &  160 &  167 &  169 & 6.1 & 0.36 & 0.14 & 0.15 \\
\noalign{\vspace{1.40pt}}
NGC 3953 &VS  &15.5  &-21.7  &.SBR4..&3.2  &19.5  &  193 &  220 &  225 &  239 &  215 &13.5 & 0.19 & 0.05 &-0.03 \\
\noalign{\vspace{1.40pt}}
NGC 3972 &VS  &15.5  &-19.8  &.SAS4..&1.7  &19.9  &   79 &  107 &  122 &  135 &  134 & 7.5 & 0.44 & 0.31 & 0.29 \\
\noalign{\vspace{1.40pt}}
UGC 6917 &VS  &15.5  &-19.2  &.SB.9..&2.8  &21.9  &   81 &  101 &  111 &   -- &  111 & 9.0 & 0.32 & 0.23 & 0.21 \\
\noalign{\vspace{1.40pt}}
NGC 3985 &VS  &15.5  &-19.0  &.SBS9$\ast$.&0.8  &19.5  &   42 &   73 &   -- &   -- &   93 & 2.3 & 0.82 &   -- &   -- \\
\noalign{\vspace{1.40pt}}
UGC 6923 &VS  &15.5  &-18.4  &.I..9$\ast$.&1.3  &21.5  &   46 &   72 &   79 &   -- &   81 & 4.6 & 0.66 & 0.23 & 0.20 \\
\noalign{\vspace{1.40pt}}
NGC 3992 &VS  &15.5  &-21.8  &.SBT4..&3.4  &19.2  &   -- &  271 &  264 &  273 &  237 &  30 &   -- &-0.06 &-0.09 \\
\noalign{\vspace{1.40pt}}
UGC 6983 &VS  &15.5  &-19.0  &.SBT6..&3.0  &22.0  &   92 &  102 &  113 &  108 &  109 &13.5 & 0.14 & 0.26 &0.085 \\
\noalign{\vspace{1.40pt}}
UGC 7089 &VS  &15.5  &-19.1  &.S..8$\ast$.&2.5  &22.5  &   45 &   65 &   77 &   -- &   79 & 7.9 & 0.53 & 0.41 & 0.41 \\
\noalign{\vspace{1.40pt}}
NGC 4100 &VS  &15.5  &-21.0  &PSAT4..&2.4  &19.2  &  125 &  185 &  195 &  192 &  159 &19.6 & 0.56 & 0.12 &-0.11 \\
\noalign{\vspace{1.40pt}}
NGC 4138 &VS  &15.5  &-20.6  &.LAR+..&1.2  &19.1  &  168 &  181 &  192 &  195 &  150 &16.0 & 0.11 & 0.15 &-0.10 \\
\noalign{\vspace{1.40pt}}
NGC 4218 &VS  &15.5  &-18.5  &.S..1\$.&0.7  &19.8  &   83 &  104 &   -- &   -- &   73 & 1.5 & 0.34 &   -- &   -- \\
\noalign{\vspace{1.40pt}}
UGC 7603 &dBB & 6.8  &-16.88 &.SBS7?\/&0.90 &20.8  &   29 &   43 &   -- &   -- &   48 & 2.5 & 0.59 &   -- &   -- \\
\noalign{\vspace{1.40pt}}
DDO 154  &B   &   4  &-14.8  &.IBS9..&0.5  &22.2  &   15 &   24 &   32 &   37 &   43 & 5.7 & 0.64 & 0.70 & 0.34 \\
\noalign{\vspace{1.40pt}}
NGC 5033 &B   &11.9  &-21.2  &.SAS5..&5.8  &20.6  &  220 &  219 &  208 &  192 &  200 &  27 &-0.01 &-0.12 &-0.11 \\
\noalign{\vspace{1.40pt}}
DDO 168  &B   & 3.5  &-16.2  &.IB.9..&0.9  &22.4  &   31 &   51 &   -- &   -- &   49 & 2.5 & 0.71 &   -- &   -- \\
\noalign{\vspace{1.40pt}}
DDO 170  &B   &  12  &-15.5  &.I..9$\ast$.&1.3  &  --  &   32 &   50 &   56 &   59 &   62 & 7.2 & 0.67 & 0.27 & 0.21 \\
\noalign{\vspace{1.40pt}}
NGC 5533 &B   &55.8  &-22.4  &.SAT2..&11.4 &21.0  &  274 &  263 &  240 &  245 &  227 &  56 &-0.06 &-0.22 &-0.16 \\
\noalign{\vspace{1.40pt}}
NGC 5585 &B   & 6.2  &-18.5  &.SXS7..&1.4  &20.9  &   54 &   77 &   89 &   92 &   89 & 7.2 & 0.49 & 0.37 & 0.16 \\
\noalign{\vspace{1.40pt}}
UGC 9211 &dBB & 12.6 &-16.21 &.I..9$\ast$.&1.32 &22.6  &   38 &   55 &   -- &   -- &   60 & 3.6 & 0.52 &   -- &   -- \\
\noalign{\vspace{1.40pt}}
NGC 6195 &SG  & 131  &-23.2  &.S..3..&6.1  &19.9  &  236 &  245 &  247 &  250 &  243 &  40 & 0.05 & 0.02 &-0.01 \\
\noalign{\vspace{1.40pt}}
NGC 6503 &B   & 5.9  &-19.7  &.SAS6..&1.7  &20.9  &   96 &  113 &  114 &  115 &  122 &15.1 & 0.23 & 0.03 & 0.05 \\
\noalign{\vspace{1.40pt}}
NGC 6674 &B   &49.3  &-22.6  &.SBR3..&8.3  &21.5  &  290 &  265 &  247 &  235 &  242 &  52 &-0.13 &-0.17 &-0.08 \\
\noalign{\vspace{1.40pt}}
NGC 7331 &B   &14.9  &-22.4  &.SAS3..&4.5  &20.5  &  257 &  246 &  233 &  238 &  238 &  28 &-0.06 &-0.13 &-0.03 \\
\noalign{\vspace{1.40pt}}
\noalign{\smallskip}\hline
\noalign{\vspace{4pt}}
\end{tabular}
}
\vbox{\hsize=\hsize
(1) Galaxy name (ordered by R.A.);
(2) source of the rotation curve data. B represents Broeils (1992),
dBB is de Blok \& Bosma (2002), SG means Spekkens \& Giovanelli (2006),
SMBB is Swaters et al.\ (2003a), and VS means Verheijen \& Sancisi (2001);
(3) the adopted distance, taken from the original papers;
(4) the absolute $R$-band magnitude;
(5) the morphological type according to the RC3 catalog (de Vaucouleurs et al.\ 1991), using the same coding;
(6) and (7) the disk scale length and the $R$-band central disk surface
brightness. Optical data are from Paper~II for the dBB and SMBB data, from
Verheijen (1997) for the VS data, from Spekkens \& Giovanelli (2006) for the
SG data, and from Broeils (1992) for the B data. The data from Broeils (1992) 
are in the $B$-band. These were converted to $R$-band assuming $B-R=1.0$.
(8) to (11) the rotation velocities at one, two, three and four disk
scale lengths;
(12) and (13) the rotation velocity at the
last measured point and the radius of the last measured point;
(14), (15) and (16) the logarithmic shape
of the rotation curve between one and two disk scale lengths, between
two and three disk scale lengths, and between two disk scale lengths
and the last measured point.
}
\end{table*}

\begin{figure}
\begin{center}
\resizebox{\hsize}{!}{\includegraphics{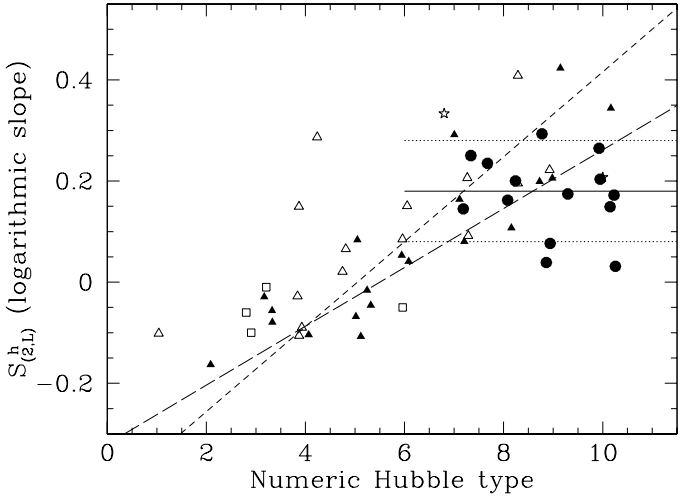}}
\end{center}
\vskip-0.02cm
\caption{ Logarithmic slope of the rotation curve between 2 disk scale
  lengths and the last measured point $S_\mathrm{(2,L)}^{\,h}$ versus
  the numeric Hubble type. Only galaxies for which the last measured
  point lies beyond 3 disk scale lengths and with inclinations between
  $39^\circ$ and $80^\circ$ have been plotted. A random value between
  -0.5 and 0.5 has been added to the numeric Hubble type to spread the
  points in the plots. The filled circles represent the late-type
  galaxies with high quality rotation curves from the sample presented
  in this paper. The open triangles represent galaxies from the Ursa
  Major sample (Verheijen 1997; Verheijen \& Sancisi 2001), filled
  triangles represent the galaxies from various sources presented in
  Broeils (1992a), open squares the galaxies from Spekkens \&
  Giovanelli (2006). The stars represent galaxies with
  \Halpha\ rotation curves from the samples of de Blok \& Bosma (2002)
  and Swaters et al.\ (2003a). The long-dashed line represents the
  relation between Hubble and rotation curve shape as reported by
  Casertano \& van Gorkom (1991), the short-dashed line the same
  relation as found by Broeils (1992a). The horizontal solid and
  dotted lines indicate the average value for $S_\mathrm{(2,L)}^{\,h}$
  and the $1\sigma$ range for galaxies with Hubble type later than Sc
  (see text for details).}
\label{figtypeshape}
\end{figure}

\subsection{Rotation curve shape and morphology}

The relation between rotation curve shape and morphology is explored
in Fig.\ref{figtypeshape}.  In this figure, the logarithmic slope
between two disk scale lengths and the last measured point,
$S_\mathrm{(2,L)}^{\,h}$, is plotted against the numeric Hubble type.
Only galaxies in which the \HI\ extends to beyond three disk scale
lengths have been plotted. For galaxies with less extended \HI\ disks
$S_\mathrm{(2,L)}^{\,h}$ is not well-determined because the rotation
velocities lie on the rising branch.  In Fig.~\ref{figtypeshape}, the
filled circles represent the high-quality sample. Also shown in these
figures are the logarithmic slopes for the spiral galaxies in the
samples of Verheijen \& Sancisi (2001, open triangles), Broeils
(1992a, filled triangles), and Spekkens \& Giovanelli (2006, open
squares). We have also plotted the logarithmic slopes for the galaxies
with \Halpha\ rotation curves, from the samples of dBB and SMvdBB
(stars).

We have chosen to use $S_\mathrm{(2,L)}^{\,h}$ rather than the slope
between two thirds of the optical radius $R_{25}$ and the last
measured point (as was done by e.g., CvG and
Broeils 1992a).  Both choices have their merits: using disk scale
lengths, the shape parameter is measured at the same relative
locations in the disks, and using the optical radius the shape is
measured at the same surface brightness.  We have also used the
optical radius to define the shape parameter, and found that the
results hardly depend on the choice of the scaling parameter, as was
also found by CvG.

\begin{figure*}[ht]
\resizebox{\hsize}{!}{\includegraphics{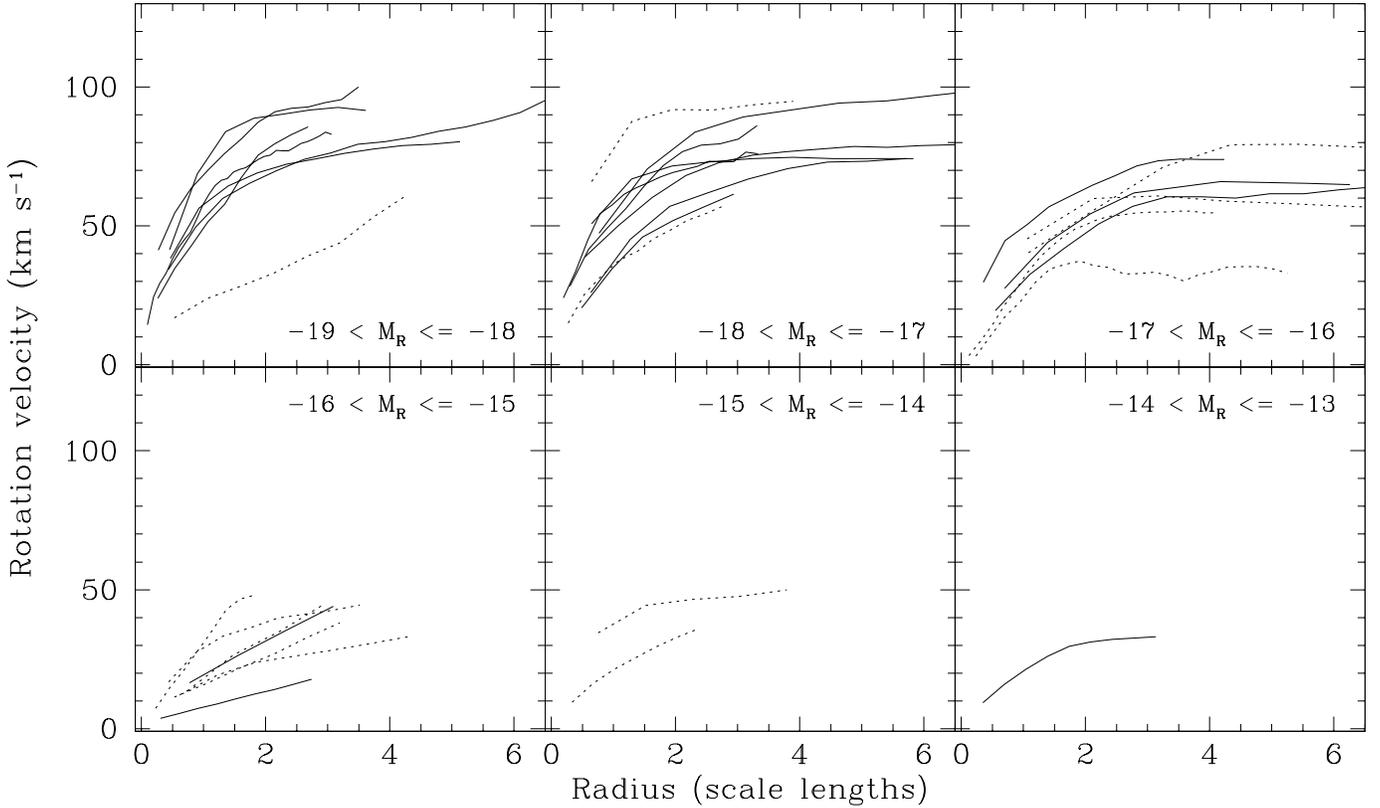}}
\caption{ Rotation curves in bins of absolute magnitude with radii
  expressed in units of disk scale length. The full lines give the
  rotation curves of the galaxies in the high quality sample, the
  dotted lines those of the galaxies in the lower quality sample. }
\label{figrcmrh}
\end{figure*}

Both CvG and Broeils (1992a) have reported
tight correlations between the shape of the rotation curves and the
Hubble type. These correlations are plotted for reference in
Fig.~\ref{figtypeshape}.

With the inclusion of the new data reported here, it now appears that
galaxies with Hubble types later than Sc all have similar outer
rotation curve slopes, with an average $S_\mathrm{(2,L)}^{\,h}$ of
$0.18\pm 0.10$ (marked in Fig.~\ref{figtypeshape} by the solid and
dotted lines). None of the galaxies in our high-quality sample have
rotation curves as shallow as seen among the dwarf galaxies in the
comparison samples. We will discuss this in Section~\ref{secdisc}.

\subsection{Rotation curve shape, amplitude, and absolute magnitude}

To investigate the relation between the rotation curve shape,
amplitude, and absolute magnitude, we have plotted the rotation curves
grouped into six intervals in absolute magnitude in
Fig.~\ref{figrcmrh}. In this figure, the radii are expressed in units
of disk scale length.  The full lines refer to galaxies in the high
quality sample and the dotted lines to galaxies in the lower quality
sample.  The most obvious relation is the decrease in amplitude
towards fainter absolute magnitudes, as expected from the Tully-Fisher
relation. Note that there is a spread in rotation curve shapes in each
luminosity bin. This spread may in part be observational in nature,
e.g., as a result of inclination errors and asymmetries in the
galaxies. The top left panel contains late-type galaxies in the
transition regime between bright galaxies and dwarf galaxies.

\begin{figure}
\begin{center}
\resizebox{0.99\hsize}{!}{\includegraphics{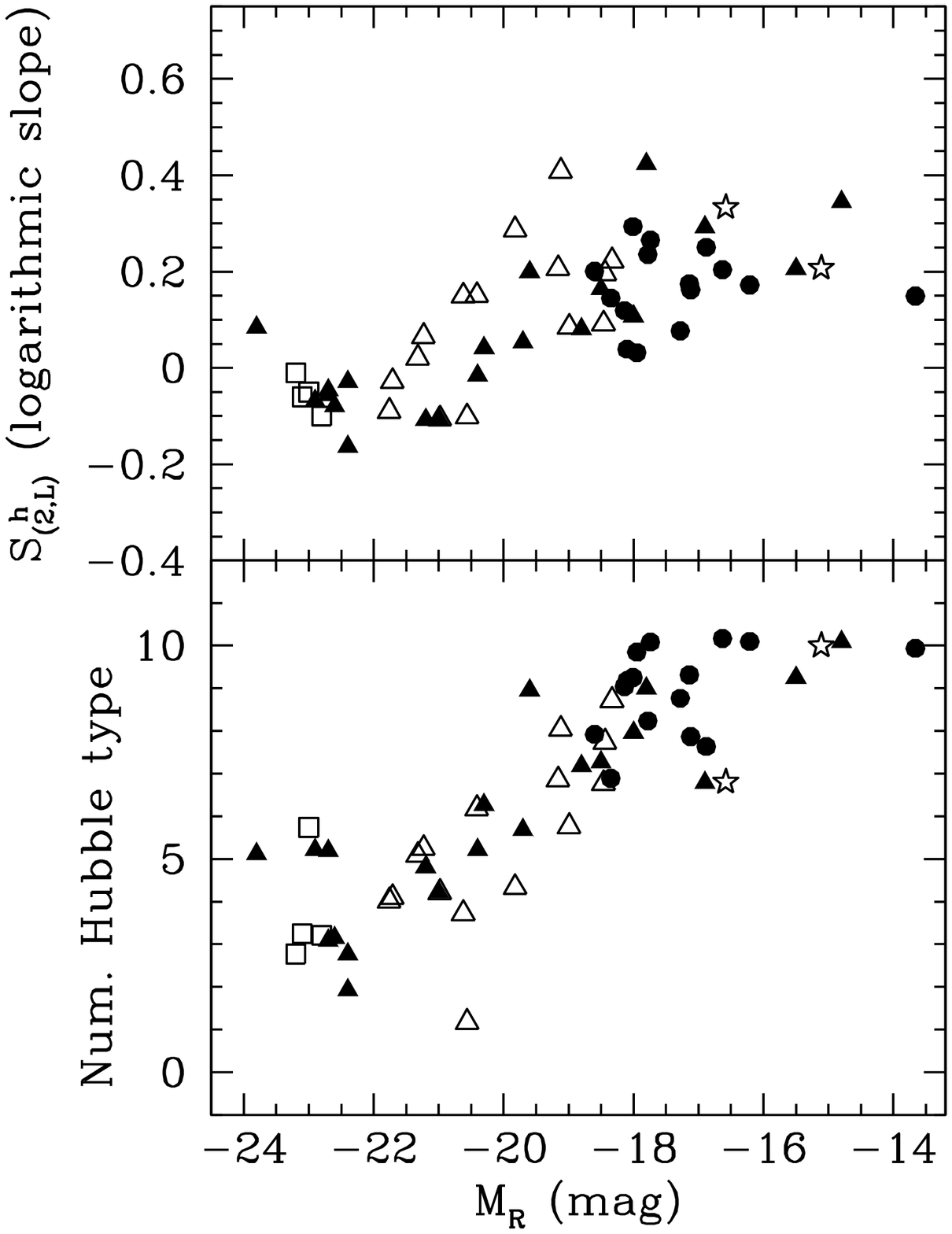}}
\end{center}
\vskip-0.02cm
\caption{ 
  Logarithmic slope of the rotation curve between 2 disk scale lengths
  and the last measured point $S_\mathrm{(2,L)}^{\,h}$ versus the
  the absolute $R$-band magnitude $M_R$ ({\it top}), and
  numeric Hubble type versus $M_R$ ({\it bottom}).  Only
  galaxies for which the last measured point lies beyond 3 disk scale
  lengths and with inclinations between $39^\circ$ and $80^\circ$ have
  been plotted. A random value between -0.5 and 0.5 has been added to
  the numeric Hubble type to spread the points in the plots. Symbol
  coding as in Fig.~\ref{figtypeshape}.
}
\label{figshapeabsmag}
\end{figure}

\begin{figure}
\resizebox{\hsize}{!}{\includegraphics{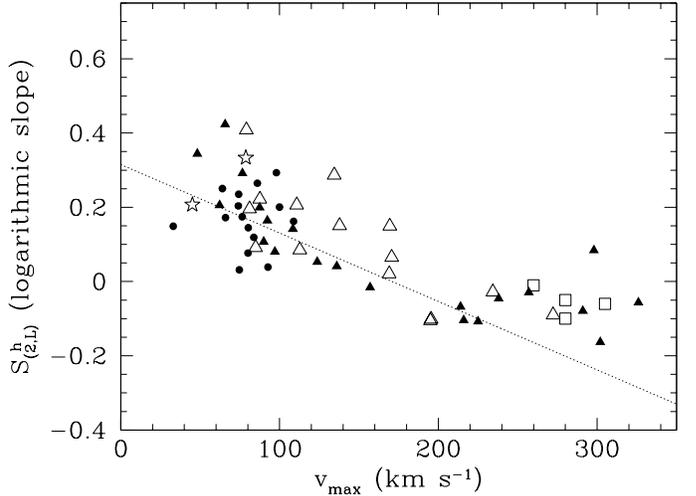}}
\caption{
  Logarithmic slope of the rotation curve between 2 disk scale lengths
  and the last measured point $S_\mathrm{(2,L)}^{\,h}$ versus the
  maximum rotation velocity. Only galaxies for which the last measured
  point lies beyond 3 disk scale lengths and with inclinations between
  $39^\circ$ and $80^\circ$ have been plotted. Symbol coding as in
  Fig.~\ref{figtypeshape}. The dotted line
  represents the correlation between rotation curve shape and maximum
  rotation velocity found by Casertano \& van Gorkom (1991).
}
\label{figshapevmax}
\end{figure}

The relations between absolute magnitude or maximum rotation velocity
on the one hand, and rotation curve shape on the other, are made more
explicit in Figs.~\ref{figshapeabsmag} (top panel)
and~\ref{figshapevmax}. 

At first sight, the correlation seen between the rotation curve shape
$S_\mathrm{(2,L)}^{\,h}$ and absolute magnitude (see
Fig.\ref{figshapeabsmag}, top panel) seems to confirm the results
found by e.g., Rubin et al.\ (1985), Persic \& Salucci (1988), Broeils
(1992a), and CvG that rotation curves rise more slowly towards lower
luminosities. However, with the addition of the dwarf galaxies
presented here, the dwarf regime is better sampled and extended down
to fainter absolute magnitudes, thus making it possible to sample this
correlation in more detail at the low luminosity end.

In Fig.~\ref{figtypeshape} it was shown that the rotation curve shape
$S_\mathrm{(2,L)}^{\,h}$ is similar for galaxies with Hubble types
later than Sc. A similar trend is seen with absolute magnitude, as can
be seen in the top panel of Fig.~\ref{figshapeabsmag}. This is not
unexpected, given the tight correlation between Hubble type and
absolute magnitude (bottom panel of Fig.~\ref{figshapeabsmag}).  From
Fig.~\ref{figshapeabsmag} it is clear that, down to the faintest
magnitudes, all the galaxies in our sample with high quality rotation
curves have logarithmic slopes between 0 and 0.3, although the scatter
may increase somewhat towards lower absolute magnitudes. For these
galaxies, the mean value of $S_\mathrm{(2,L)}^{\,h}$ is $0.17\pm
0.08$, where the uncertainty is the dispersion around the mean.  The
logarithmic slopes of late-type spiral galaxies with $-20.5<M_R<-18$
have similar values. For these galaxies, the average
$S_\mathrm{(2,L)}^{\,h}$ is $0.15\pm 0.11$.  Thus, the results in
Fig.~\ref{figshapeabsmag} indicate that the outer rotation curve
shapes, as measured by $S_\mathrm{(2,L)}^{\,h}$, are similar for
late-type spiral galaxies and the late-type dwarf galaxies in our
sample. Therefore, for galaxies fainter than $M_R=-20$, the outer
rotation curve shape does not appear to depend on luminosity
down to $M_R\sim-16$ and possibly even $M_R\sim-14$, although we have
little information at the lowest luminosities.

At the brighter end, as was already pointed out by CvG and Broeils
(1992a), the most luminous galaxies tend to have declining rotation
curves. This is also seen in Fig.~\ref{figshapeabsmag}.  Galaxies
brighter than about $M_R\sim-21$ mostly have negative values of
$S_\mathrm{(2,L)}^{\,h}$, indicating lower rotation velocities at the
last measured points. Virtually all of these galaxies have bulges, as
is demonstrated in the bottom panel of Fig.~\ref{figshapeabsmag},
where the $S_\mathrm{(2,L)}^{\,h}$ is plotted against the numerical
Hubble type. A larger sample of early-type spiral and high-luminosity
galaxies is needed to study the details of the correlations between
rotation curve shape and morphology or luminosity.

Following CvG, we have also plotted $S_\mathrm{(2,L)}^{\,h}$ against
the maximum rotation velocity in Fig.~\ref{figshapevmax}.  The
relation between these quantities found by CvG, represented by the
dotted line, is given for comparison. The galaxies in the sample
presented here follow the same relation. This may seem to indicate
that the results presented here are in good agreement with the
relation found by CvG. However, at the lowest rotation velocities, the
galaxies in the comparison sample tend to fall above the CvG
relation. In fact, 4 out 5 galaxies with $\varv_\mathrm{max}<80$ \kms\
fall well above the dotted line.  Similarly, the slowest rotators in
Fig.~6 of CvG also fall above their relation.  Thus, even though the
values for $S_\mathrm{(2,L)}^{\,h}$ for the galaxies in our sample are
consistent with the CvG relation, the $S_\mathrm{(2,L)}^{\,h}$ values
for our high-quality sample are significantly different from those for
our reference sample. We point out, however, that there are only a
small number of galaxies with $\varv_\mathrm{max}<80$ \kms. To fully
sample the relation between rotation curve shape and
$\varv_\mathrm{max}$, a larger number of beam-smearing corrected
rotation curves is needed.  We also note that at the higher mass end
the galaxies in our sample fall above the CvG relation. This was also
noted by Broeils (1992a), who remarked that there is a clear
correlation between the slope and the maximum velocity, but that it is
not a simple linear relation.

\begin{figure}
\resizebox{\hsize}{!}{\includegraphics{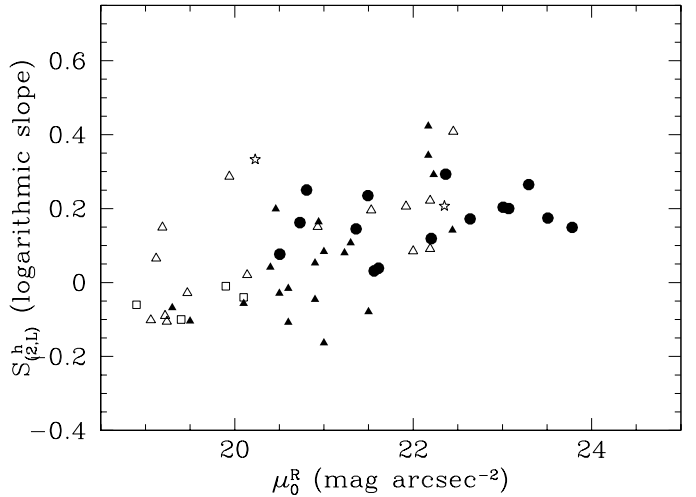}}
\caption{
  Logarithmic slope of the rotation curve between 2 disk scale lengths
  and the last measured point $S_\mathrm{(2,L)}^{\,h}$ versus the
  central extrapolated disk surface brightness $\mu_0^R$. Only
  galaxies for which the last measured point lies beyond 3 disk scale
  lengths and with inclinations between $39^\circ$ and $80^\circ$ have
  been plotted. Symbol coding as in Fig.~\ref{figtypeshape}.
}
\label{figshapemu}
\end{figure}

The slope of the rotation curve is weakly correlated with surface
brightness, as can be seen in Fig.~\ref{figshapemu}. At intermediate
surface brightnesses, there is a large spread in slopes. At the high
surface brightness end, galaxies tend to have flat or declining
rotation curves, and at the low surface brightness end, galaxies tend
to have slowly rising rotation curves.

\subsection{Rotation curve shape and light distribution}

To investigate a possible link between the light distribution and the
rotation curve, we compare the central concentration of light and the
inner rotation curve shape.  A straightforward measure of the central
concentration of light is given by the parameter $\Delta\mu_R$, which
we defined as:
\begin{equation}
\Delta\mu_R=\mu^R_0-\mu^R_c,
\label{eqdeltamu}
\end{equation}
where $\mu_c^R$ is the observed central surface brightness. If a bulge
or a central concentration of light is present, $\Delta\mu_R$ is
positive, and if the light profile is flat-topped, $\Delta\mu_R$ is
negative.

A significant fraction of the late-type dwarf galaxies in our sample
have non-exponential light profiles in the inner regions.  Some have a
central concentration of light, and in others the light profile in the
center falls below that of an exponential fit to the outer parts of
the light profile, making the light profile flat-topped. The galaxies
in the comparison samples of Verheijen \& Sancisi (2001) and Broeils
(1992a) include both galaxies with strong bulges and galaxies that are
dominated by exponential profiles. The galaxies in the sample of
Spekkens \& Giovanelli (2006) all have bulges or central
concentrations of light.

\begin{figure}
\resizebox{\hsize}{!}{\includegraphics{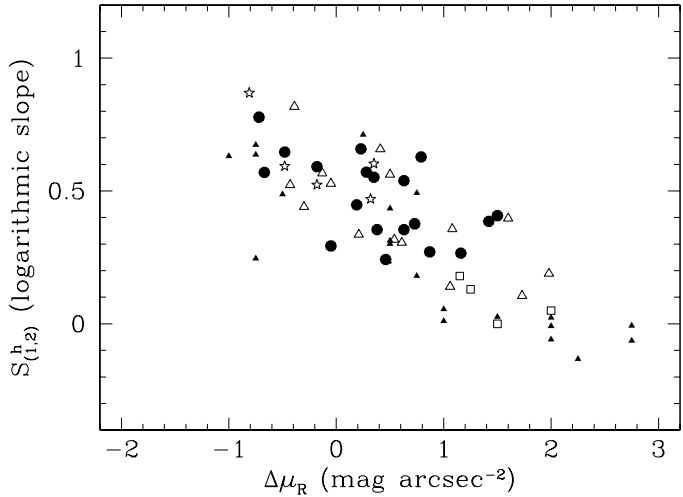}}
\caption{%
  Logarithmic slope of the rotation curve between 1 and 2 disk scale
  lengths $S_\mathrm{(1,2)}^{\,h}$ versus the difference between
  extrapolated disk surface brightness $\mu_0^R$ and the observed
  central surface brightness $\mu_c^R$. Only galaxies with
  inclinations between $39^\circ$ and $80^\circ$ have been
  plotted. Symbol coding as in Fig.~\ref{figtypeshape}.
}
\label{figinnershapemu}
\end{figure}

In Fig.~\ref{figinnershapemu}, $\Delta\mu_R$ is plotted against
$S_\mathrm{(1,2)}^{\,h}$, the logarithmic slope between one and two
disk scale lengths, for the late-type dwarf galaxies in our sample and
for the galaxies in the comparison samples. A clear correlation is
found between $\Delta\mu_R$ and $S_\mathrm{(1,2)}^{\,h}$.  Rotation
curves of galaxies with a central concentration of light are
relatively flat between one and two disk scale lengths, and therefore
they must rise steeply within one disk scale length. Galaxies with a
flat-topped light profile have rotation curves that rise more slowly
in the inner parts. This correlation between $\Delta\mu_R$ and
$S_\mathrm{(1,2)}^{\,h}$ remains even if only LSB galaxies with
$\mu_R>22$ mag arcsec$^{-2}$ are considered, although the range in
$\Delta\mu_R$ is reduced to $-1<\Delta\mu_R<1$ for those galaxies.  As
can be seen in Fig.~\ref{figinnershapemu}, the correlation between
$\Delta\mu_R$ and $S_\mathrm{(1,2)}^{\,h}$ shows considerable
scatter. It is likely this is partly due to observational
uncertainties, given that for two thirds of the galaxies the
resolution of the \HI\ observations is similar to the scale lengths of
these galaxies.

\section{Discussion and conclusions}
\label{secdisc}

In this paper, we have presented rotation curves derived from \HI\
observations for a sample of 62 galaxies.  These rotation curves have
been derived by interactively fitting model data cubes to the observed
cubes. This procedure takes the rotation curve shape, the \HI\
distribution, the inclination, and the size of the beam into account,
and makes it possible to correct for the effects of beam smearing.
Comparison of the derived rotation curves with high resolution
H$\alpha$ rotation curves shows good general agreement, indicating
that the scheme used to correct the rotation curves for beam smearing
works well.

From our sample of 62 galaxies, we selected 18 systems with rotation
curves of high quality.  These have been used to study the shapes and
amplitudes of the rotation curves and their relation with the
distribution of light.

We have found that the rotation curve shapes of late-type dwarf
galaxies in our sample are similar to those of higher luminosity,
late-type spiral galaxies, in the sense that for most of the dwarf
galaxies in our sample the rotation curves, when expressed in units of
disk scale lengths, rise as steeply in the inner parts and start to
flatten beyond about two disk scale lengths. Moreover, there are
several galaxies with relatively flat rotation curves, and with
amplitudes as low as 60 \kms.

We found no galaxies with pure solid body rotation curves that extend
beyond 3 disk scale lengths. Thus, whether the rotation curve becomes
flat to depends mainly on the extent of the rotation curve,
independent of its amplitude.

This similarity between the rotation curve shapes of late-type dwarf
and late-type spiral galaxies is also borne out by the similar
logarithmic slopes between two disk scale lengths and the last
measured point. For the late-type dwarf galaxies in our high-quality
sample, $S_\mathrm{(2,L)}^{\,h}=0.17\pm 0.08$, and for the spiral
galaxies in our reference sample $S_\mathrm{(2,L)}^{\,h}=0.15\pm
0.11$. The similarity between the rotation curve shapes of dwarf and
spiral galaxies continues down to $M_R \sim -16$~mag, and possibly
even down to $M_R \sim -14$~mag, although we have little information
at the faintest luminosities.  The ongoing FIGGS survey (e.g., Begum
et al.\ 2008) will likely fill this gap.

The results presented here confirm earlier findings by CvG and Broeils
(1992) that the rotation curve shape depends on morphology and
luminosity (at least for galaxies brighter than approximately
$M_R=-19$~mag). However, with the inclusion of the galaxies from the
sample presented here, and a larger number of spiral galaxies from the
sample of Verheijen \& Sancisi (2001), the correlation between
rotation curve shape and morphology and luminosity is better defined,
especially at the low luminosity end. Based on these new data, we find
that the outer rotation curves of galaxies with Hubble types later
than Sc or with absolute magnitudes fainter than approximately
$M_R=-19$~mag all have similar shapes. In other words, we find that
the outer rotation curve shape does not depend on luminosity or
morphology for galaxies fainter than $M_R\sim-19$ for the sample
presented here.

The main reason for this difference in the reported correlation
between rotation curve shape on the one hand and Hubble type or
absolute magnitude on the other lies in the derived rotation curve
shapes for the late-type dwarf galaxies presented here compared to
those derived from rotation curves as published in the
literature. Both beam smearing (see Section~\ref{secrc} and references
therein) and intensity-weighted velocities (see Paper~I and references
therein) will lead to a shallower derived rotation curves, especially
where the rotation curve is rising or turning over. These differences
may well explain the differences in derived rotation curves.

Considering the bright galaxies that have outer rotation-curve slopes
different from those of late-type dwarf galaxies, it is interesting to
note that the vast majority of these have a bulge (or a strong central
concentration of light). These bulges have a strong effect on the
shape of the inner rotation shape, causing a steep rise in the
center. But the effect of these bulges may extend to larger radii as
well and explain the falling rotation curves seen in some galaxies
(e.g. Noordermeer et al.\ 2007).  Given that we found that the outer
rotation curve shapes for galaxies fainter than $M_R\sim-19$ (which
overwhelmingly do not have bulges) do not depend on luminosity, it is
tempting to speculate that the change in rotation curve shapes for
brighter galaxies is the result from the increasing contribution of
the bulge, while the rotation curve shape of the underlying disk
remains unchanged.

We have found a relation between the shape of the central distribution
of light and the inner rise of the rotation curve. Galaxies with a
stronger central concentration of light have a more steeply rising
rotation curve. This correlation between the light distribution and
the inner rise of the rotation curve is well known for spiral galaxies
(e.g., Kent 1987, Corradi \& Capaccioli 1990), but in this paper we
have found that dwarf galaxies with a central concentration of light
also have rotation curves that rise more steeply in the center than
the rotation curves of dwarf galaxies that do not have a central
concentration of light.

The observed correlation between the light distribution and the inner
rotation curve shape, as seen in both spiral and late-type dwarf
galaxies, implies that galaxies with stronger central concentrations
of light also have higher central mass densities, and it suggests that
the luminous mass dominates the gravitational potential in the central
regions, even in low surface brightness dwarf galaxies.

\begin{acknowledgements}
We are grateful to Martin Vogelaar, Hans Terlouw and Kor Begeman for
the implementation of the programs used here in the {\sc gipsy}
software package.
We thank Marc Verheijen and Erwin de Blok for making their data
available in electronic format.
This research has made use of the NASA/IPAC Extragalactic Database
(NED) which is operated by the Jet Propulsion Laboratory, California
Institute of Technology, under contract with the National Aeronautics
and Space Administration.
\end{acknowledgements}

\appendix

\section{Rotation curves}
\label{appfigures}

In the next pages overviews are presented of the rotation curves for
all galaxies in this study. For each galaxy a figure is given with two
panels. In the left panel, the position-velocity diagram along the
major axis is shown. Contour levels are at $-4\sigma$ and $-2\sigma$
(dotted), and $2\sigma$, $4\sigma$, ... .  Overlayed on the
position-velocity diagram is the rotation curve. If the rotation curve
has been derived for the approaching and the receding side separately,
then these are shown.  In the right panel, the derived rotation curve
(represented by the thick solid line and the filled circles) is
shown. If the rotation curves for the approaching and receding sides
were derived separately, both are shown, together with the average.
The thin solid line represents the the approaching side, the dotted
line the receding side.  The arrow at the bottom of the panel
indicates a radius of 2 optical disk scale lengths.

\clearpage

\onecolumn

\vskip-0.15cm
\noindent
\resizebox{0.45\hsize}{!}{\includegraphics{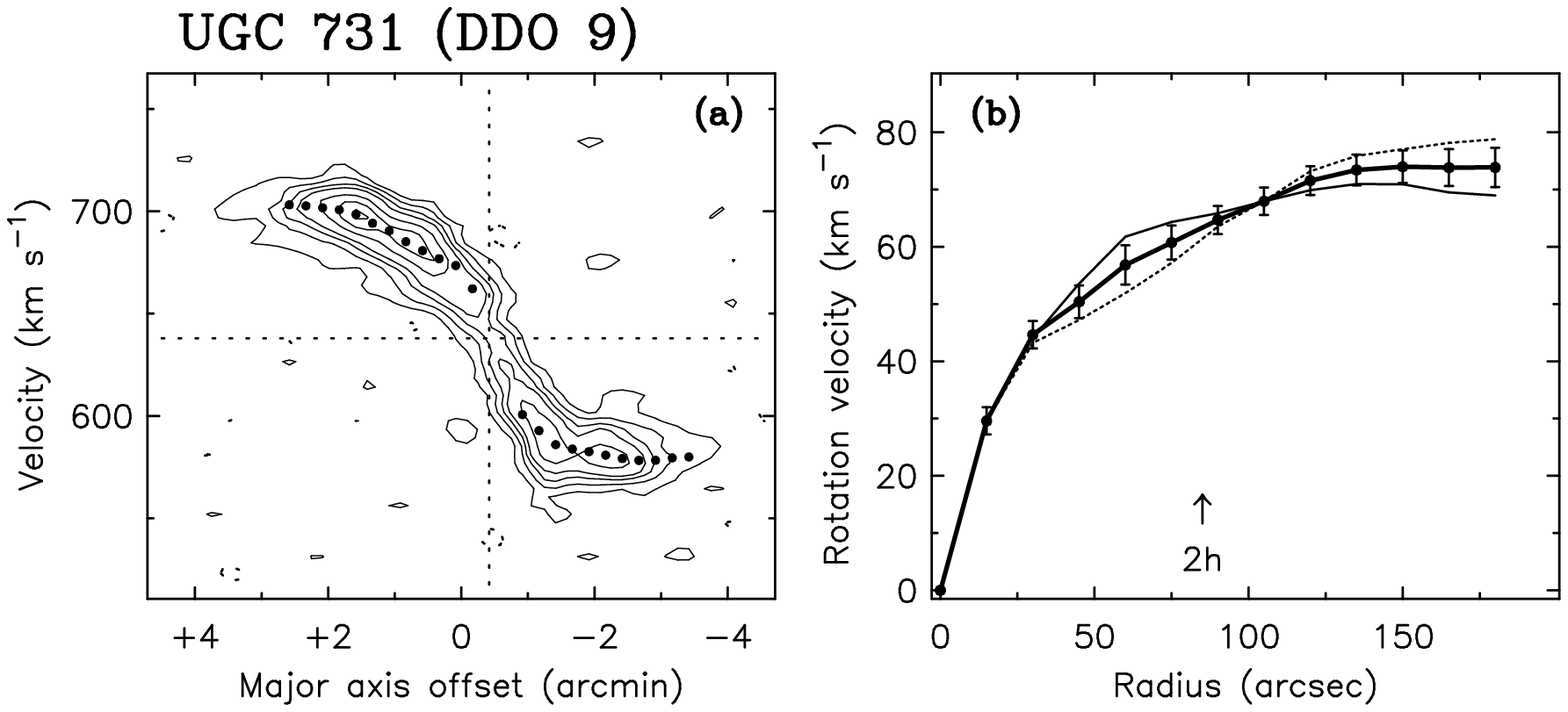}}
\kern0.5cm
\resizebox{0.45\hsize}{!}{\includegraphics{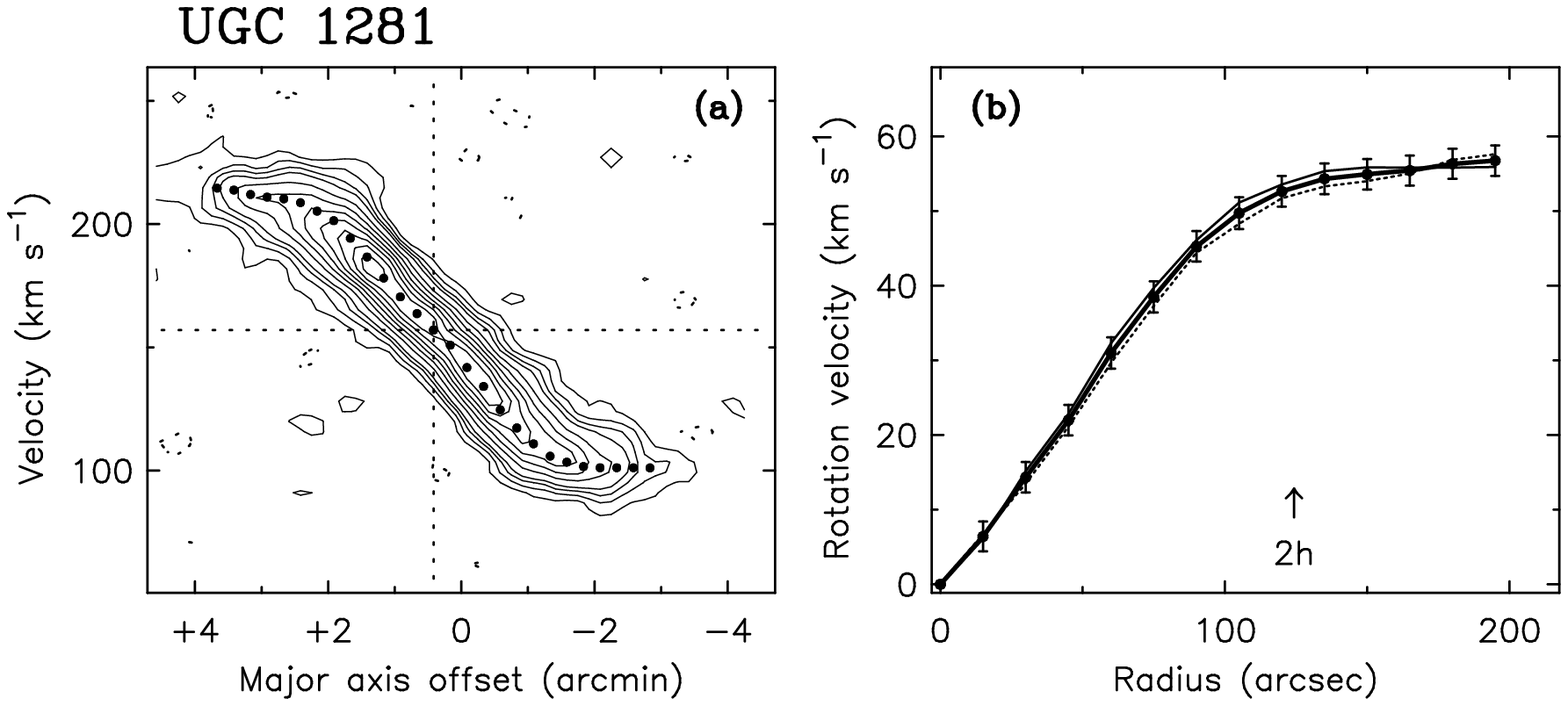}}
\newline
\vskip-0.15cm
\noindent
\resizebox{0.45\hsize}{!}{\includegraphics{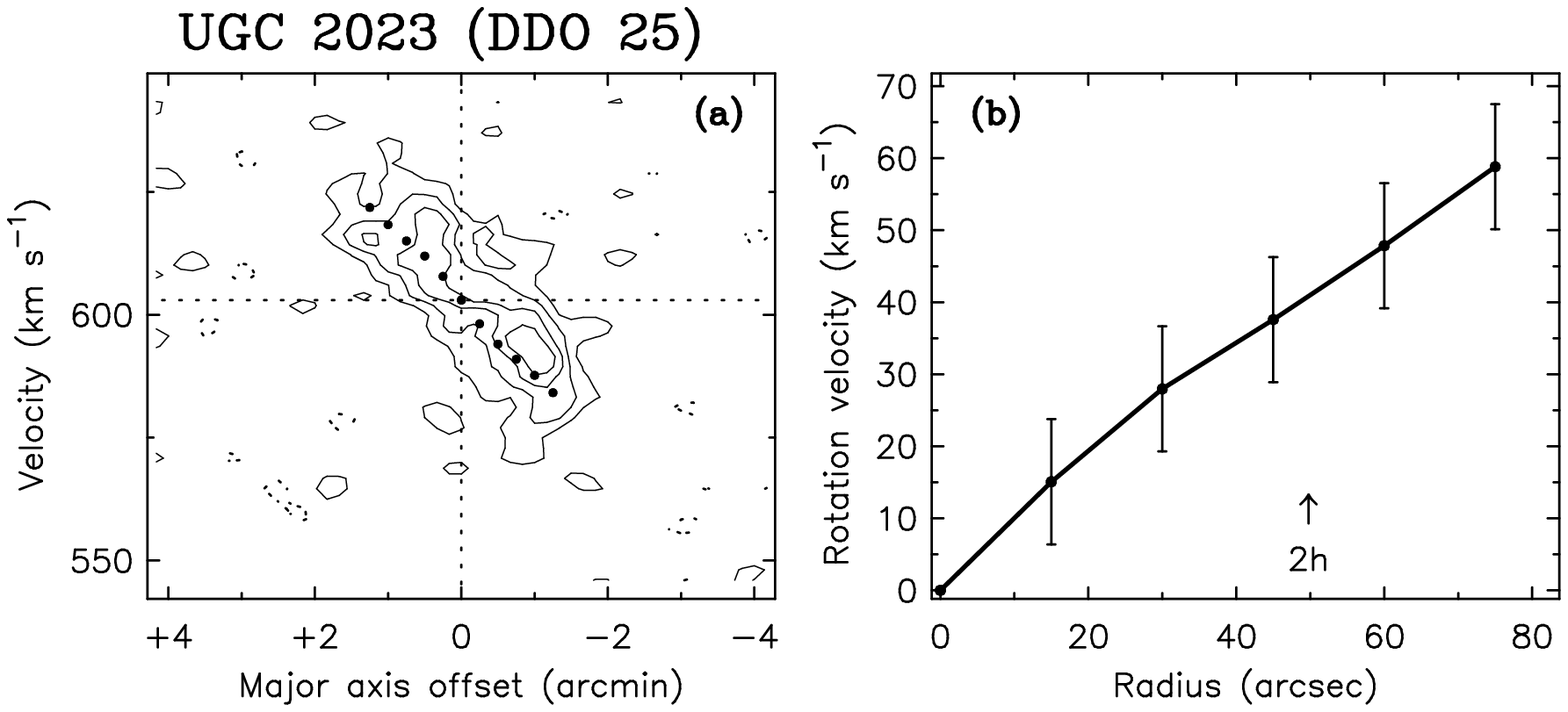}}
\kern0.5cm
\resizebox{0.45\hsize}{!}{\includegraphics{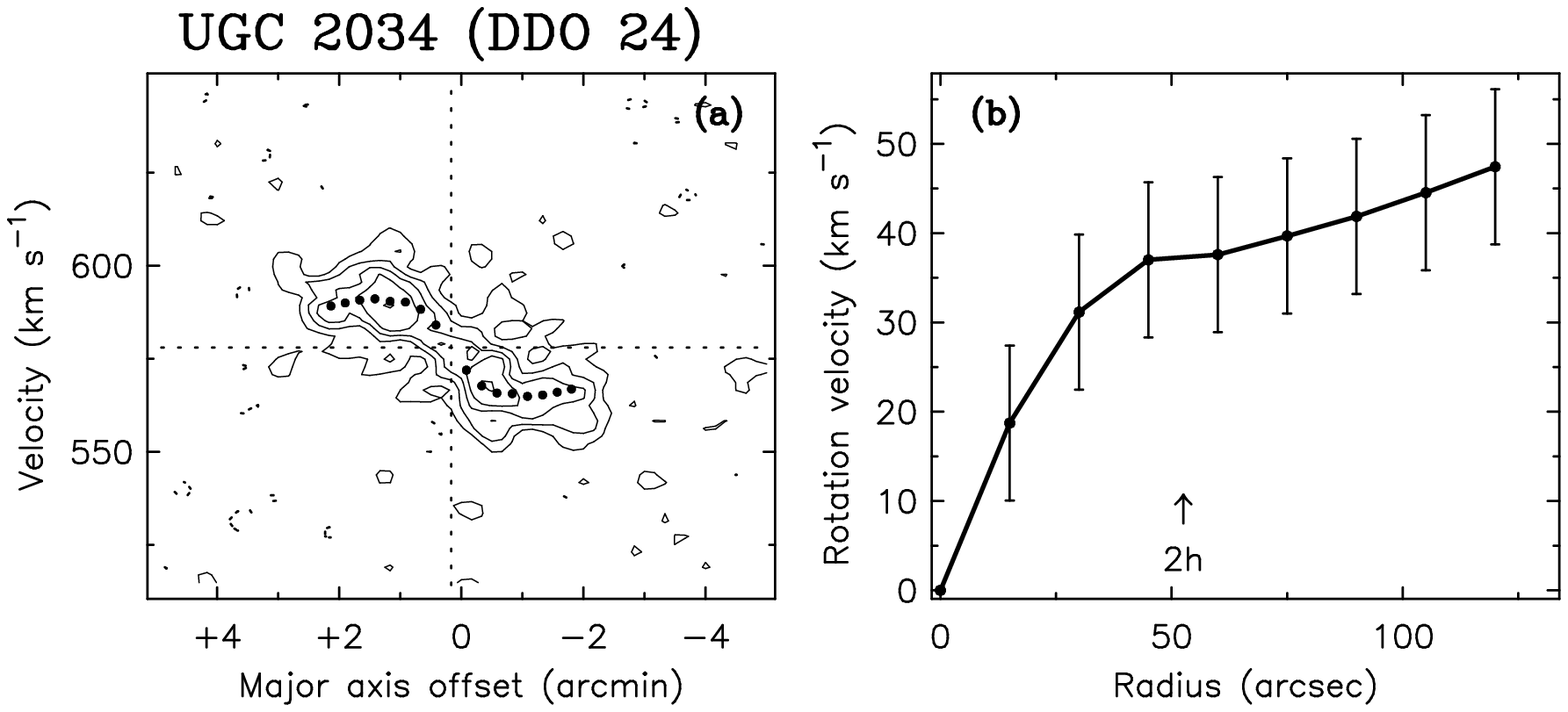}}
\newline
\vskip-0.15cm
\noindent
\resizebox{0.45\hsize}{!}{\includegraphics{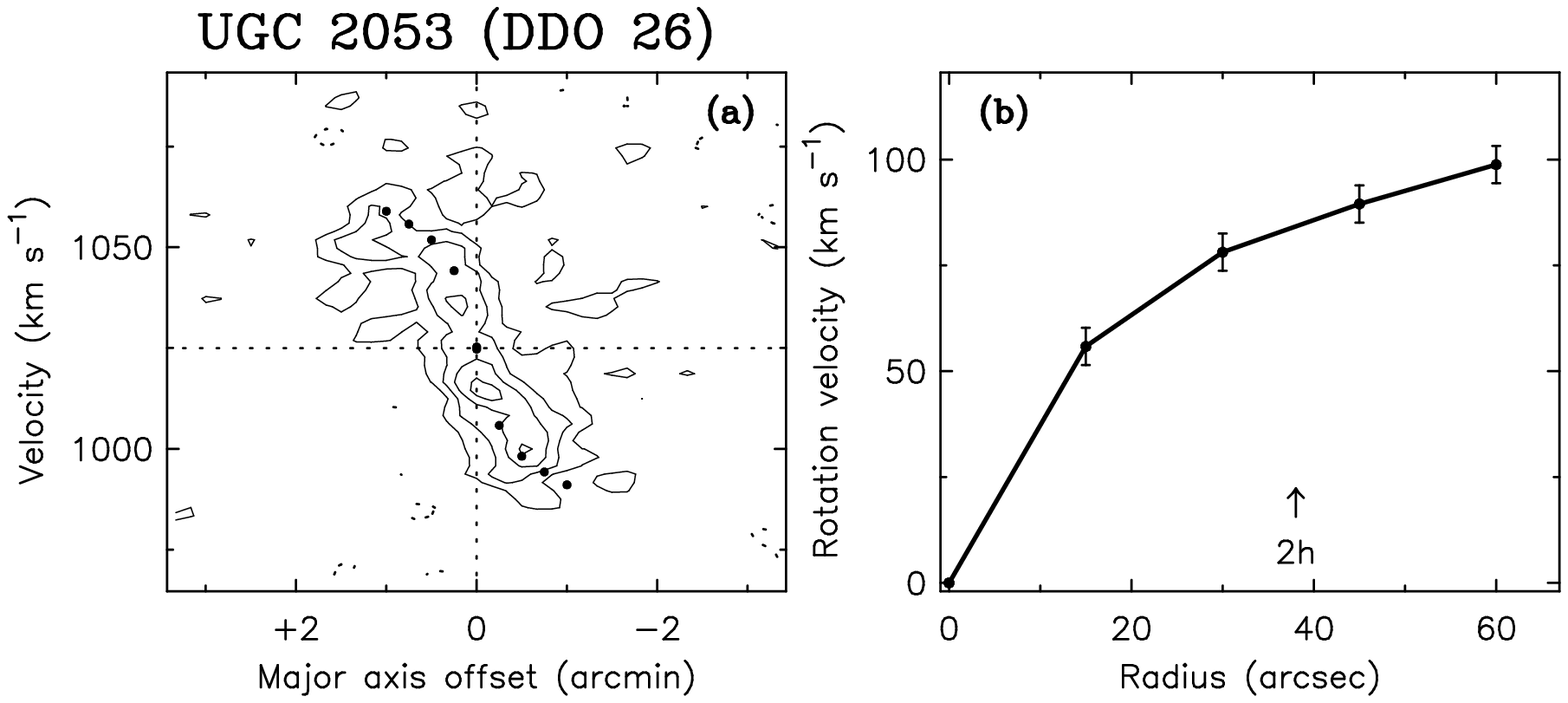}}
\kern0.5cm
\resizebox{0.45\hsize}{!}{\includegraphics{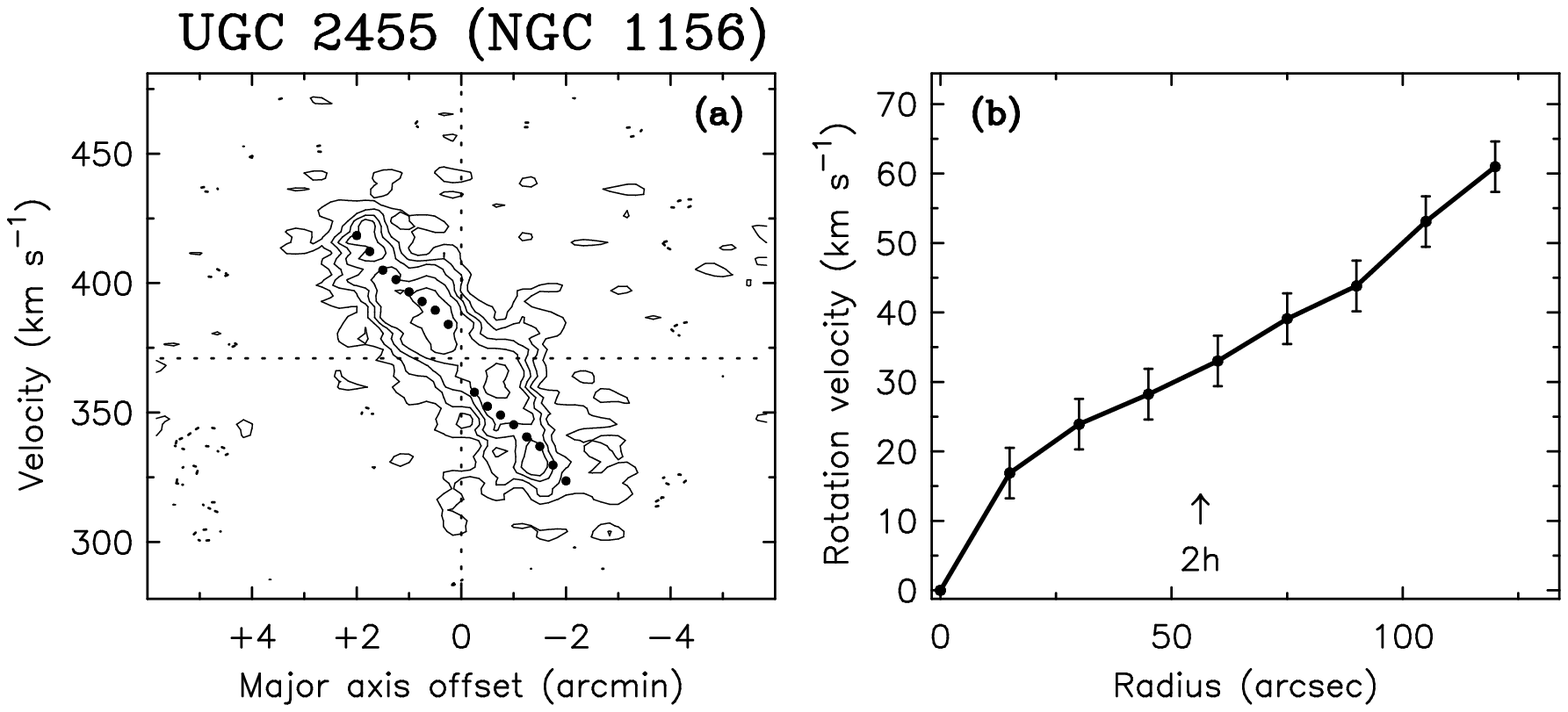}}
\newline
\vskip-0.15cm
\noindent
\resizebox{0.45\hsize}{!}{\includegraphics{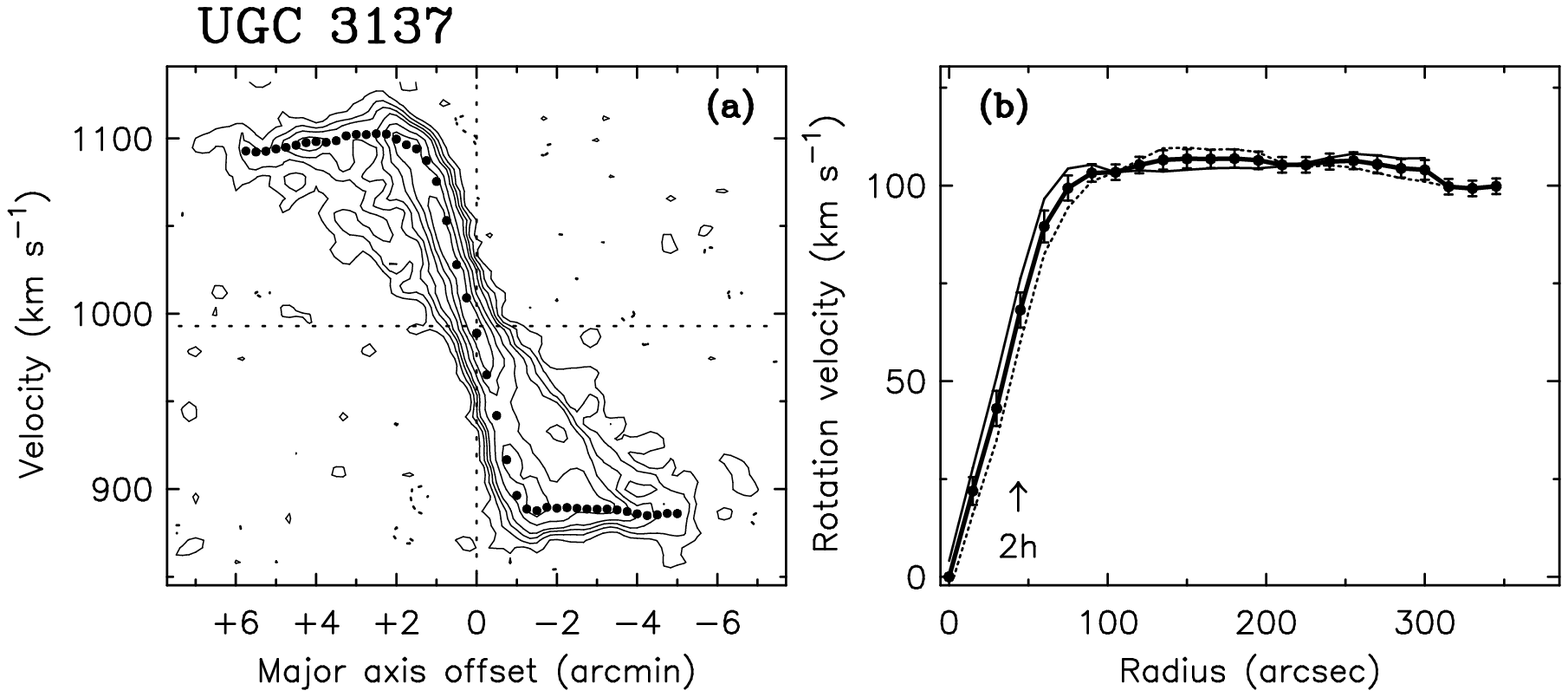}}
\kern0.5cm
\resizebox{0.45\hsize}{!}{\includegraphics{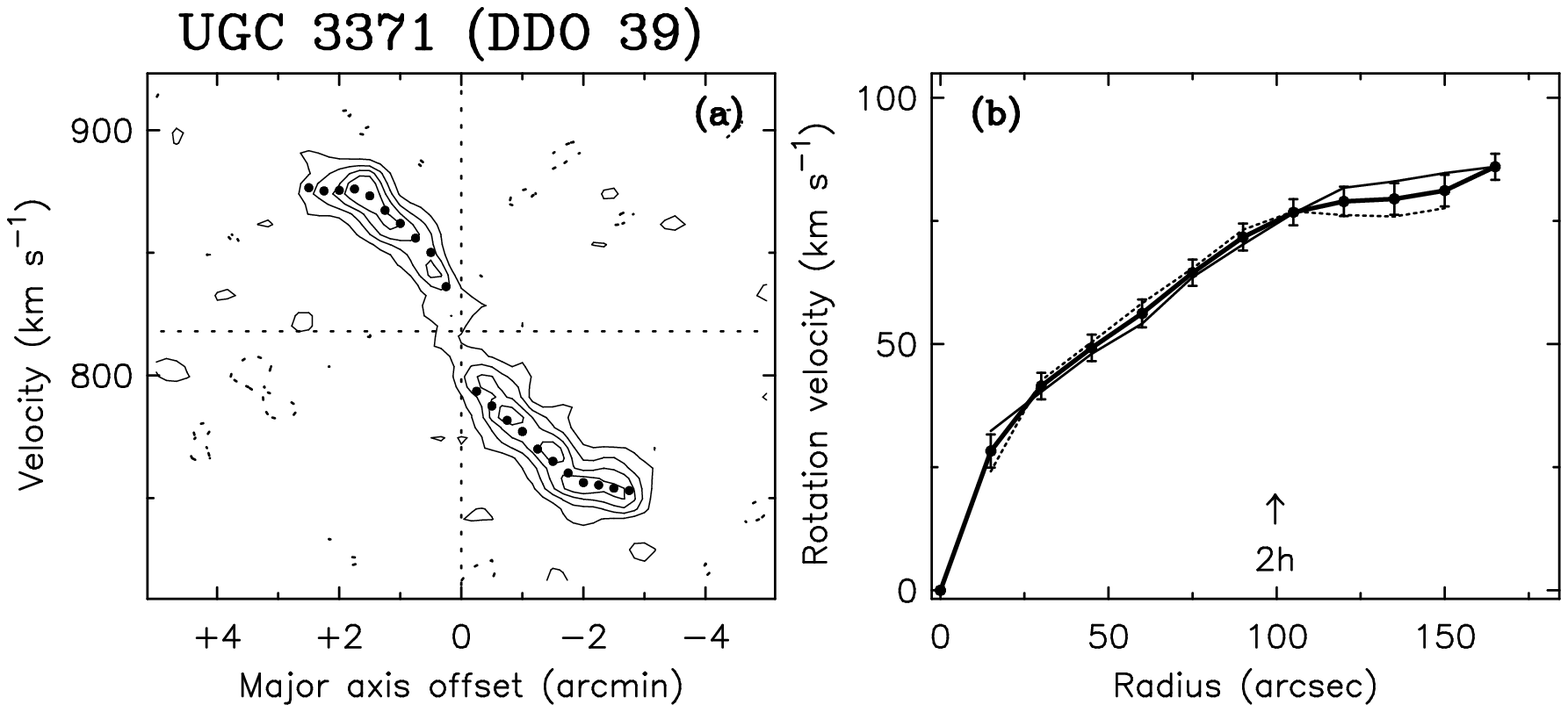}}
\newline
\vskip-0.15cm
\noindent
\resizebox{0.45\hsize}{!}{\includegraphics{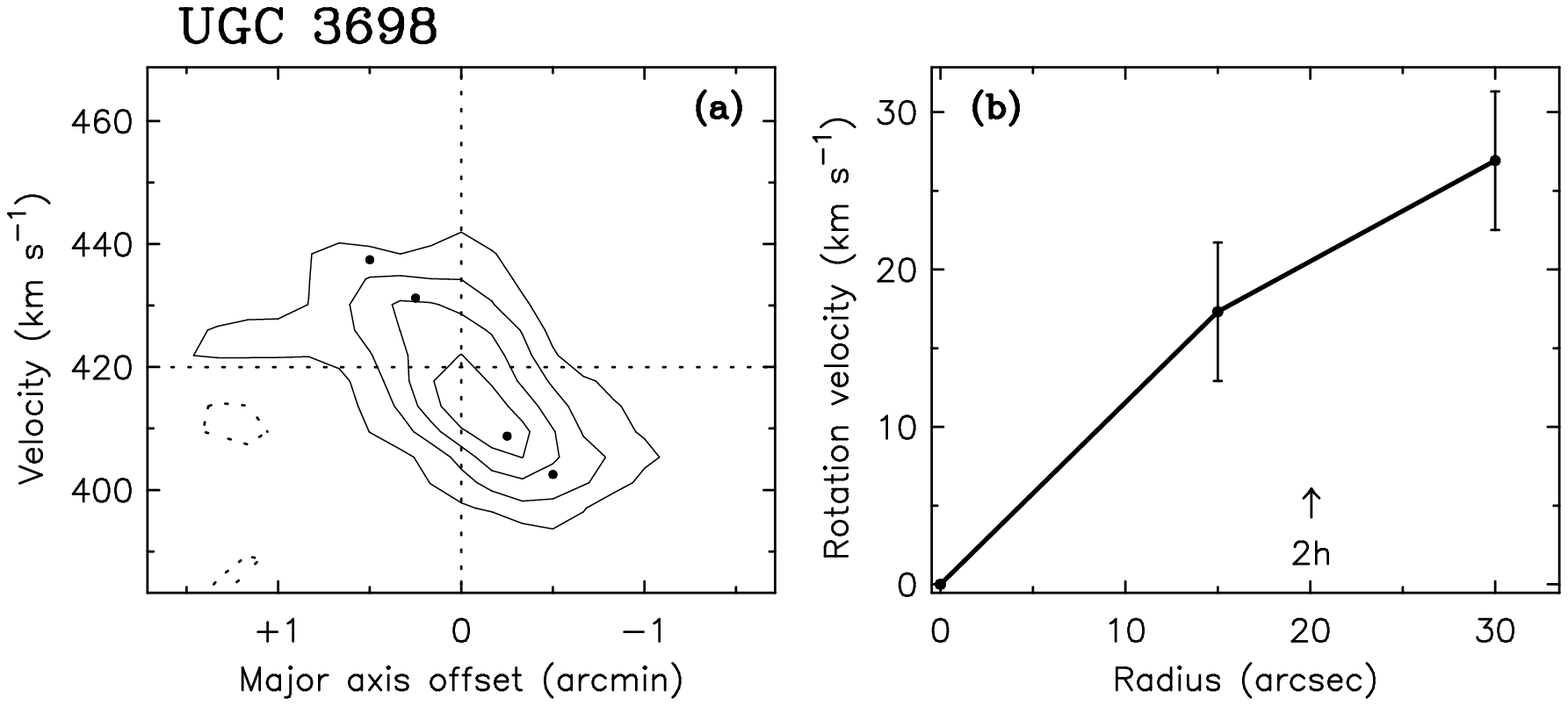}}
\kern0.5cm
\resizebox{0.45\hsize}{!}{\includegraphics{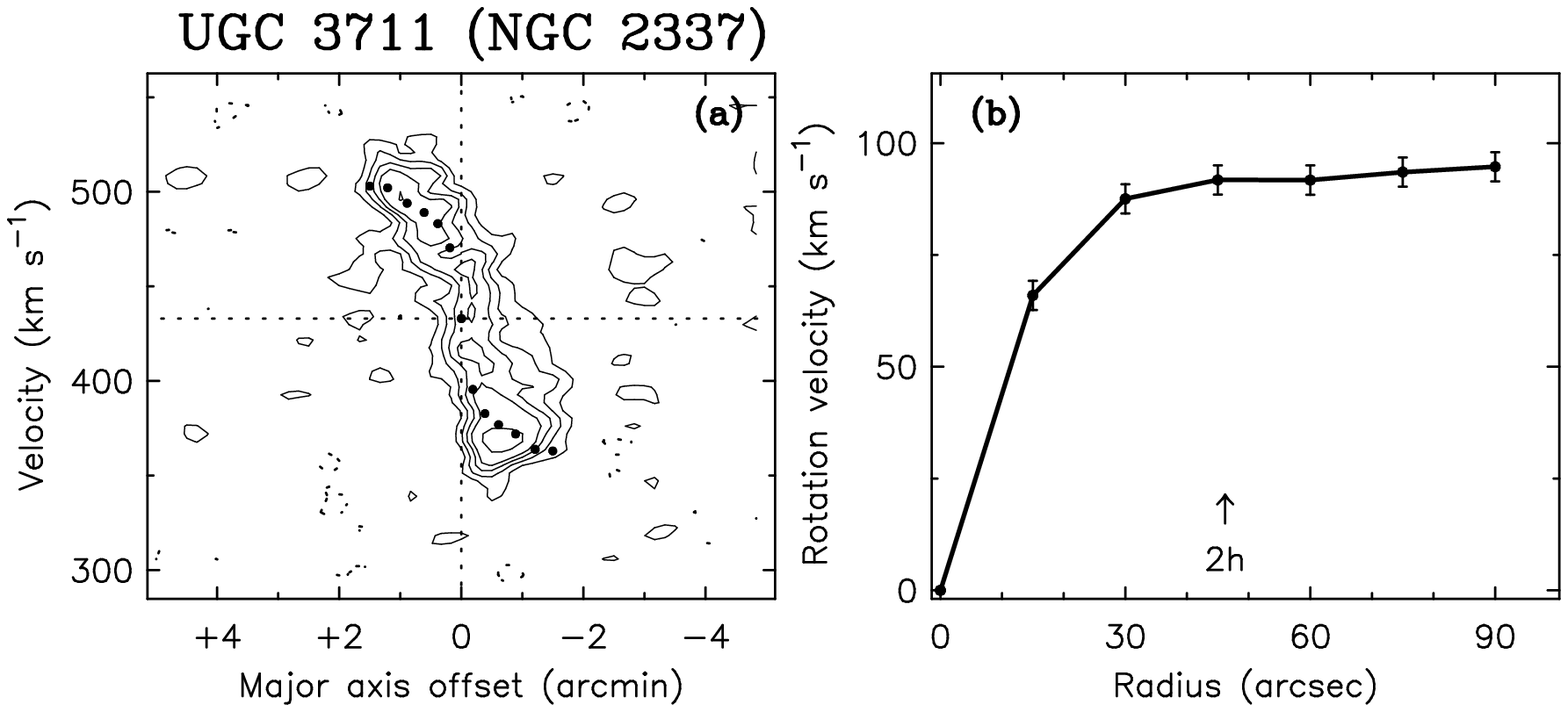}}
\newline
\vskip-0.15cm
\noindent
\resizebox{0.45\hsize}{!}{\includegraphics{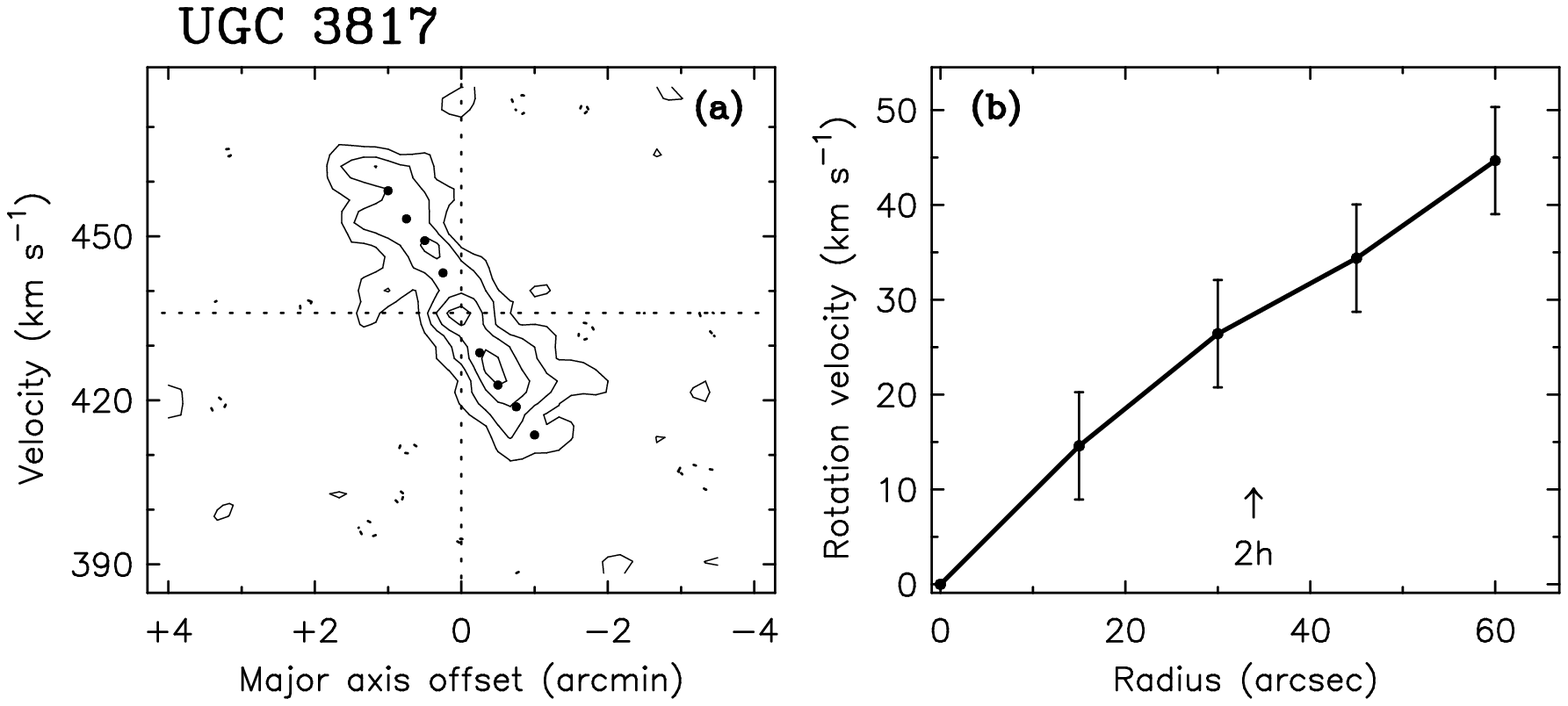}}
\kern0.5cm
\resizebox{0.45\hsize}{!}{\includegraphics{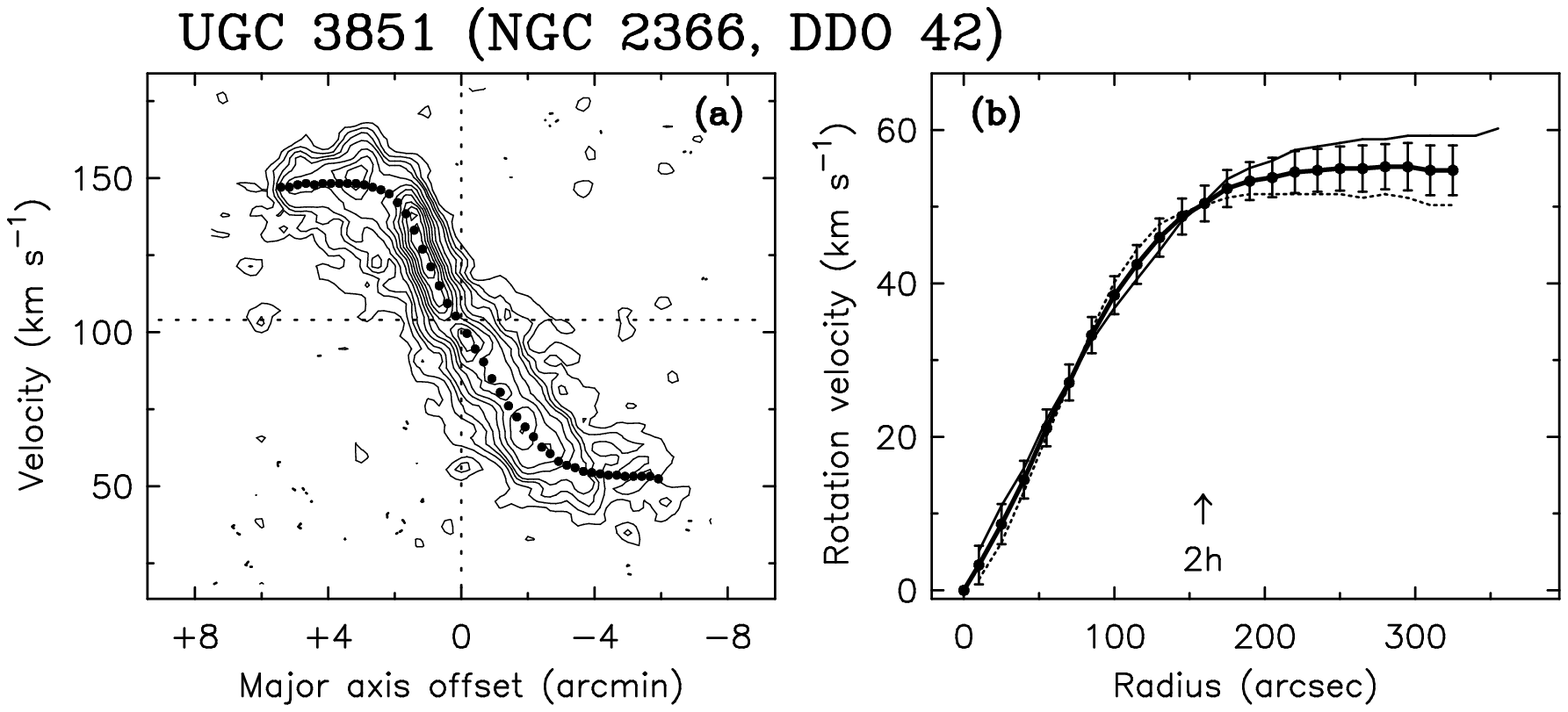}}
\clearpage
\vskip-0.15cm
\noindent
\resizebox{0.45\hsize}{!}{\includegraphics{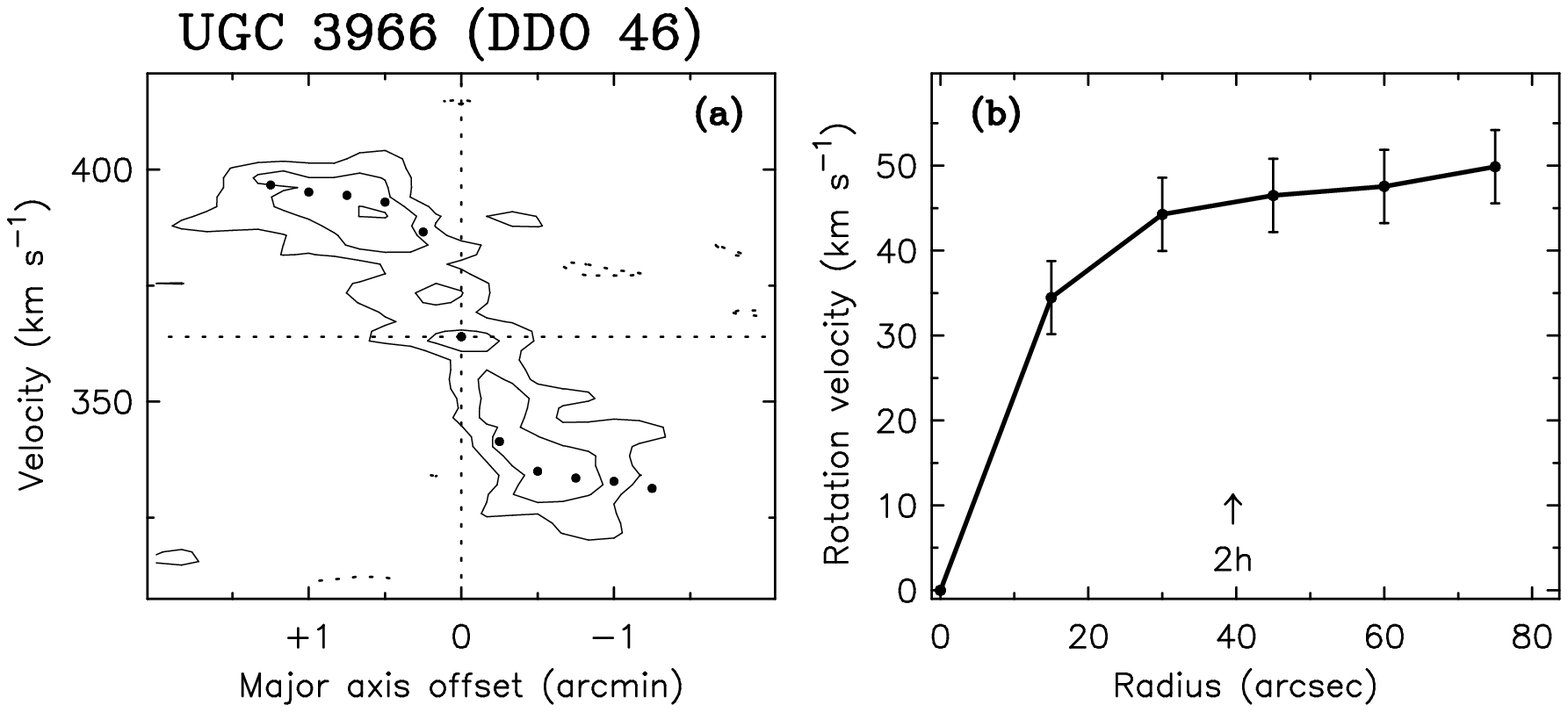}}
\kern0.5cm
\resizebox{0.45\hsize}{!}{\includegraphics{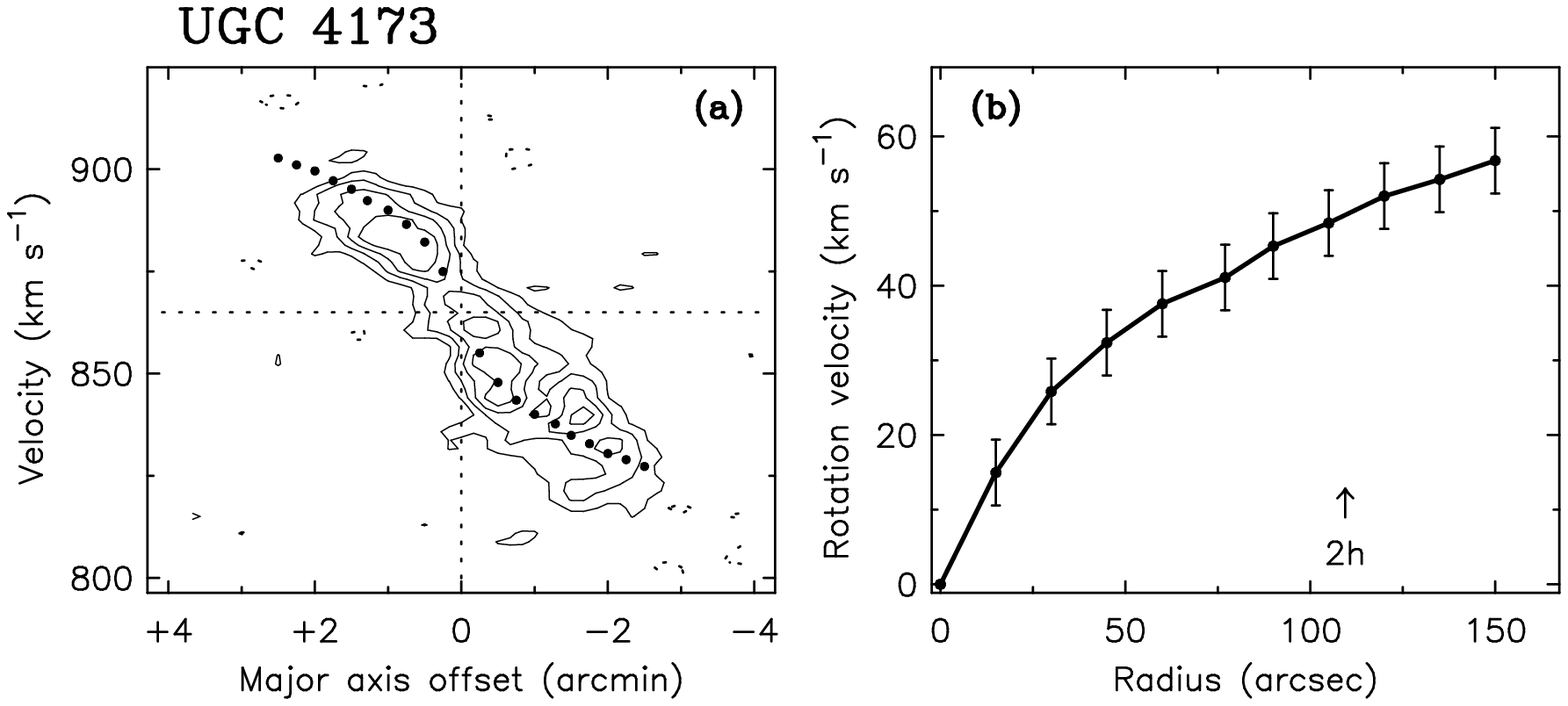}}
\newline
\vskip-0.15cm
\noindent
\resizebox{0.45\hsize}{!}{\includegraphics{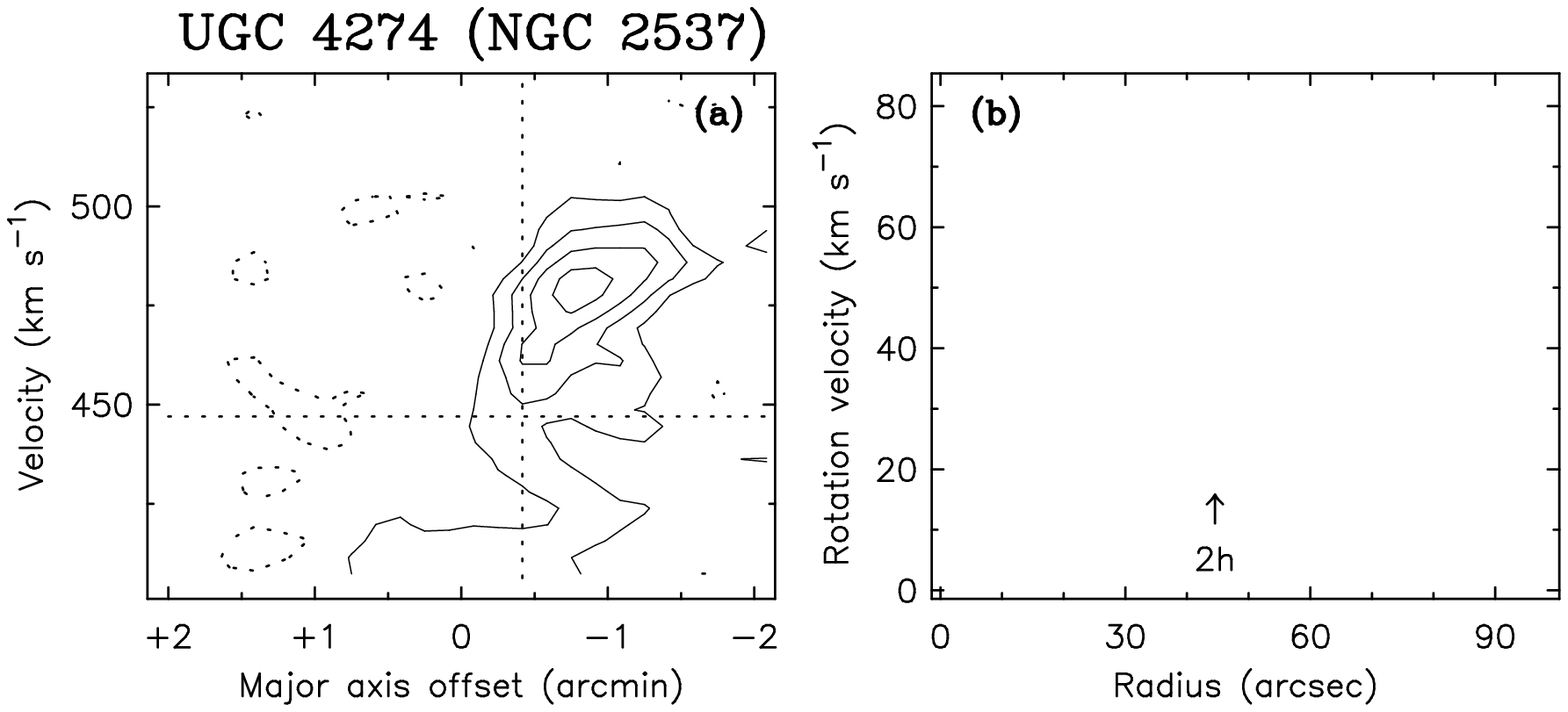}}
\kern0.5cm
\resizebox{0.45\hsize}{!}{\includegraphics{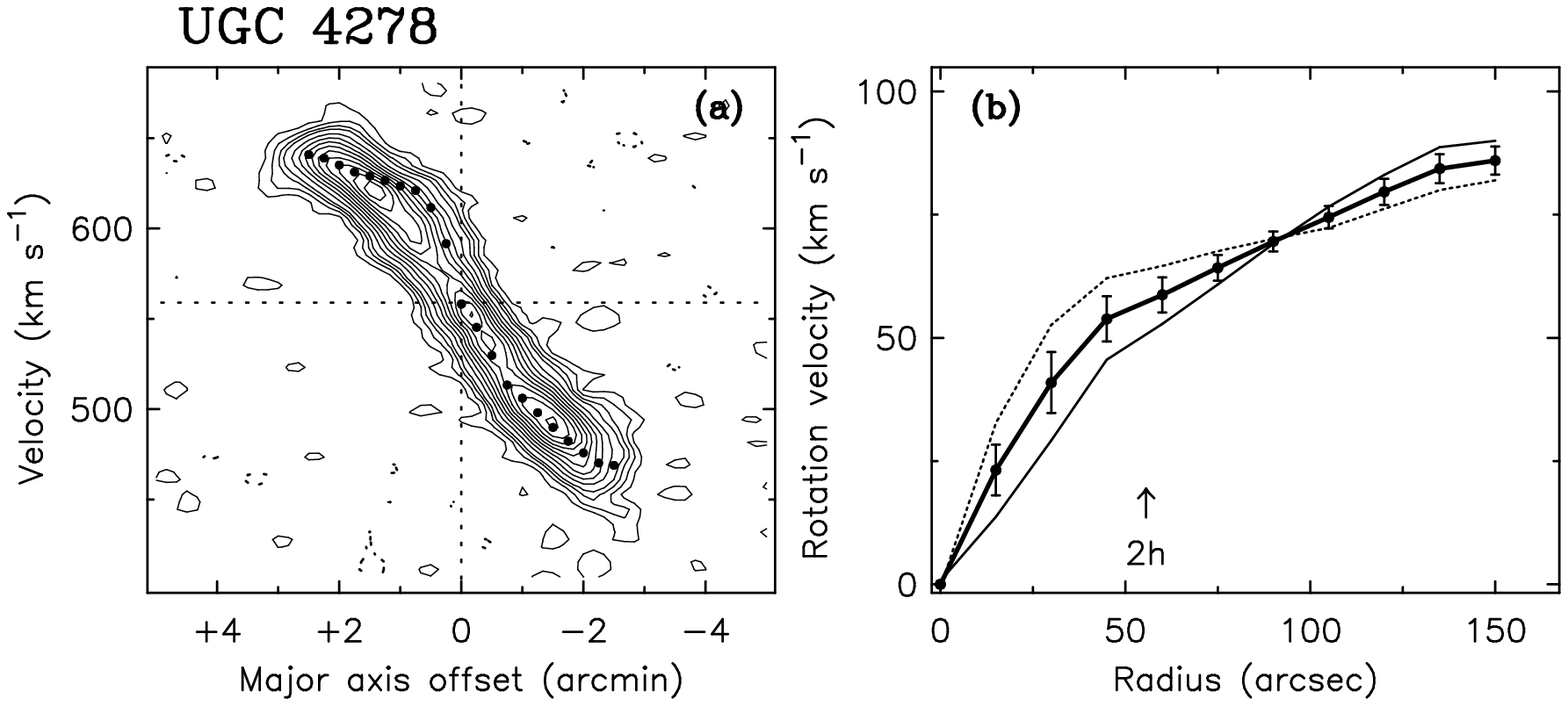}}
\newline
\vskip-0.15cm
\noindent
\resizebox{0.45\hsize}{!}{\includegraphics{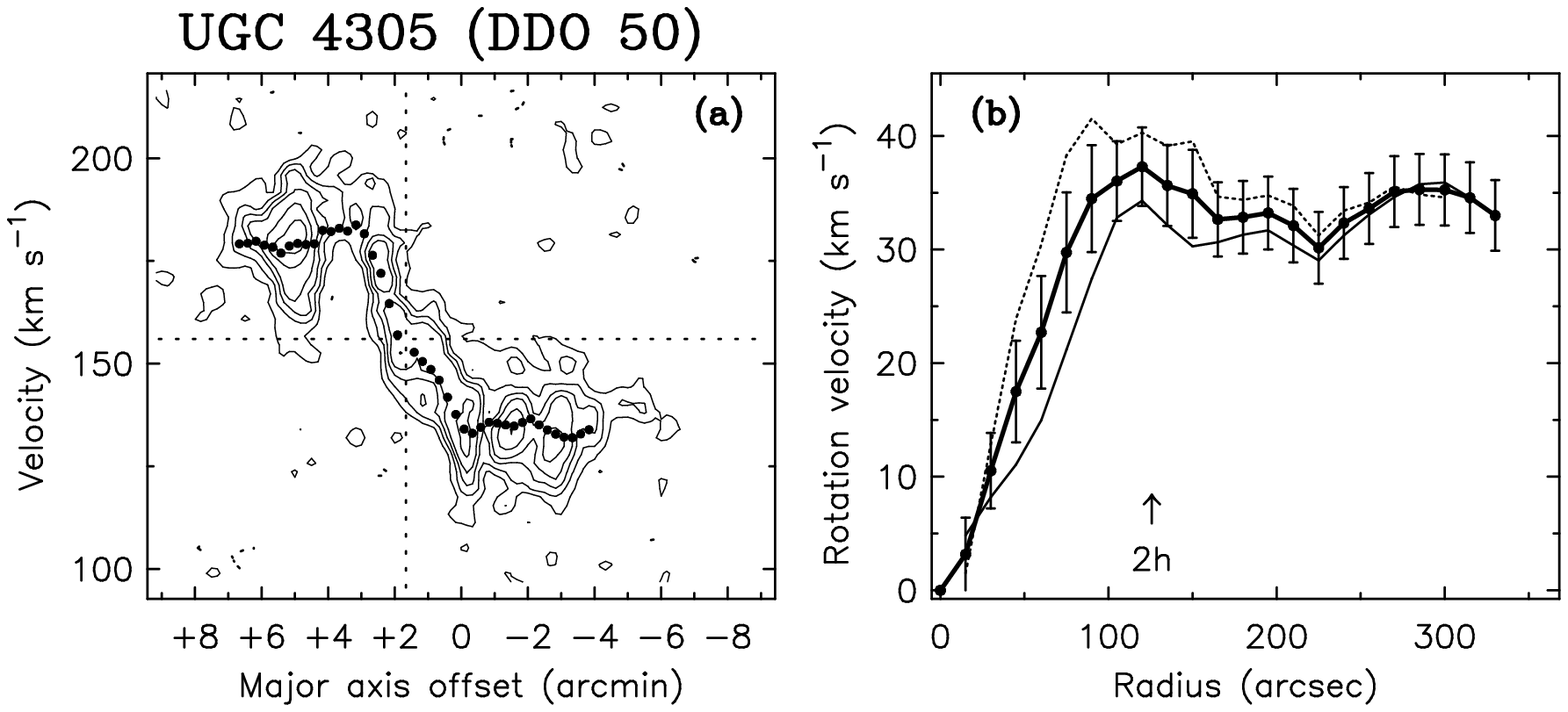}}
\kern0.5cm
\resizebox{0.45\hsize}{!}{\includegraphics{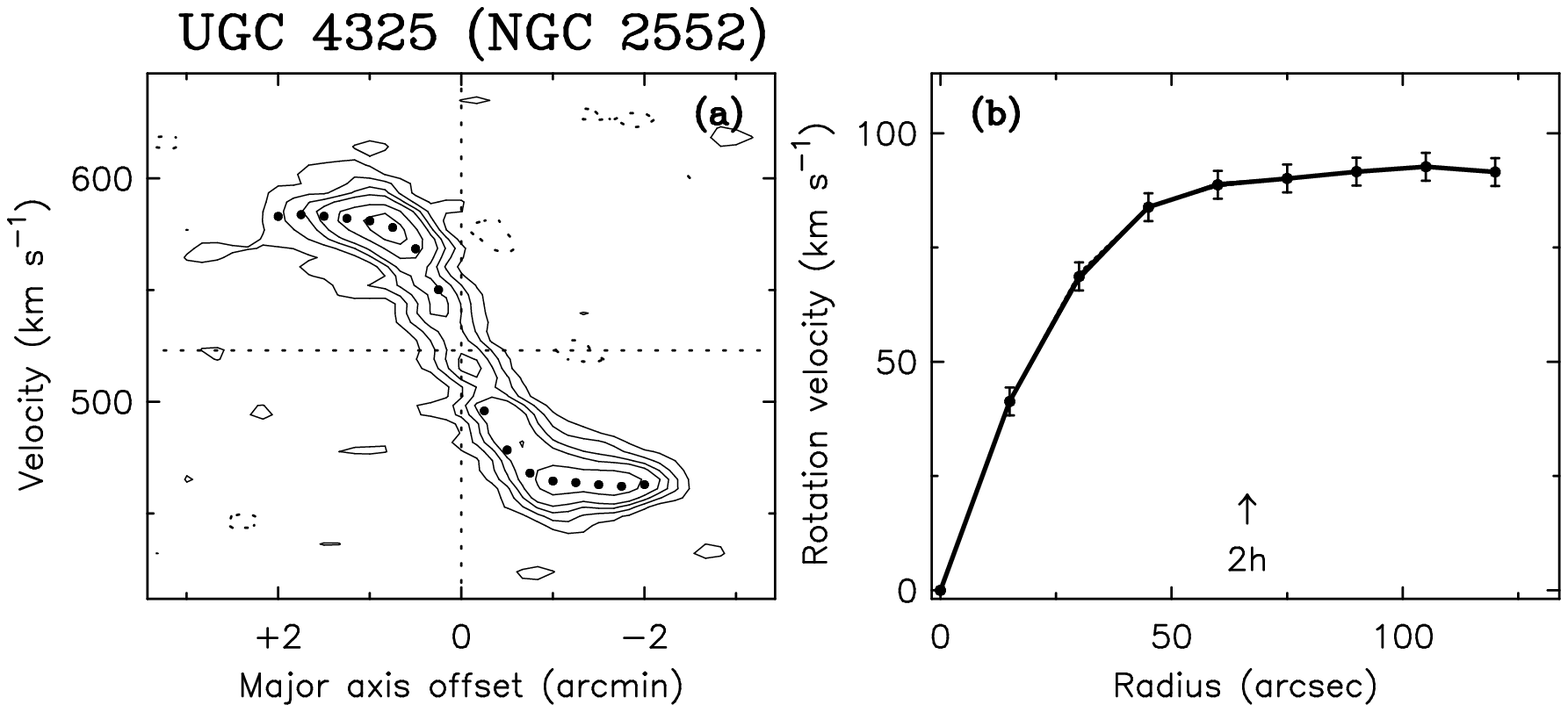}}
\newline
\vskip-0.15cm
\noindent
\resizebox{0.45\hsize}{!}{\includegraphics{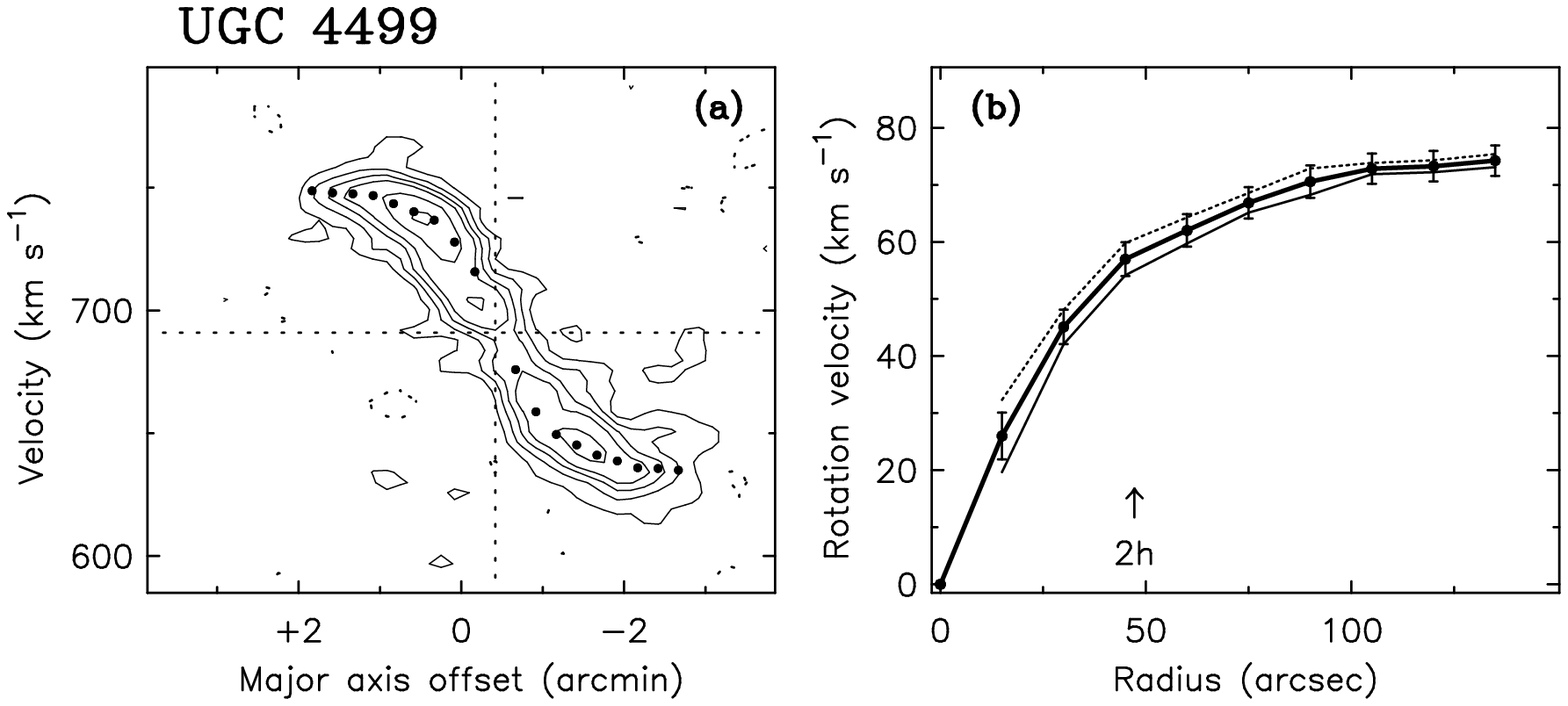}}
\kern0.5cm
\resizebox{0.45\hsize}{!}{\includegraphics{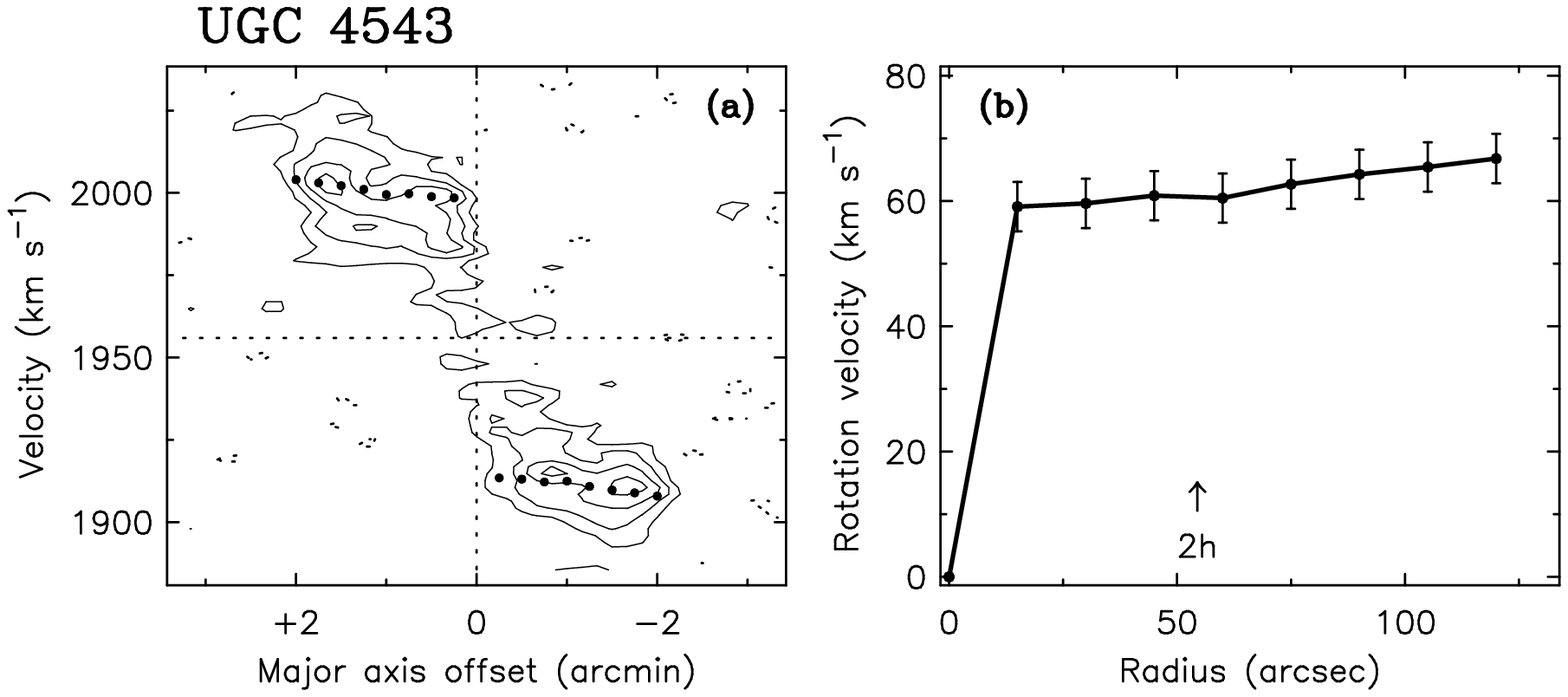}}
\newline
\vskip-0.15cm
\noindent
\resizebox{0.45\hsize}{!}{\includegraphics{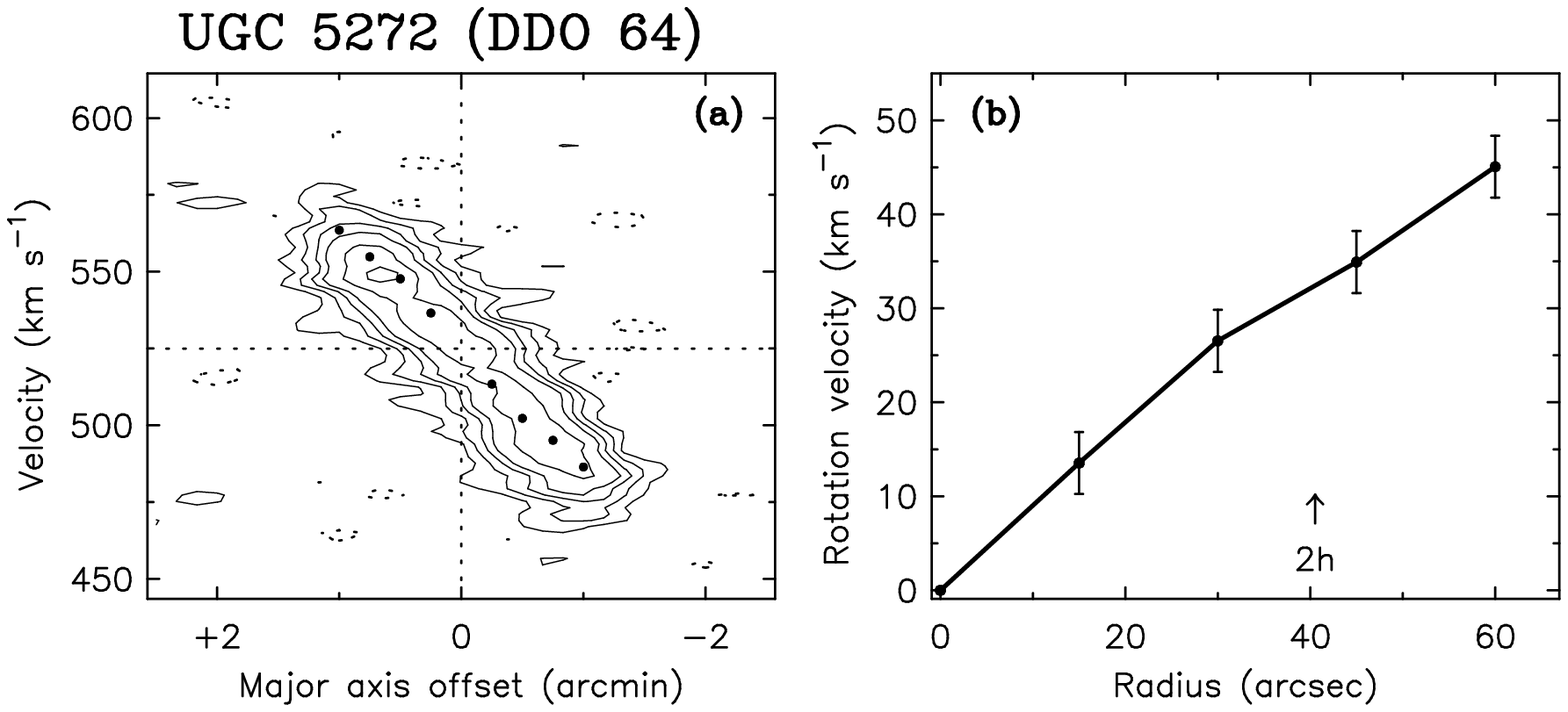}}
\kern0.5cm
\resizebox{0.45\hsize}{!}{\includegraphics{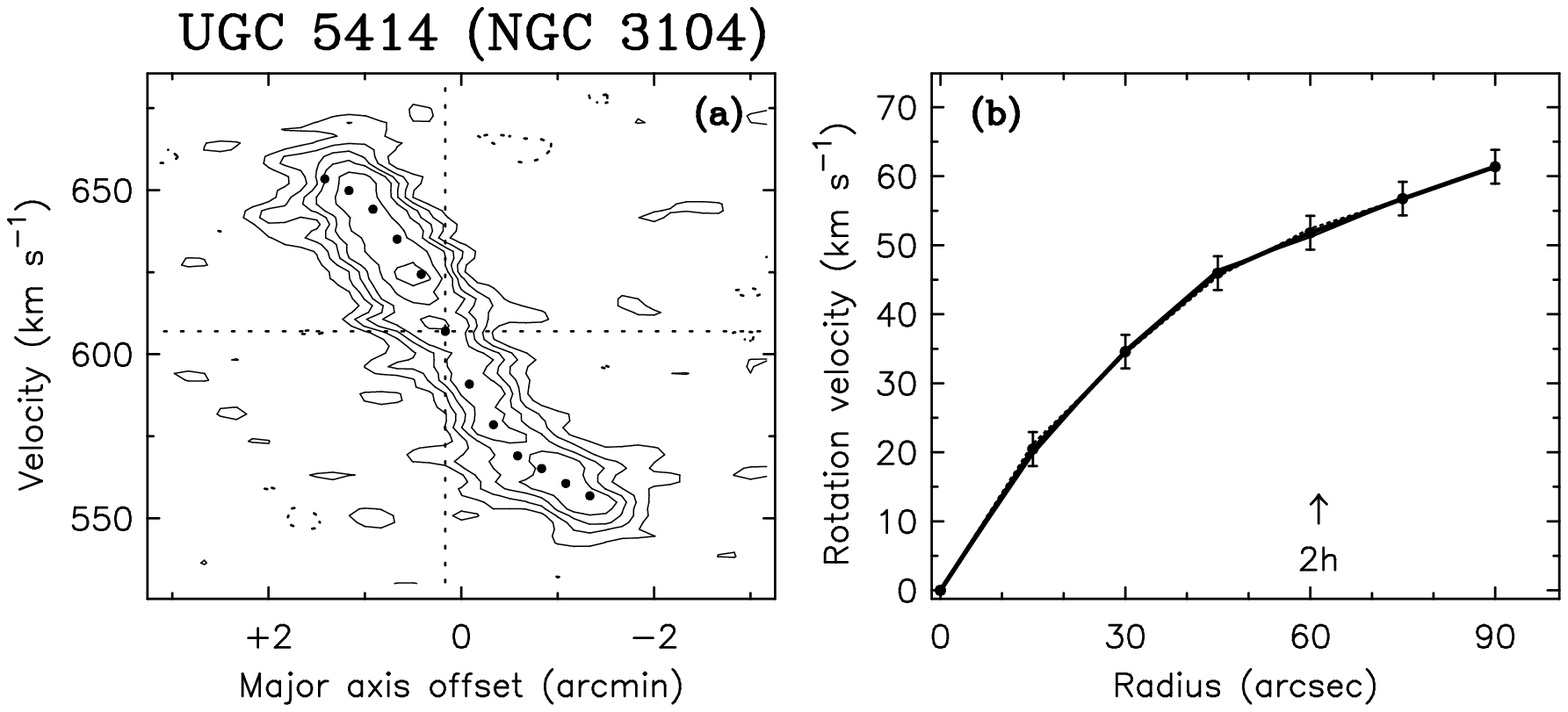}}
\newline
\vskip-0.15cm
\noindent
\resizebox{0.45\hsize}{!}{\includegraphics{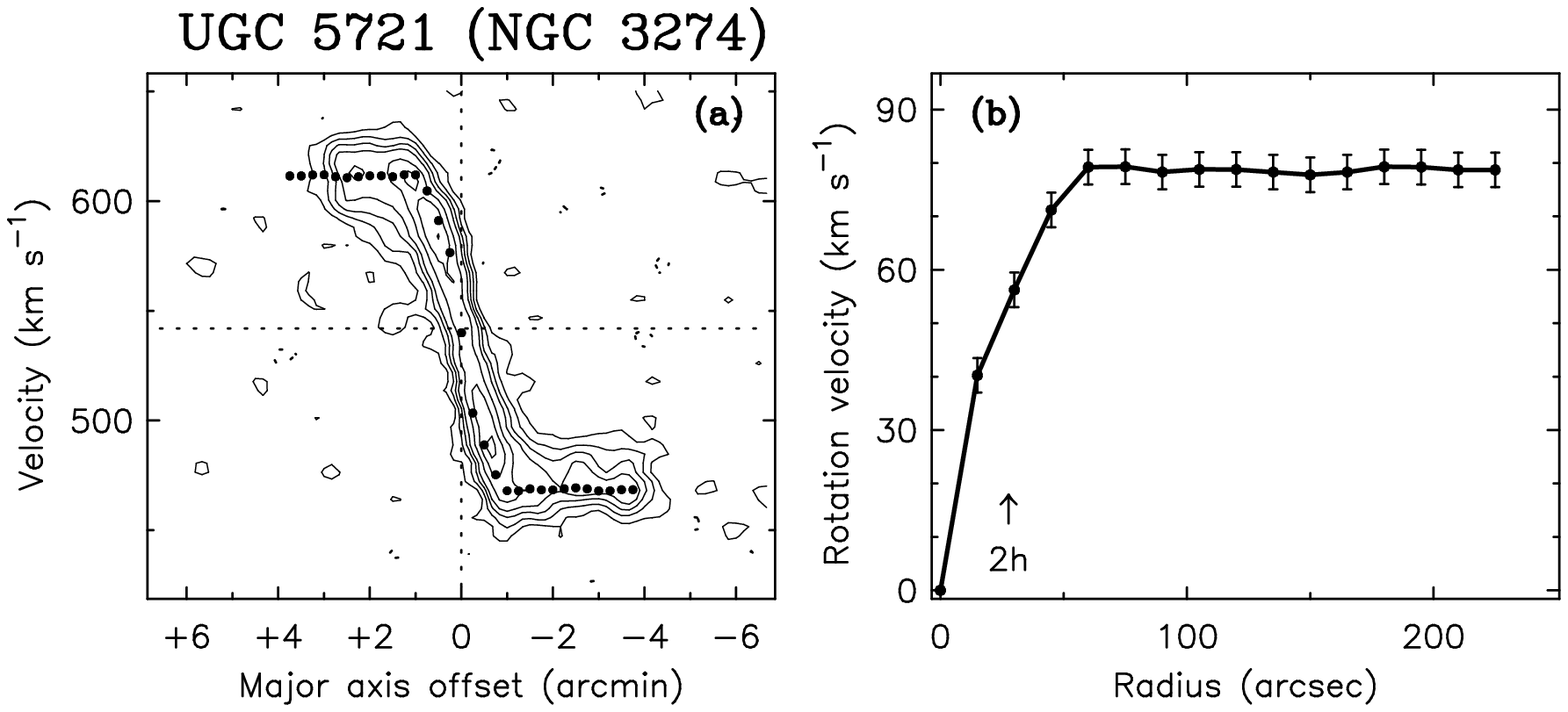}}
\kern0.5cm
\resizebox{0.45\hsize}{!}{\includegraphics{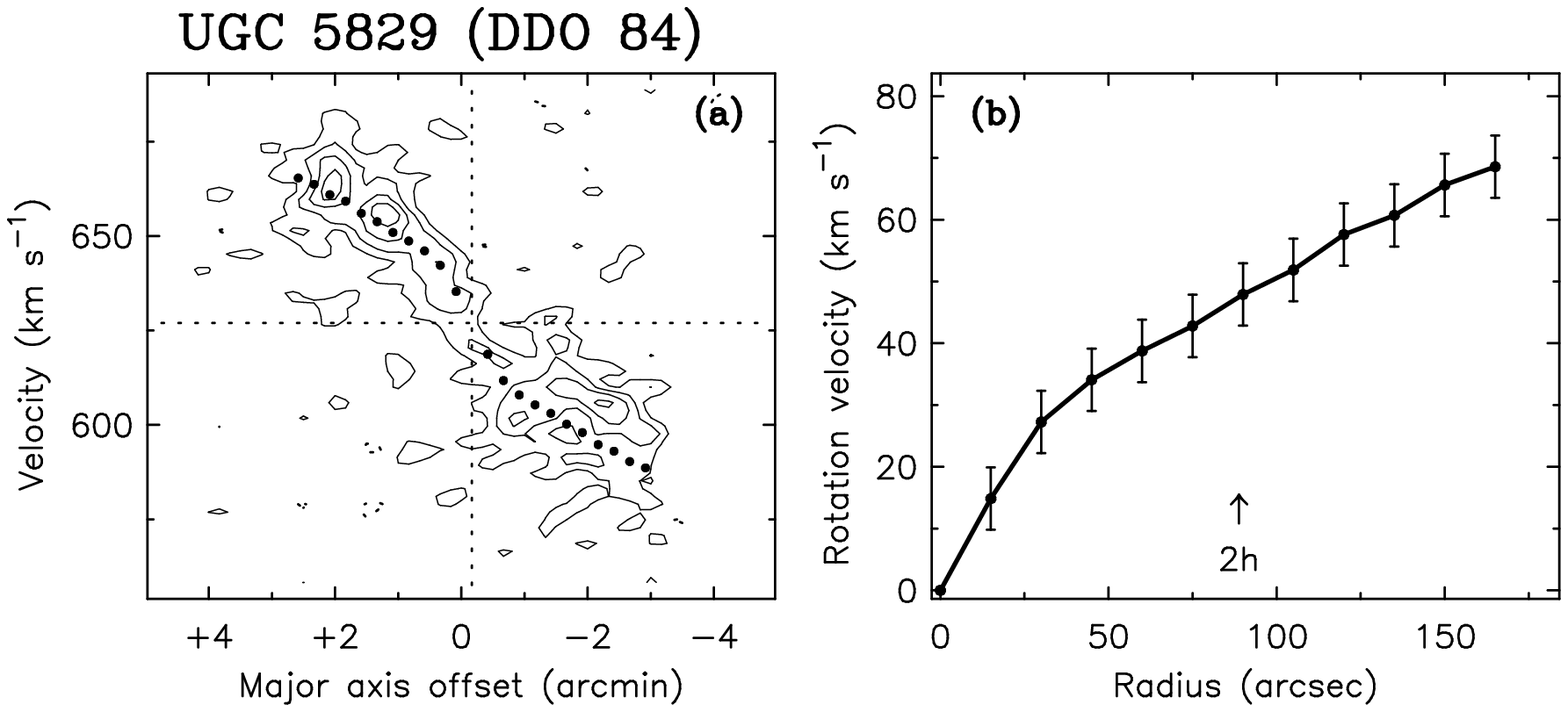}}
\clearpage
\vskip-0.15cm
\noindent
\resizebox{0.45\hsize}{!}{\includegraphics{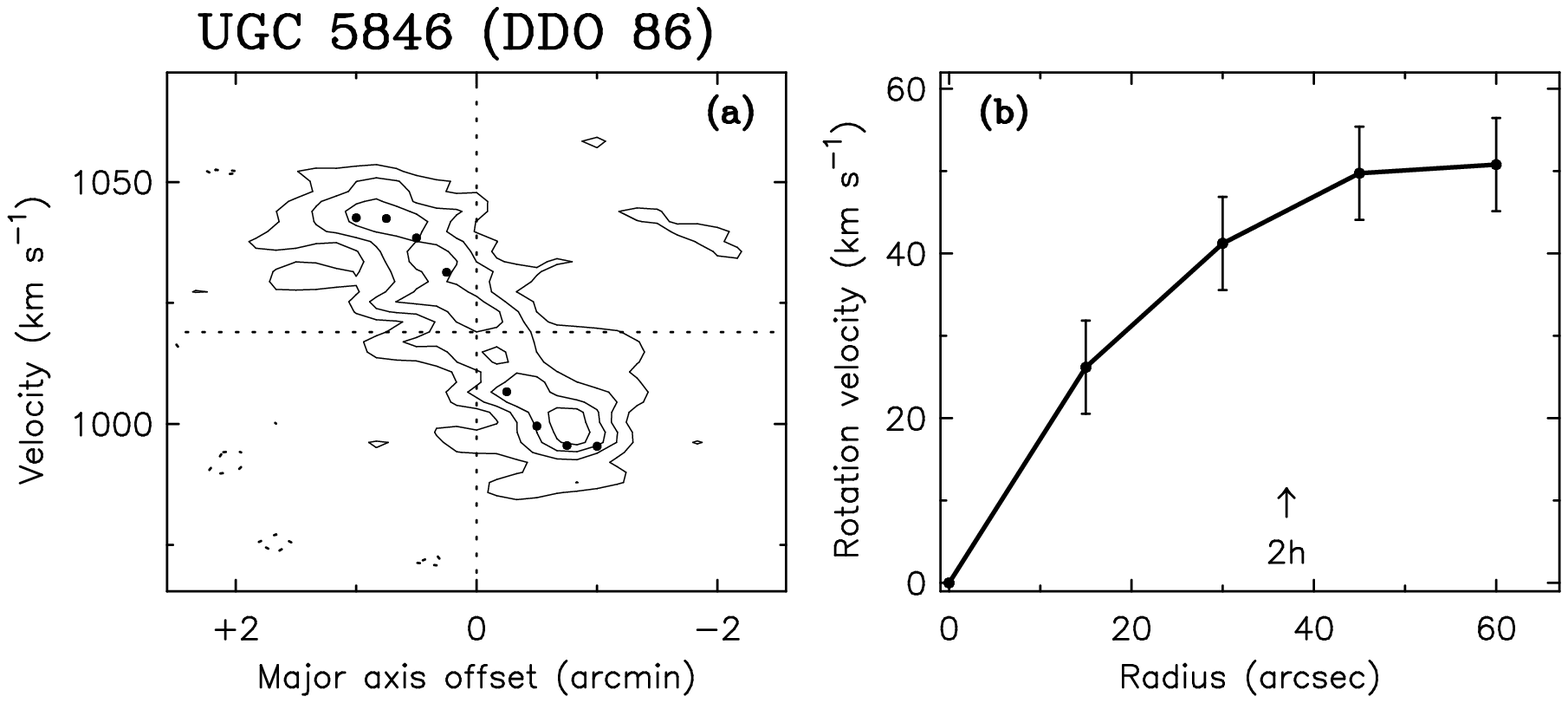}}
\kern0.5cm
\resizebox{0.45\hsize}{!}{\includegraphics{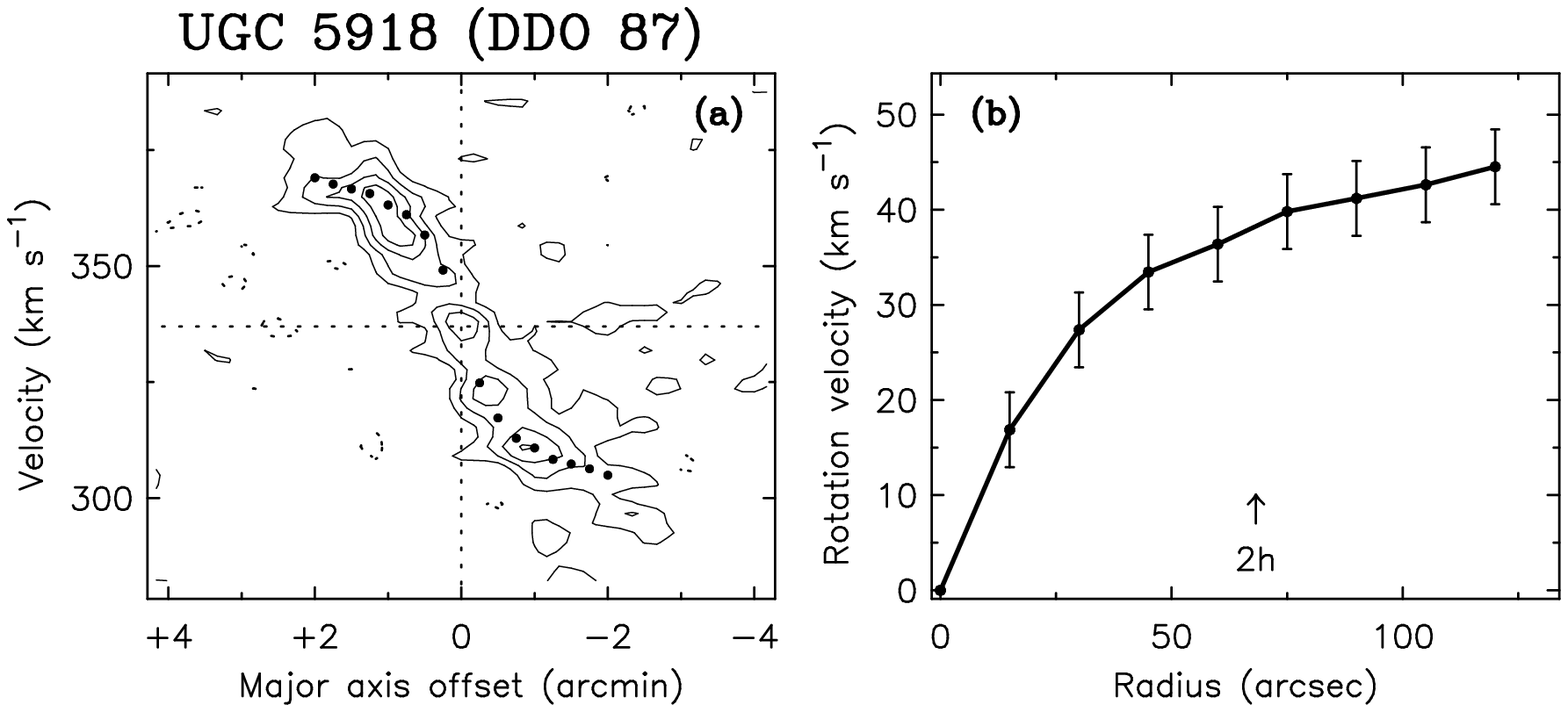}}
\newline
\vskip-0.15cm
\noindent
\resizebox{0.45\hsize}{!}{\includegraphics{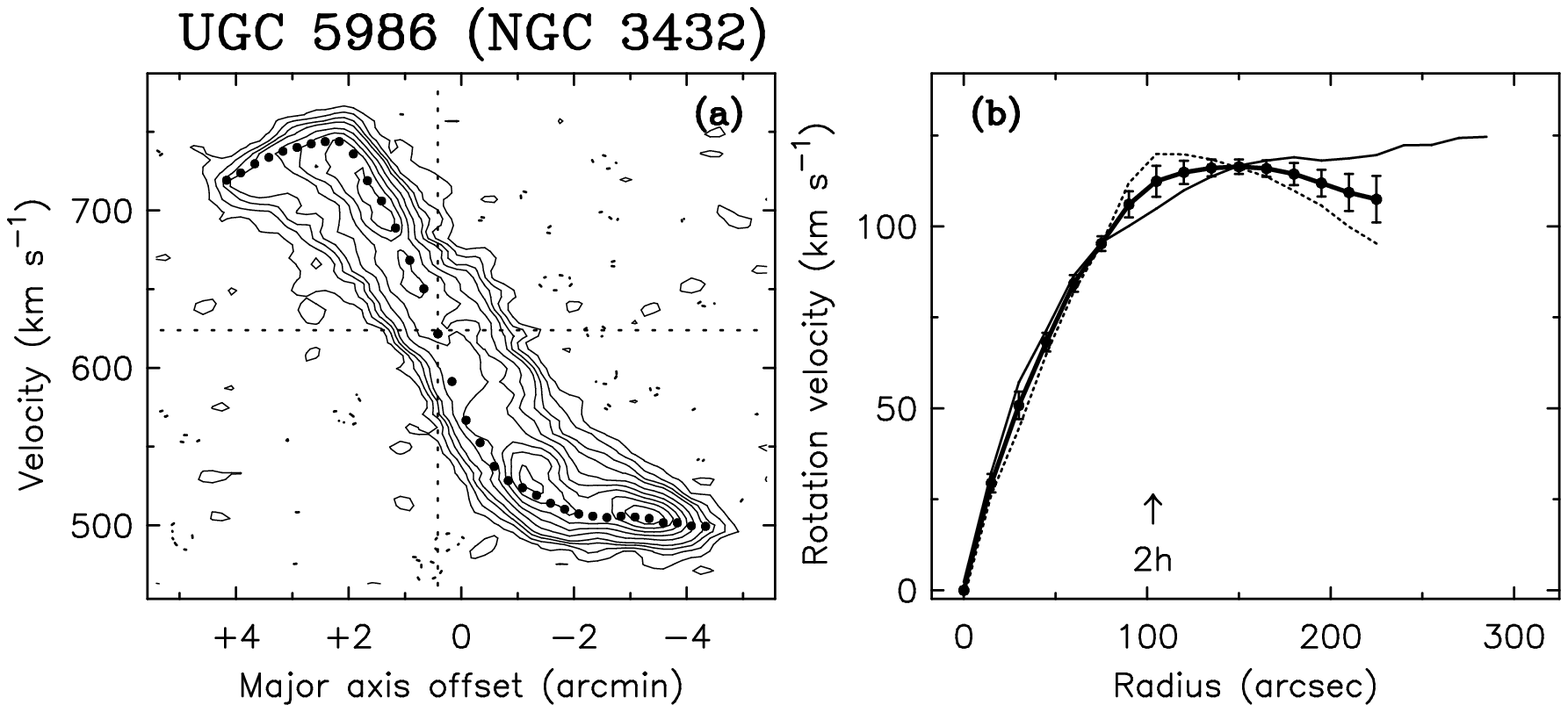}}
\kern0.5cm
\resizebox{0.45\hsize}{!}{\includegraphics{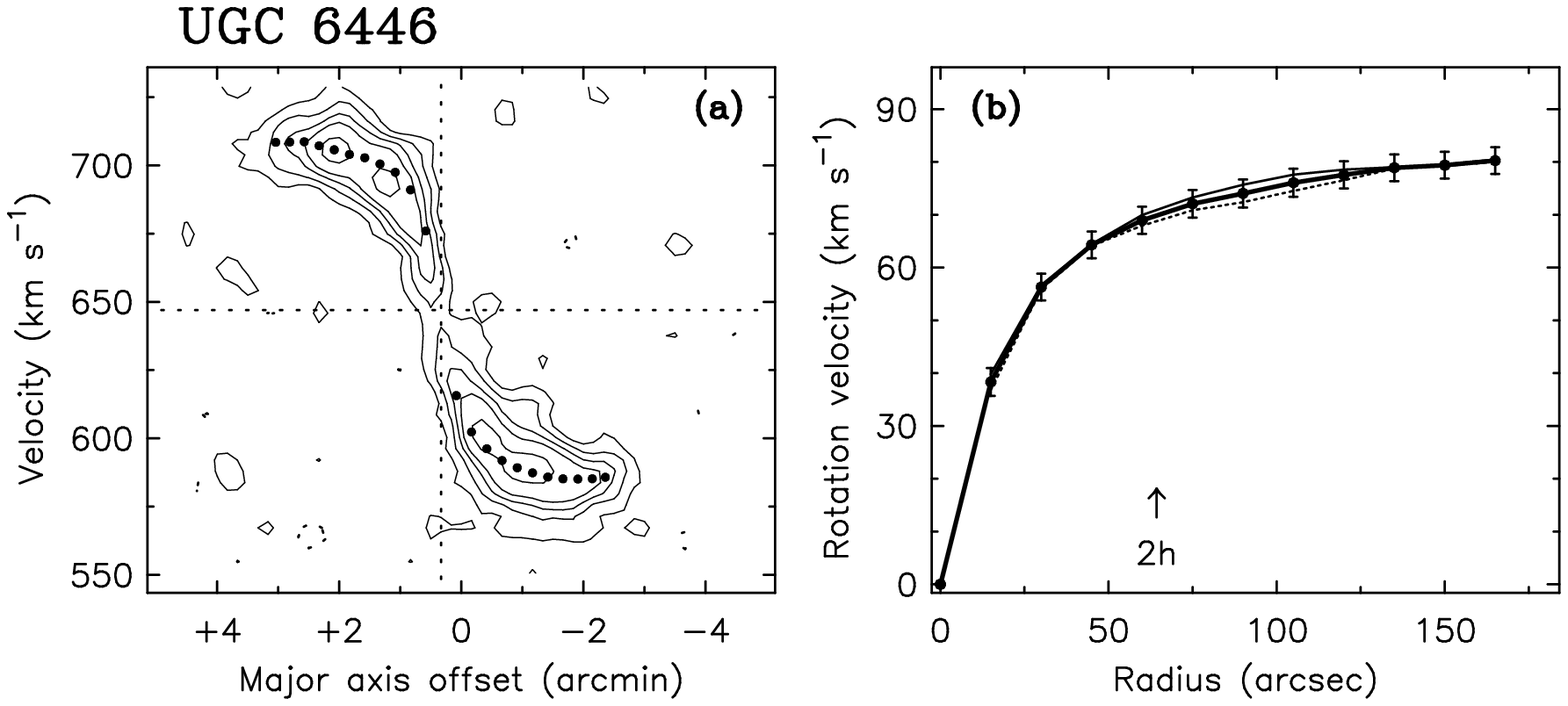}}
\newline
\vskip-0.15cm
\noindent
\resizebox{0.45\hsize}{!}{\includegraphics{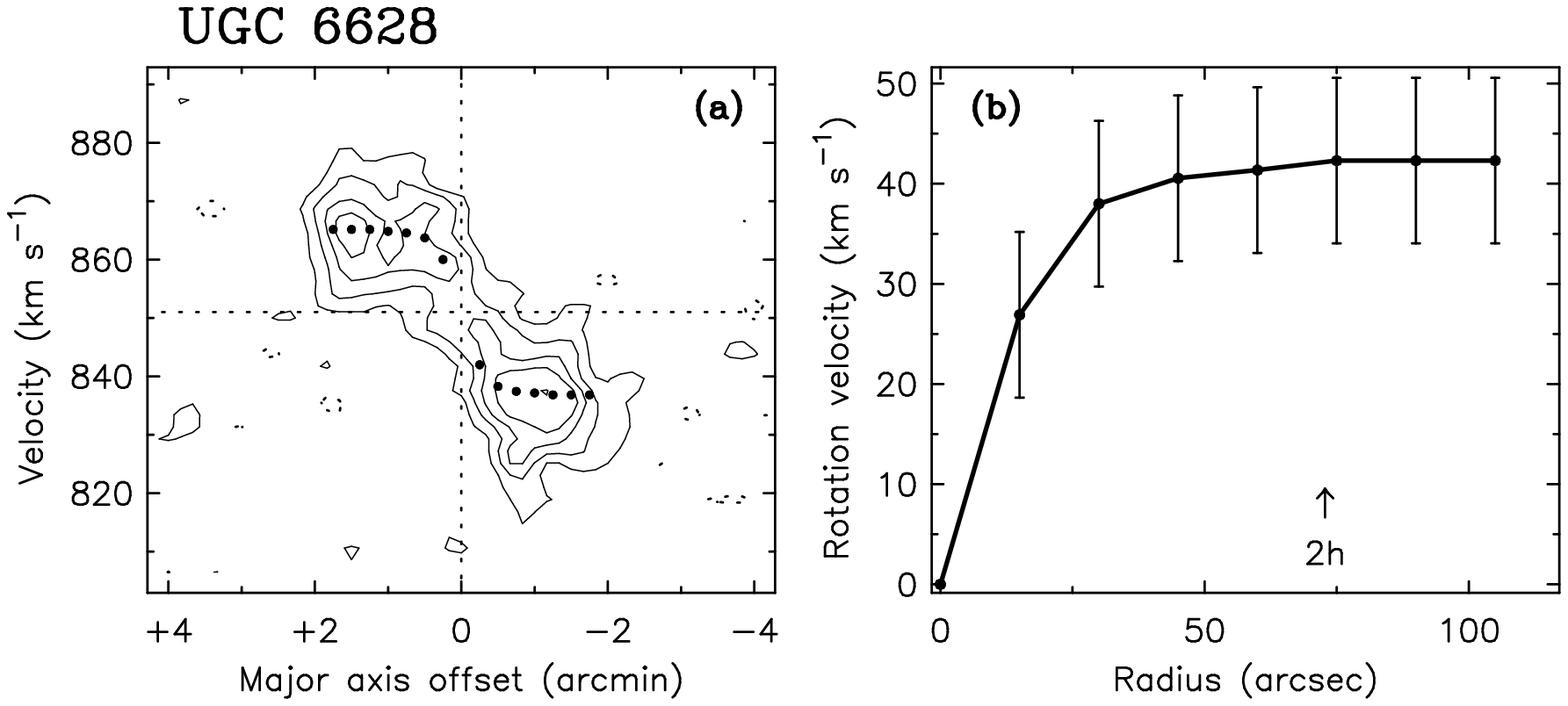}}
\kern0.5cm
\resizebox{0.45\hsize}{!}{\includegraphics{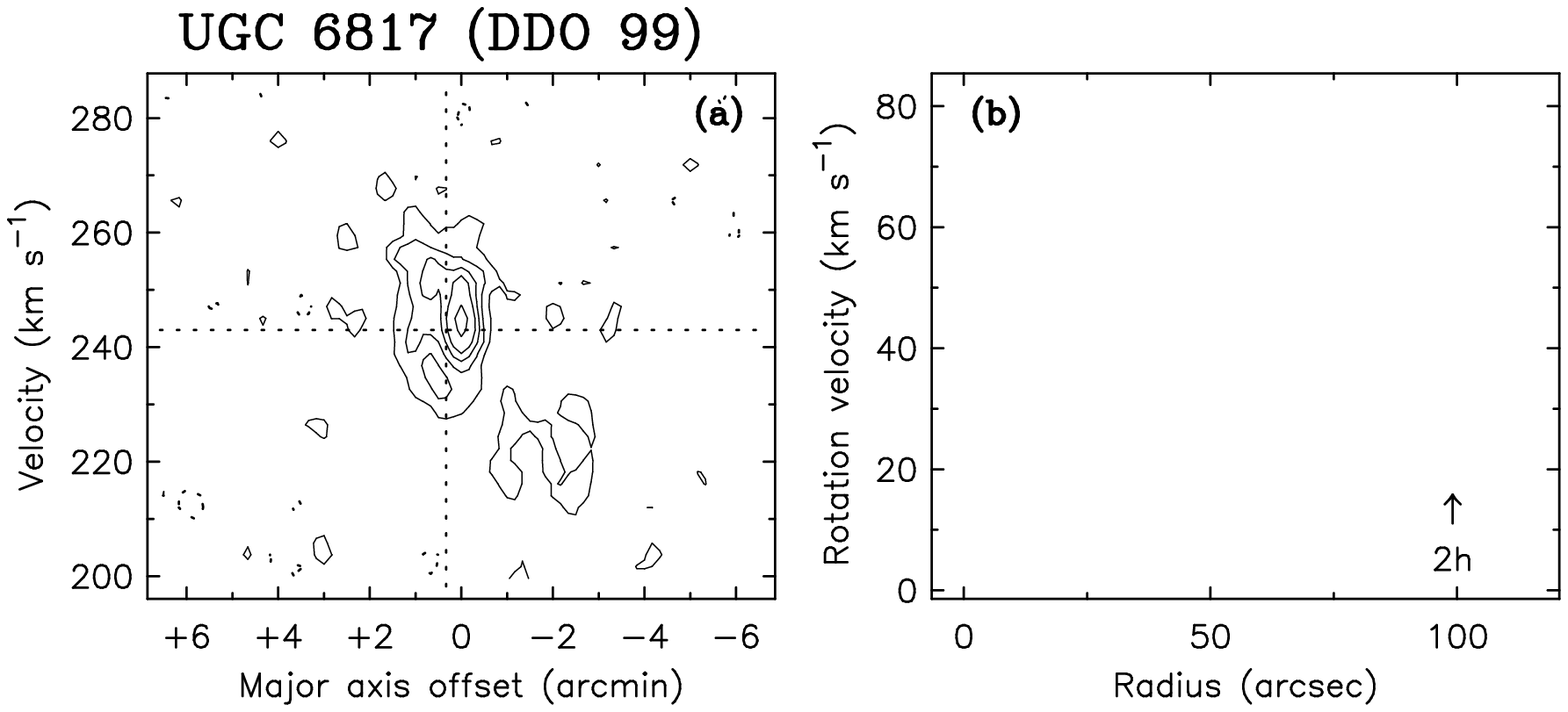}}
\newline
\vskip-0.15cm
\noindent
\resizebox{0.45\hsize}{!}{\includegraphics{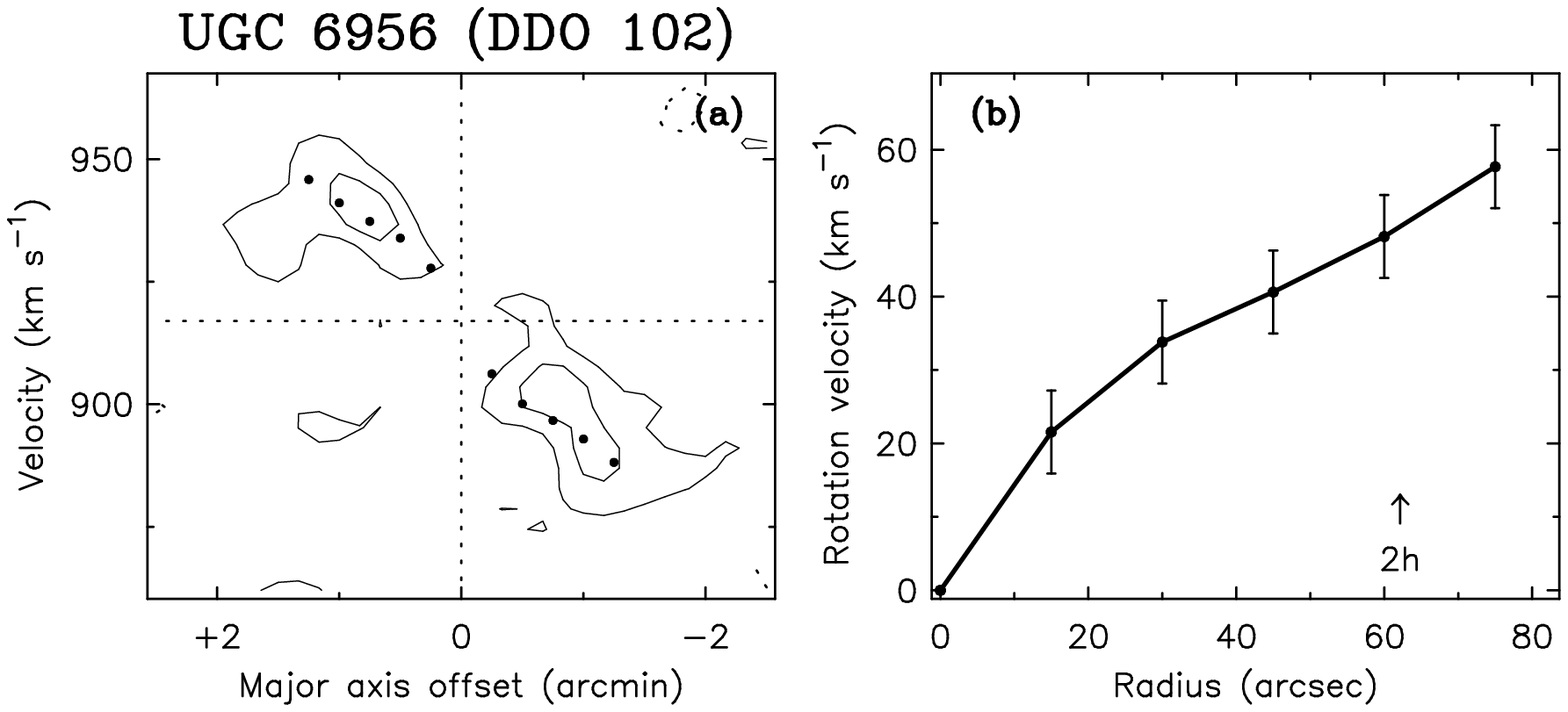}}
\kern0.5cm
\resizebox{0.45\hsize}{!}{\includegraphics{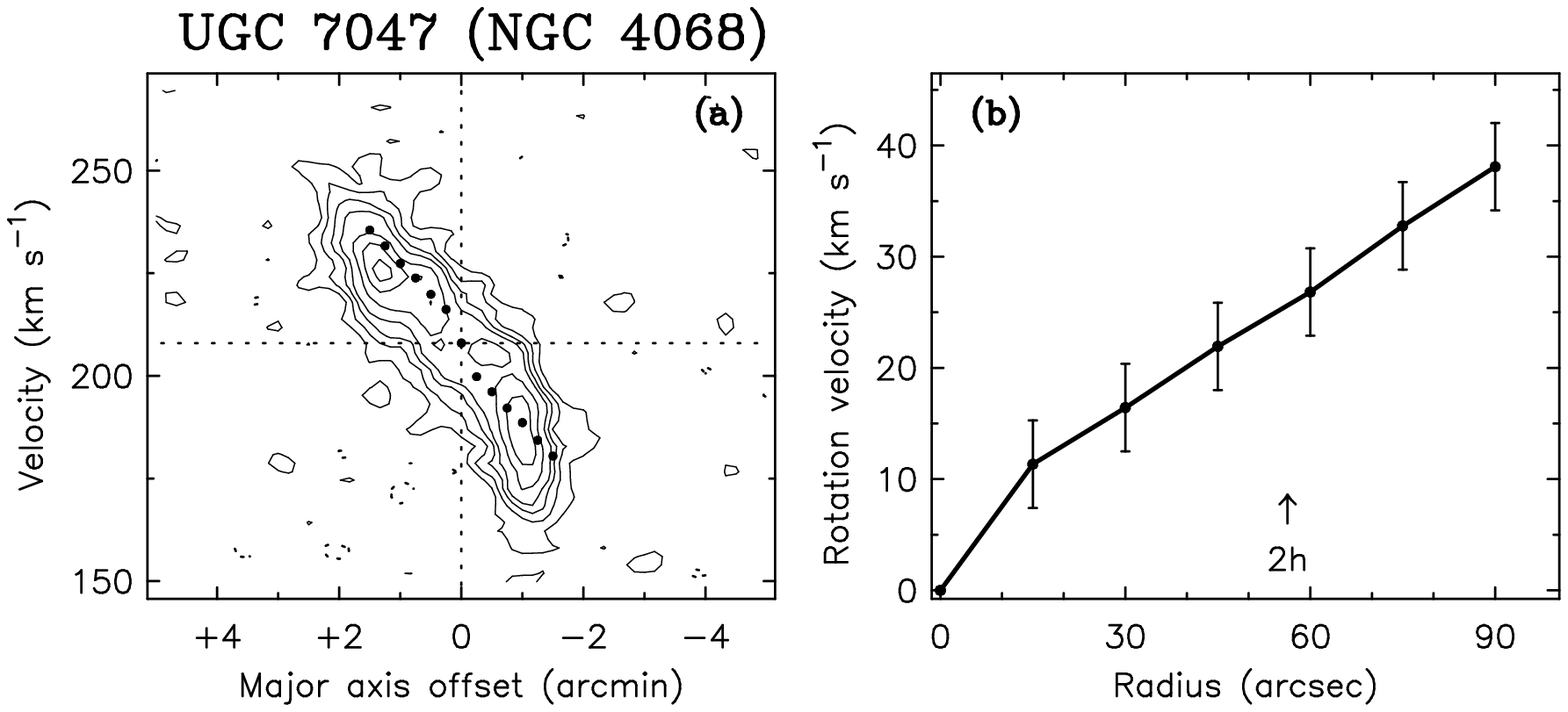}}
\newline
\vskip-0.15cm
\noindent
\resizebox{0.45\hsize}{!}{\includegraphics{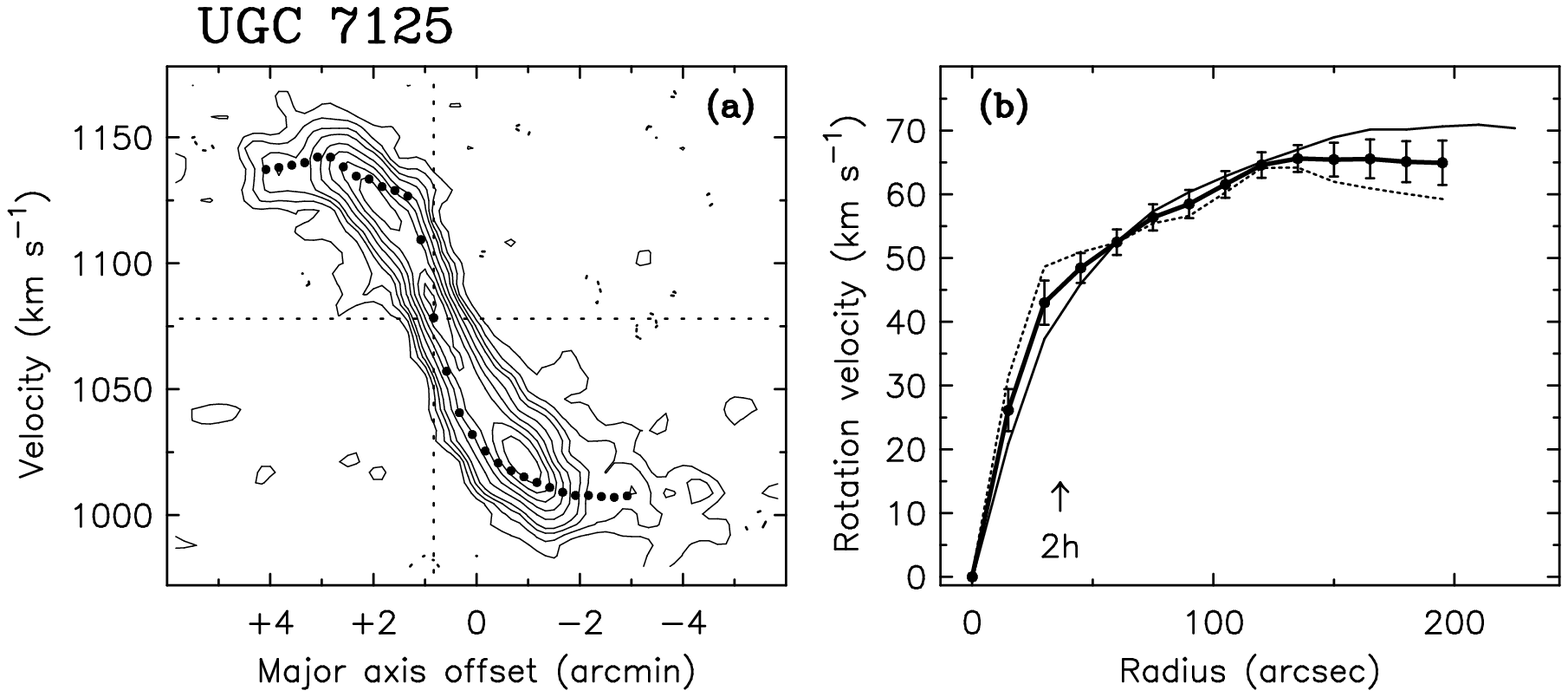}}
\kern0.5cm
\resizebox{0.45\hsize}{!}{\includegraphics{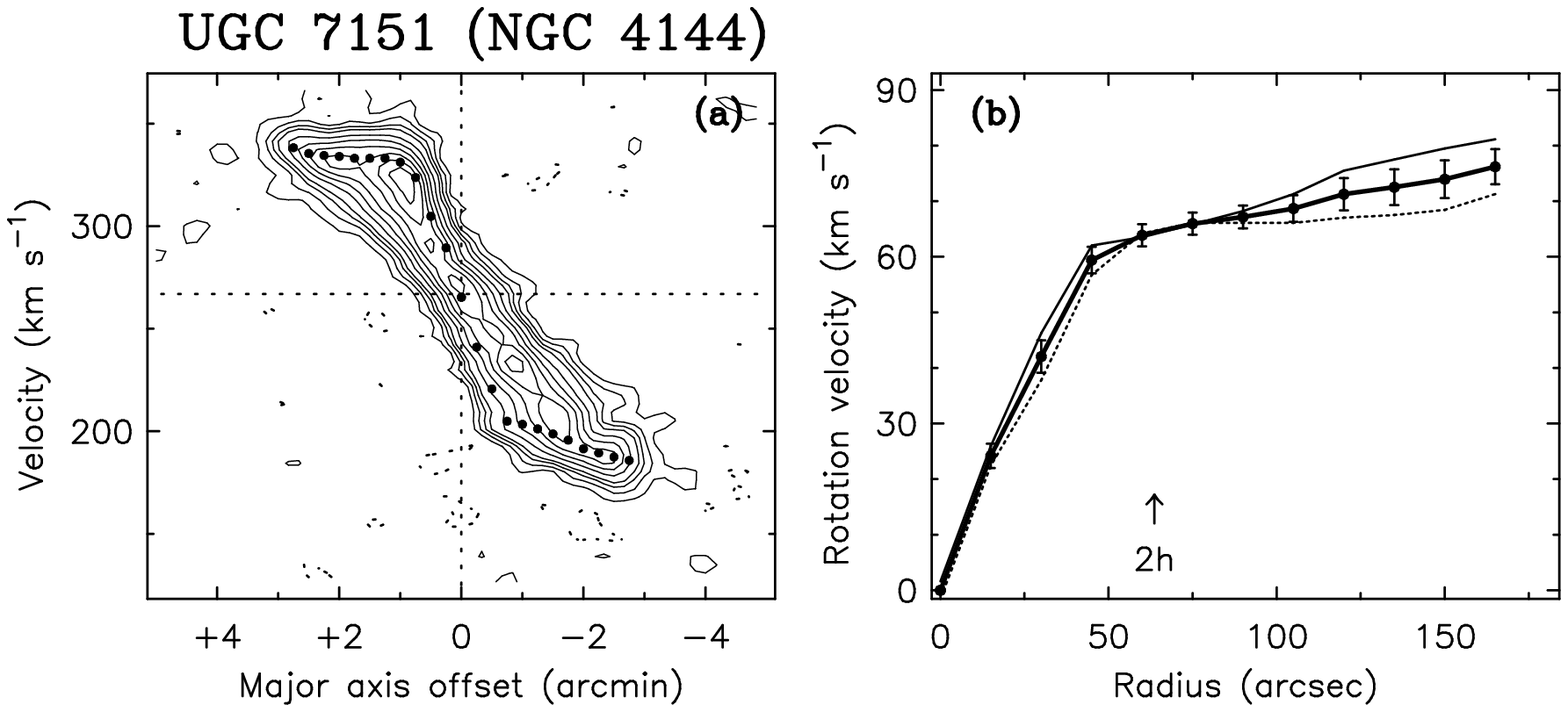}}
\newline
\vskip-0.15cm
\noindent
\resizebox{0.45\hsize}{!}{\includegraphics{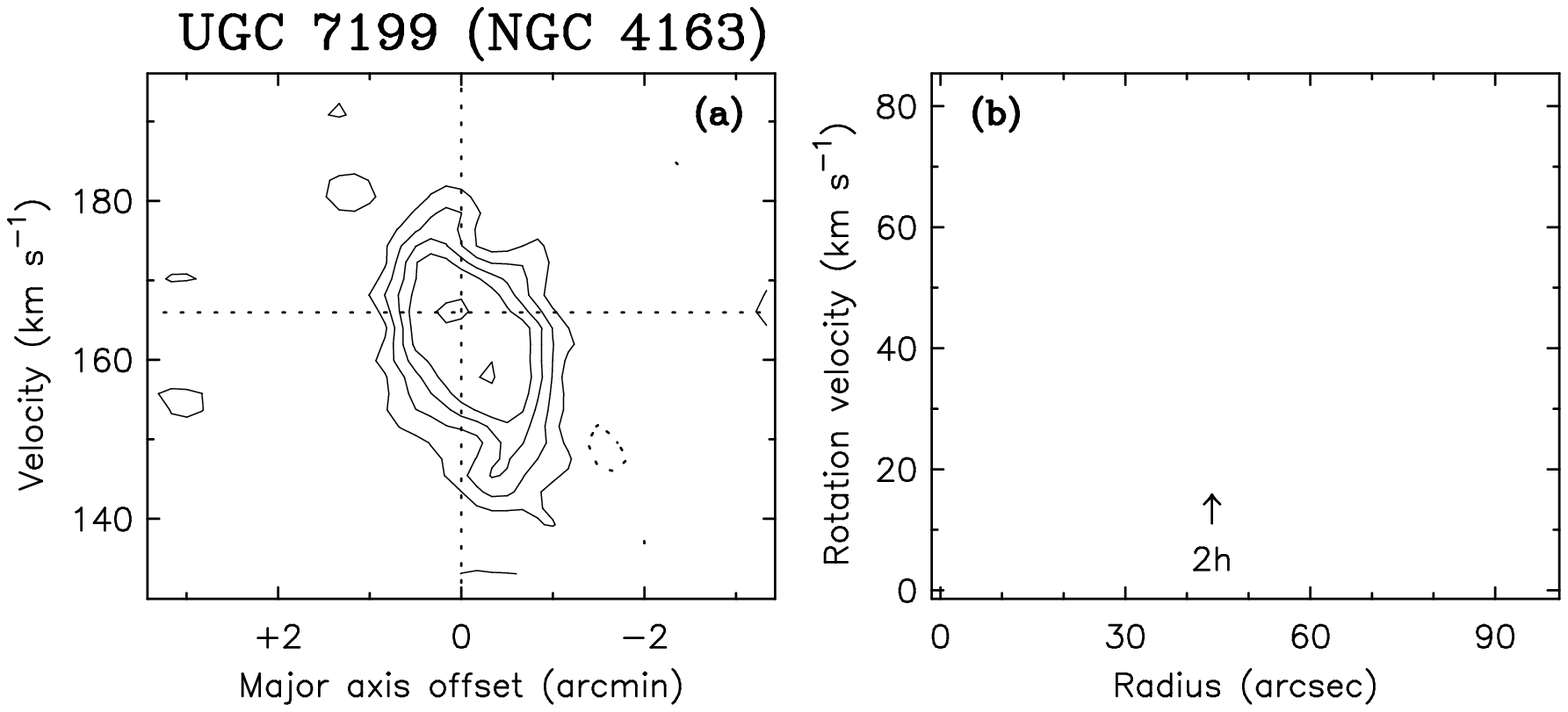}}
\kern0.5cm
\resizebox{0.45\hsize}{!}{\includegraphics{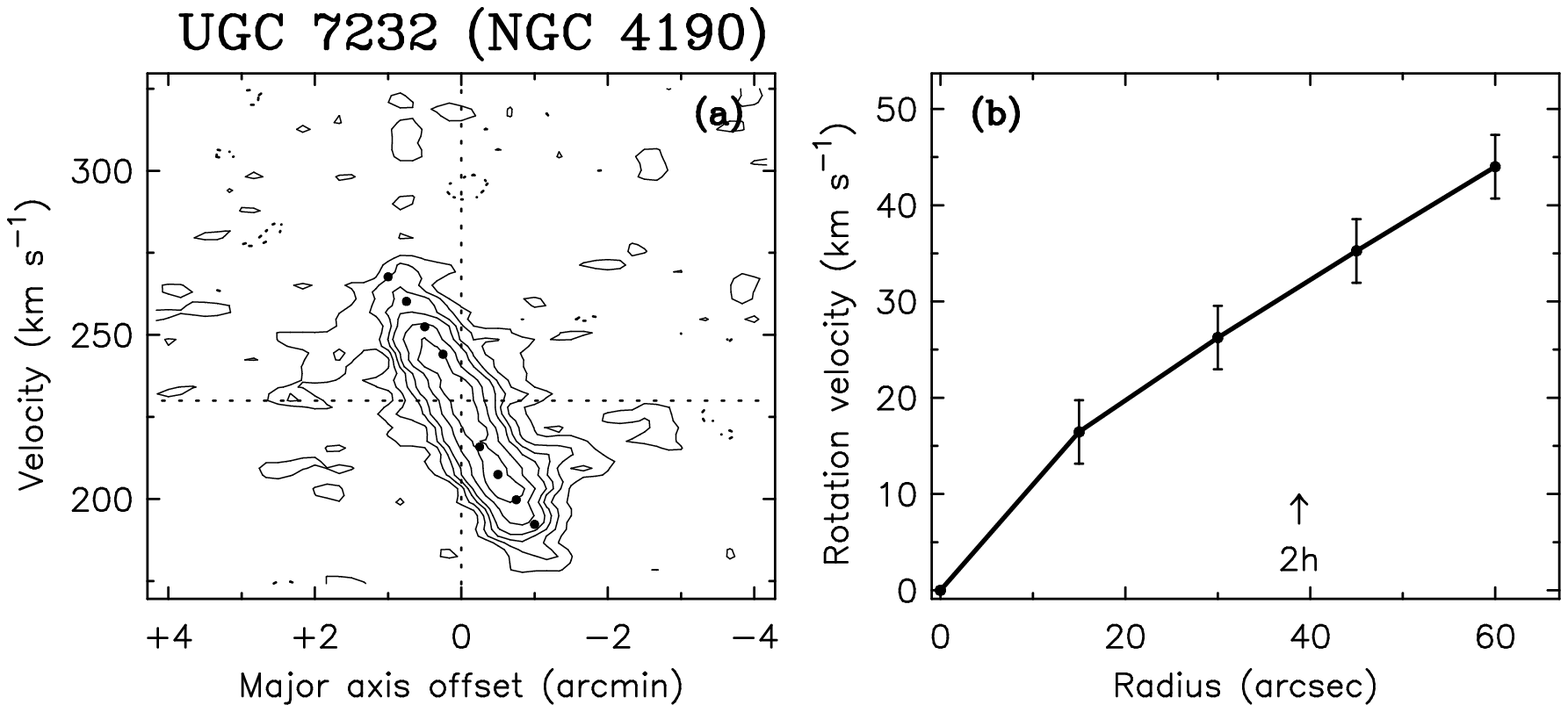}}
\clearpage
\vskip-0.15cm
\noindent
\resizebox{0.45\hsize}{!}{\includegraphics{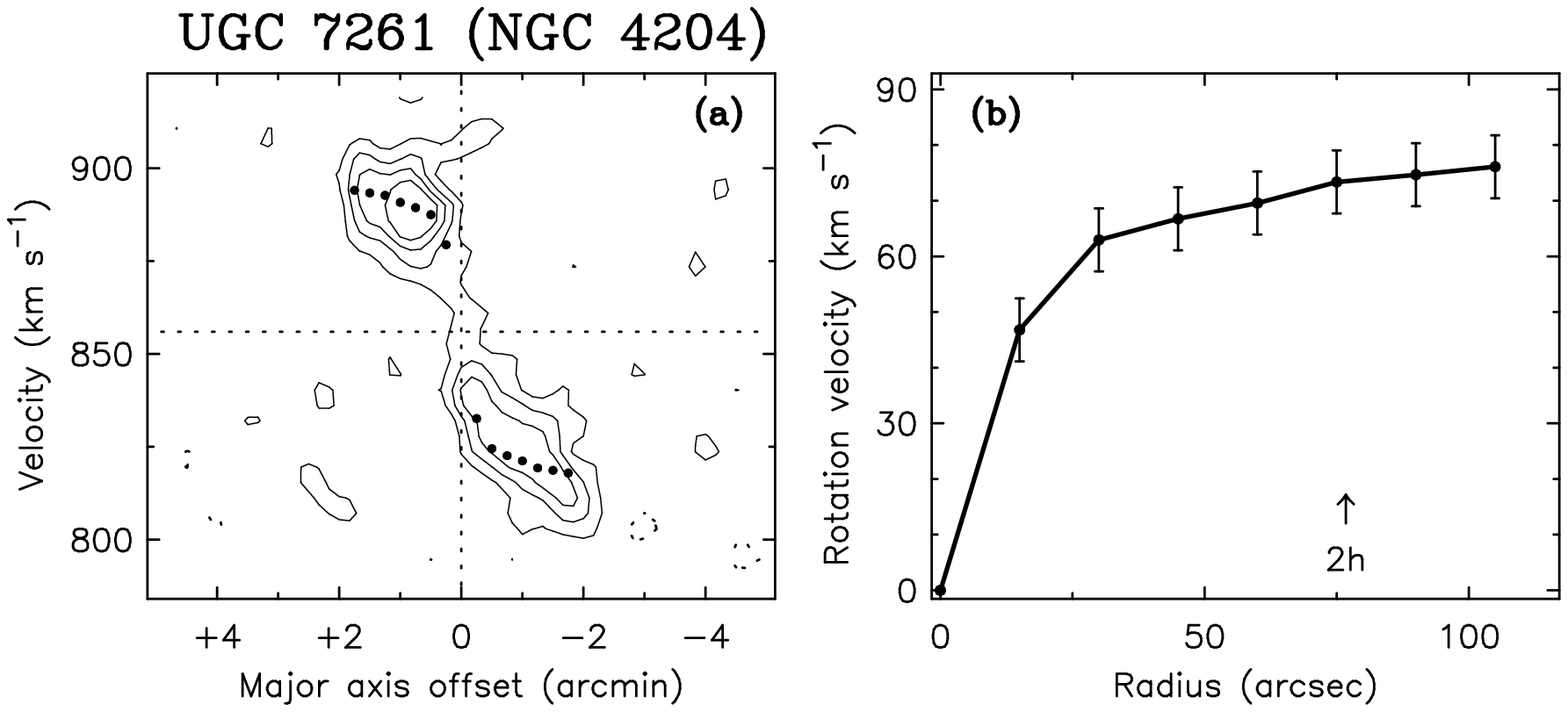}}
\kern0.5cm
\resizebox{0.45\hsize}{!}{\includegraphics{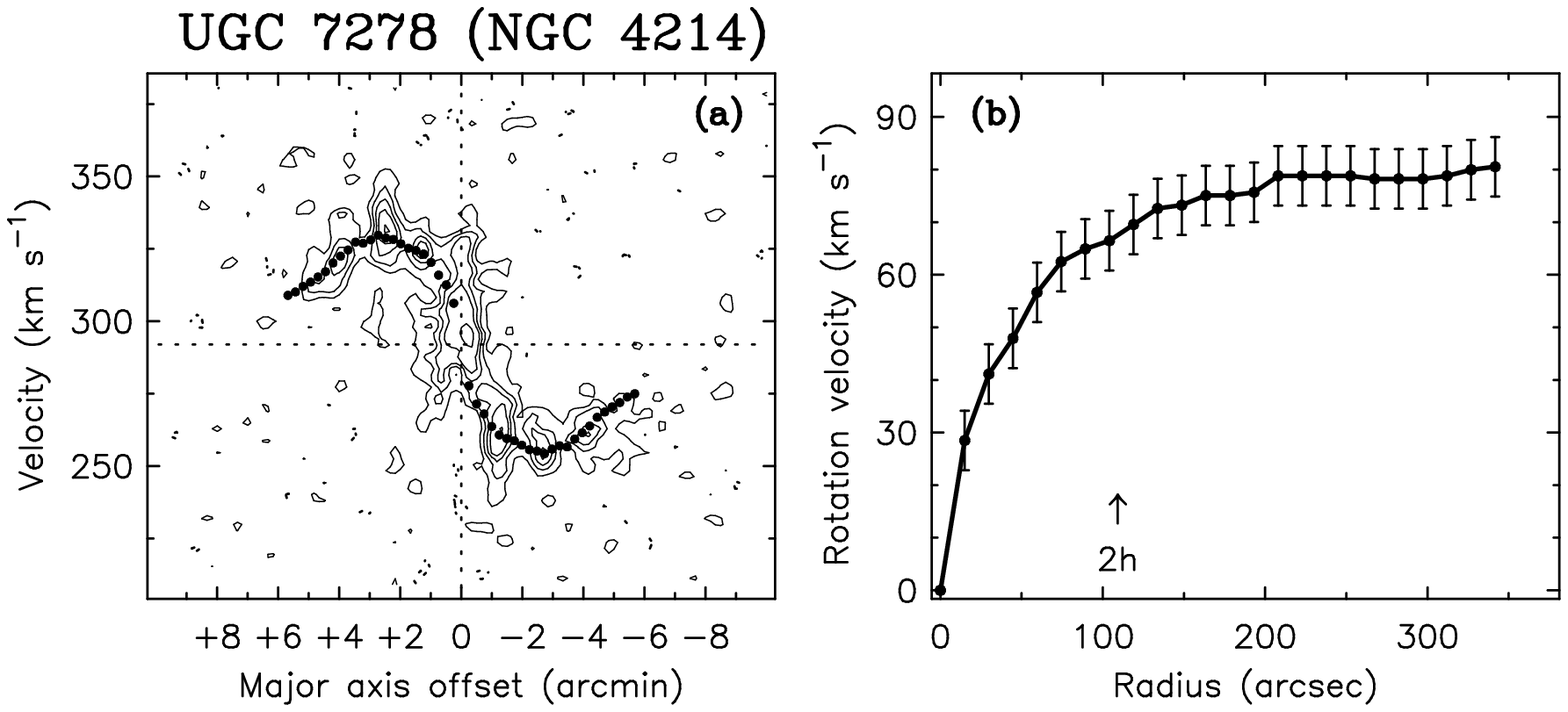}}
\newline
\vskip-0.15cm
\noindent
\resizebox{0.45\hsize}{!}{\includegraphics{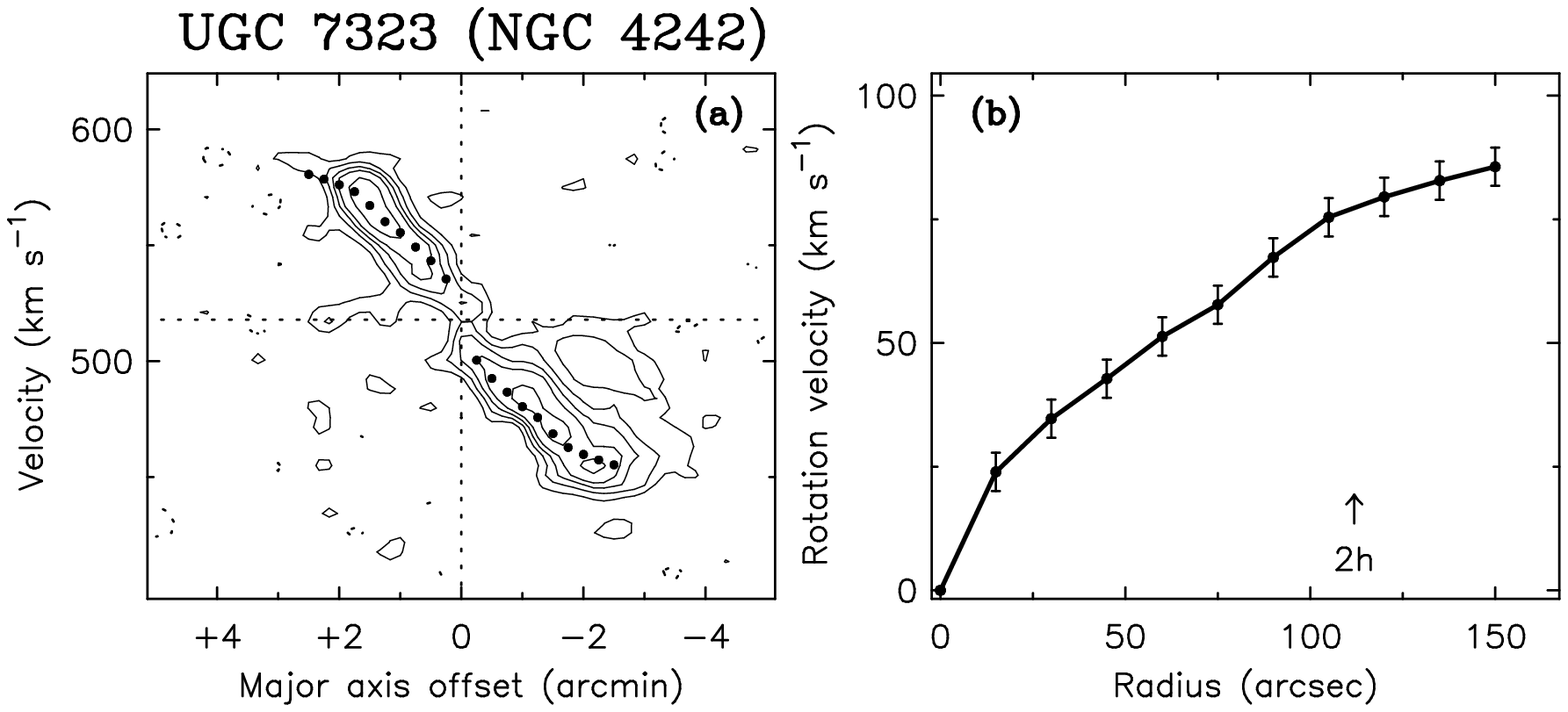}}
\kern0.5cm
\resizebox{0.45\hsize}{!}{\includegraphics{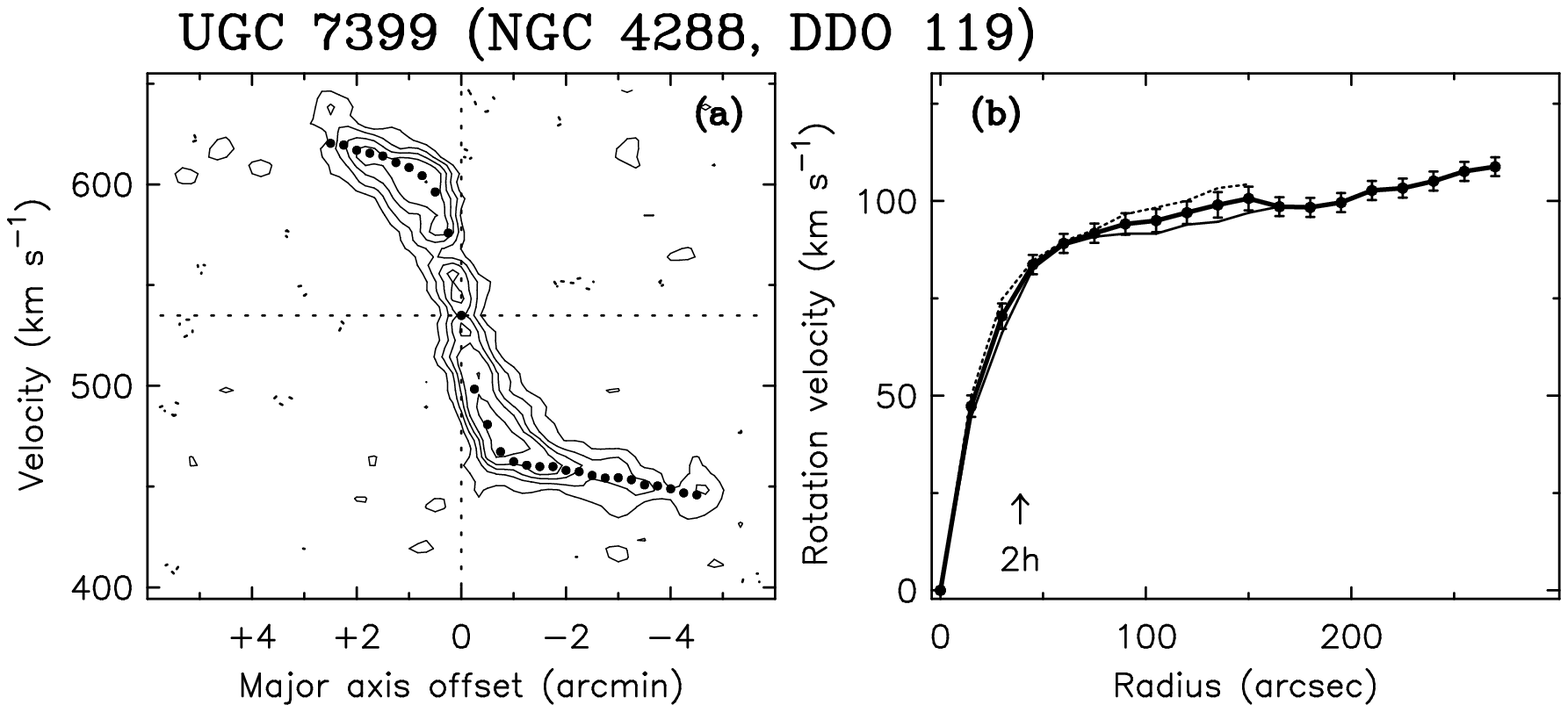}}
\newline
\vskip-0.15cm
\noindent
\resizebox{0.45\hsize}{!}{\includegraphics{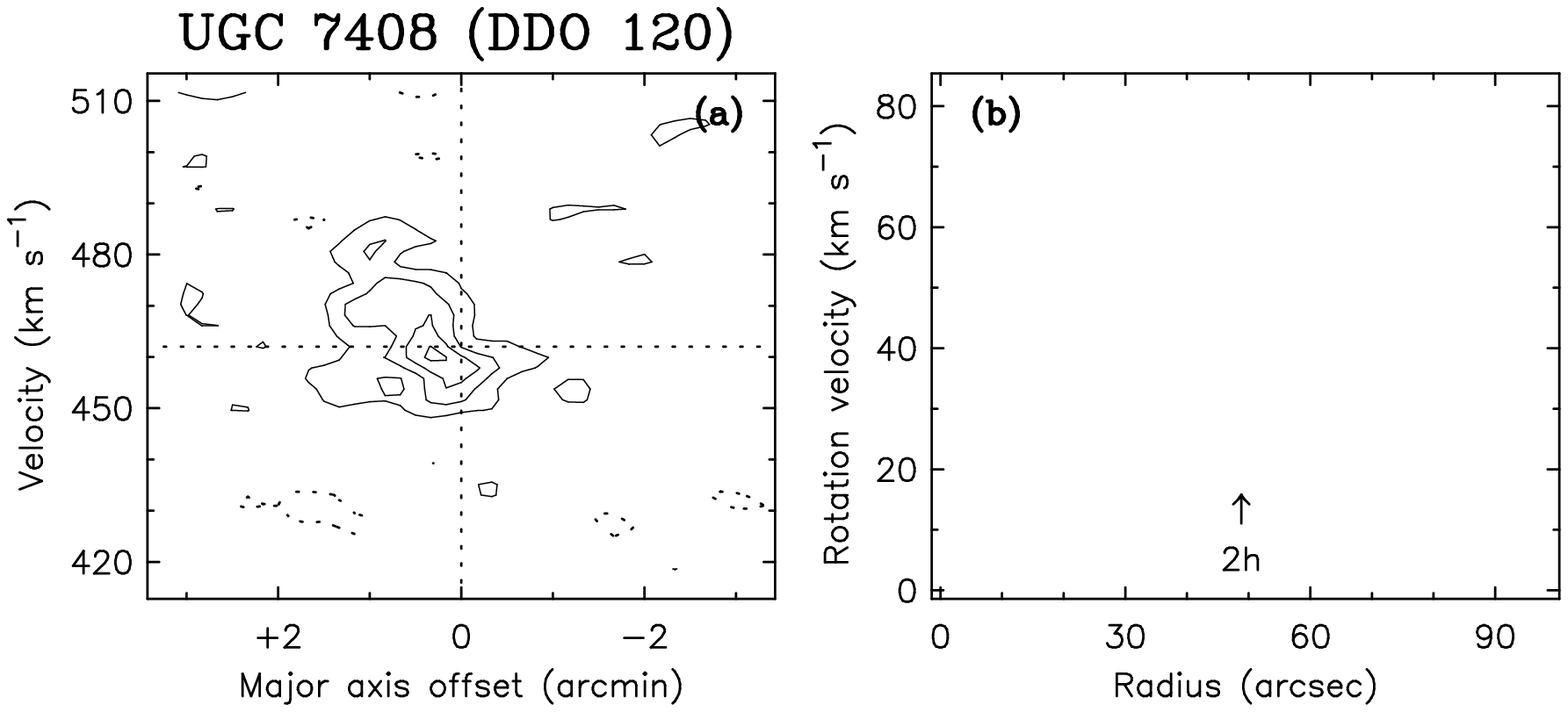}}
\kern0.5cm
\resizebox{0.45\hsize}{!}{\includegraphics{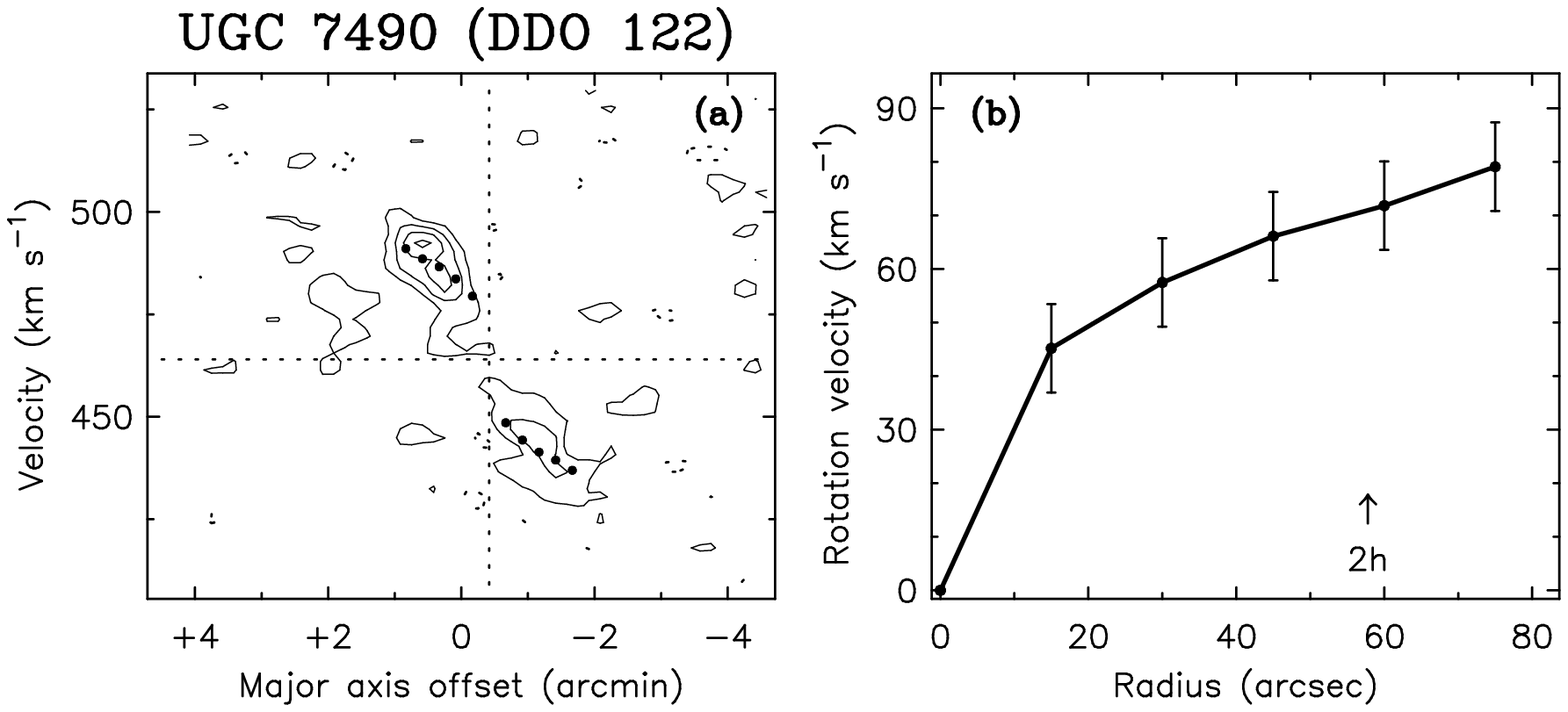}}
\newline
\vskip-0.15cm
\noindent
\resizebox{0.45\hsize}{!}{\includegraphics{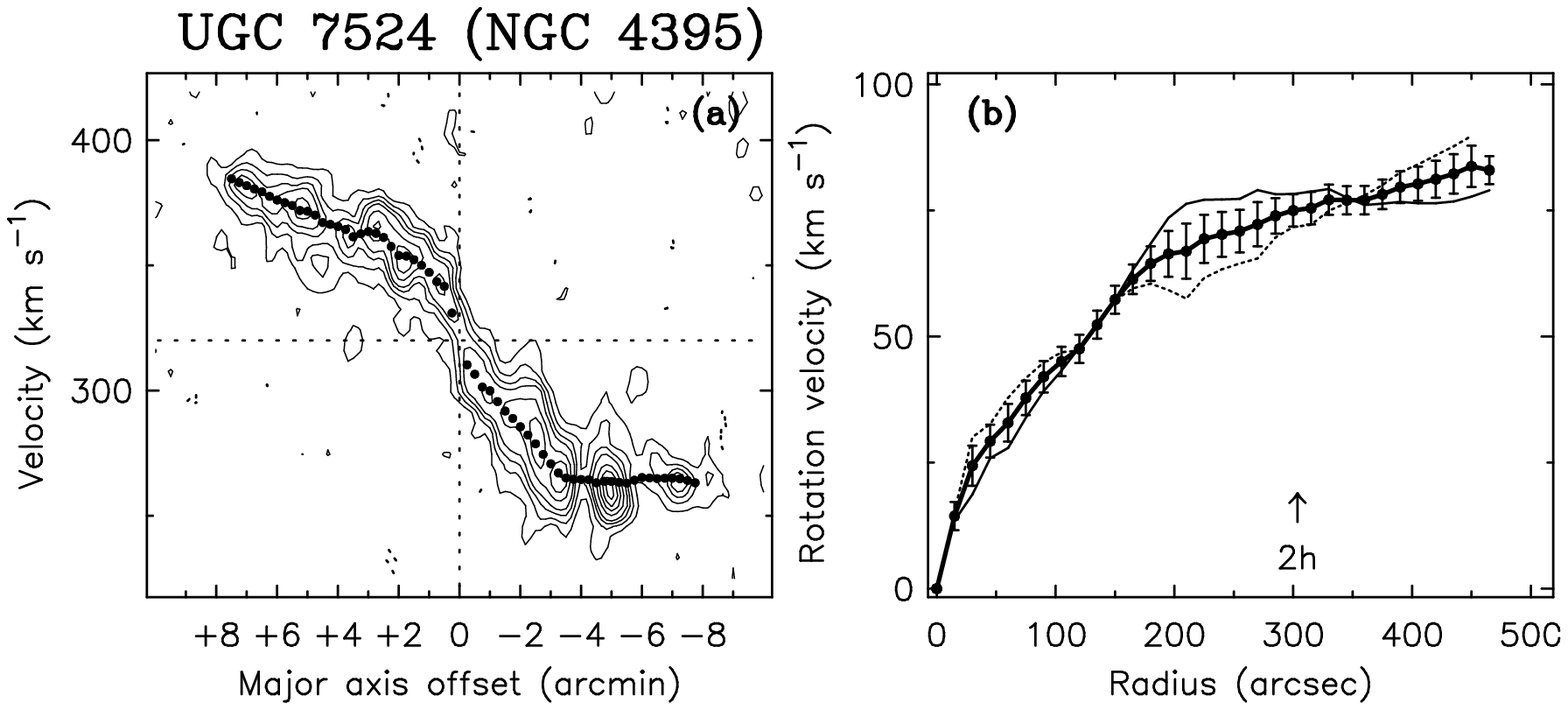}}
\kern0.5cm
\resizebox{0.45\hsize}{!}{\includegraphics{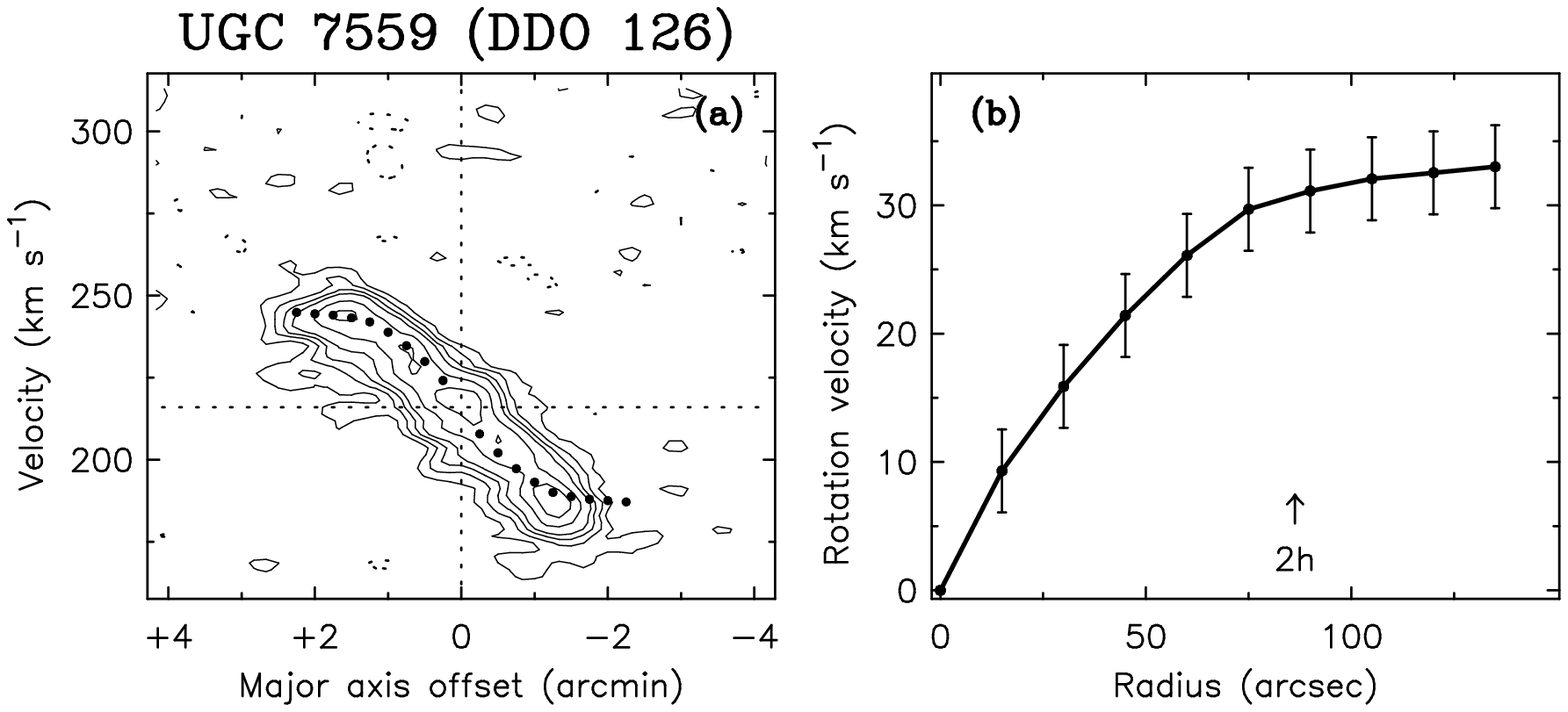}}
\newline
\vskip-0.15cm
\noindent
\resizebox{0.45\hsize}{!}{\includegraphics{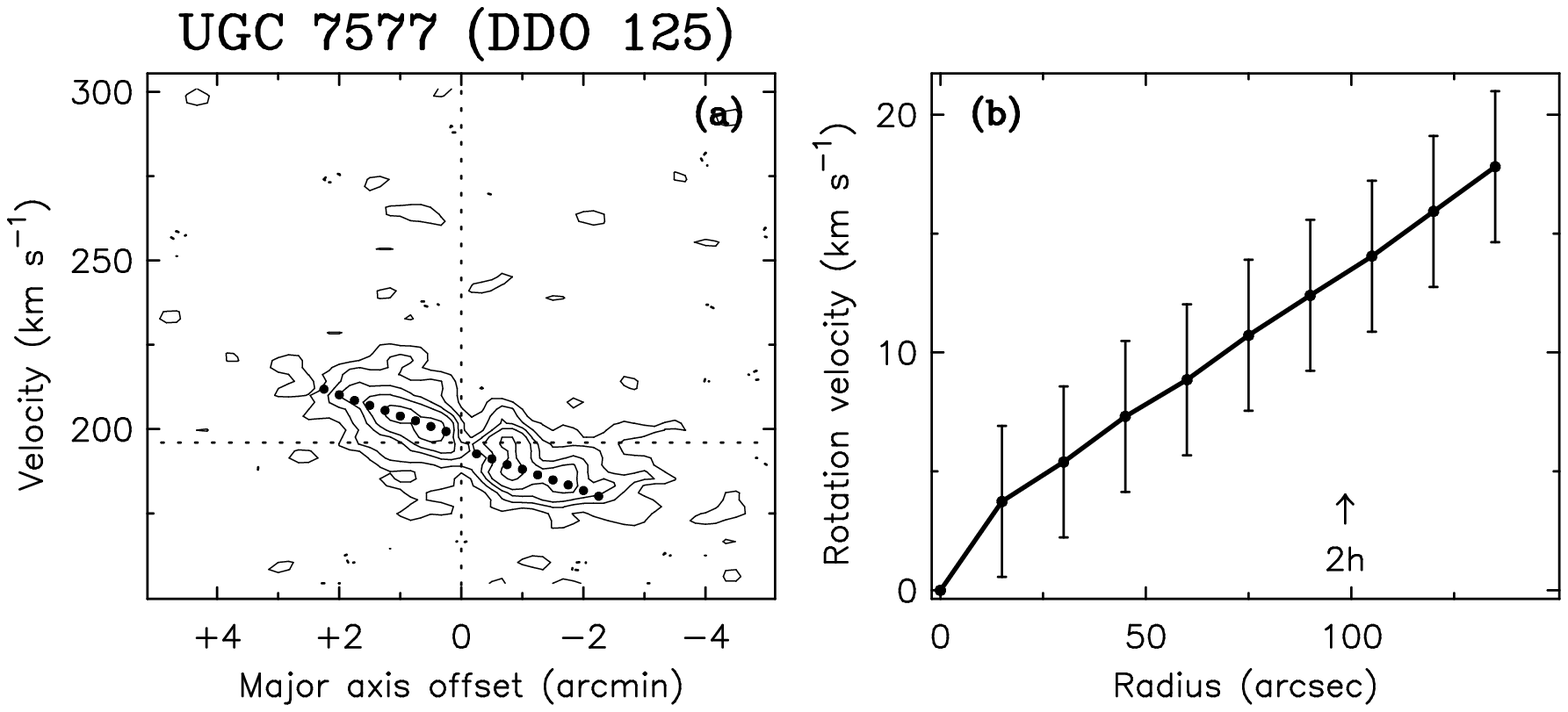}}
\kern0.5cm
\resizebox{0.45\hsize}{!}{\includegraphics{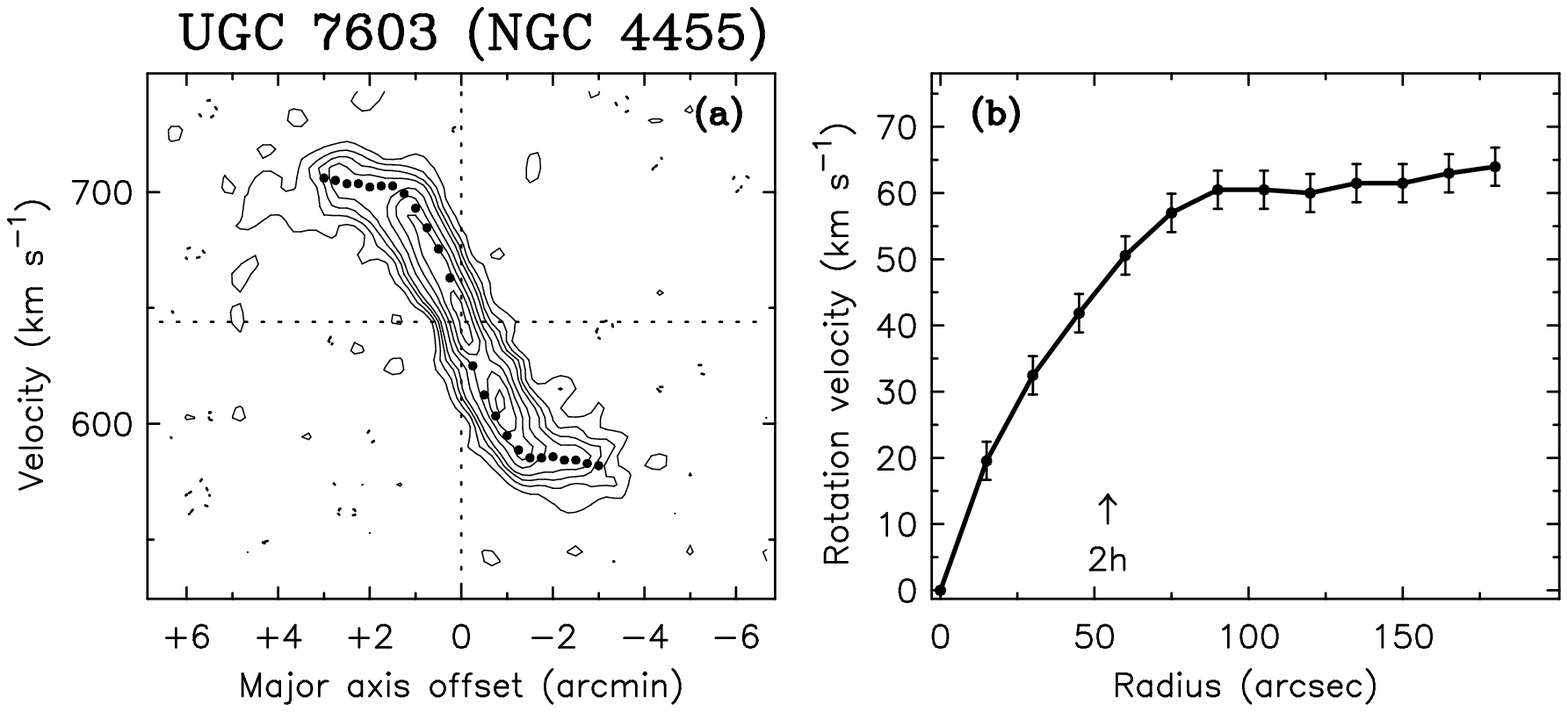}}
\newline
\vskip-0.15cm
\noindent
\resizebox{0.45\hsize}{!}{\includegraphics{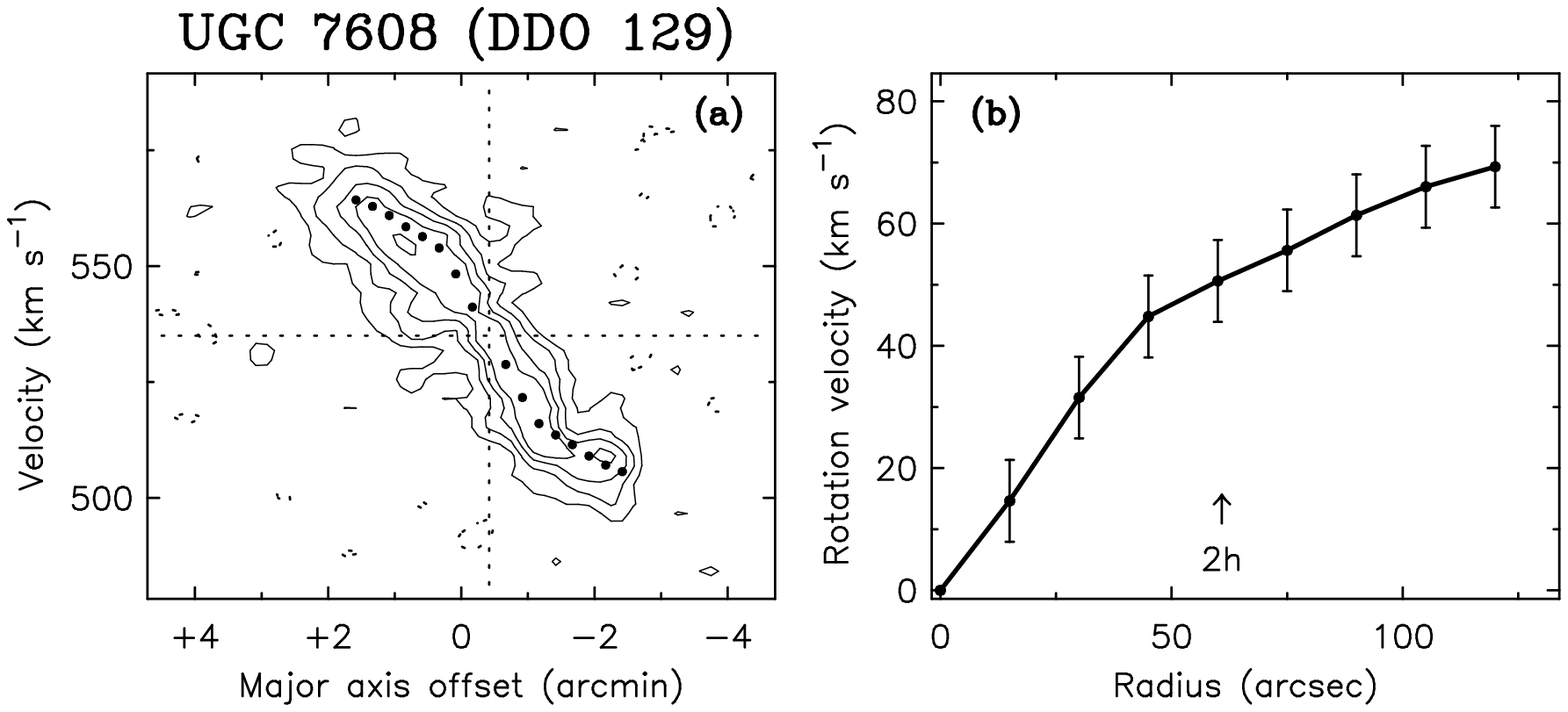}}
\kern0.5cm
\resizebox{0.45\hsize}{!}{\includegraphics{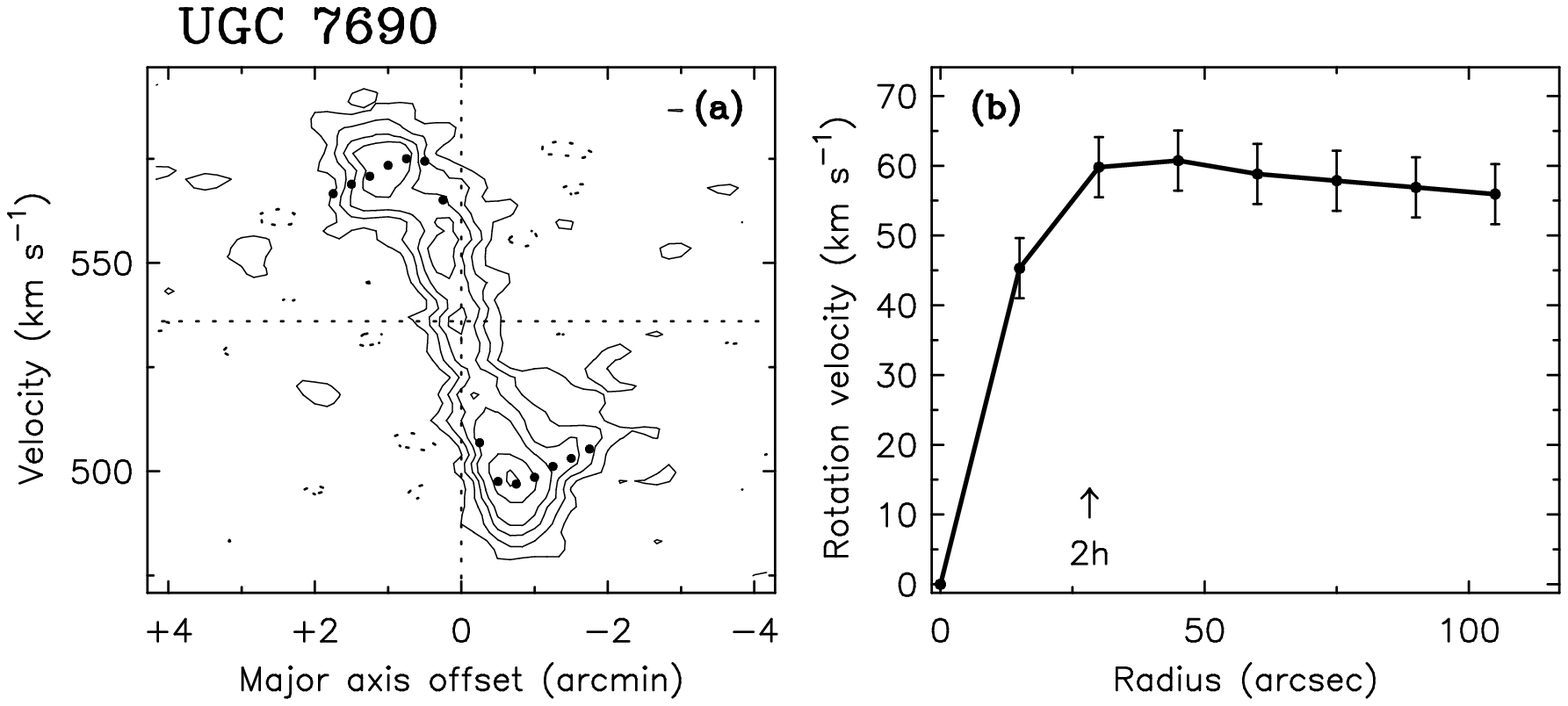}}
\clearpage
\vskip-0.15cm
\noindent
\resizebox{0.45\hsize}{!}{\includegraphics{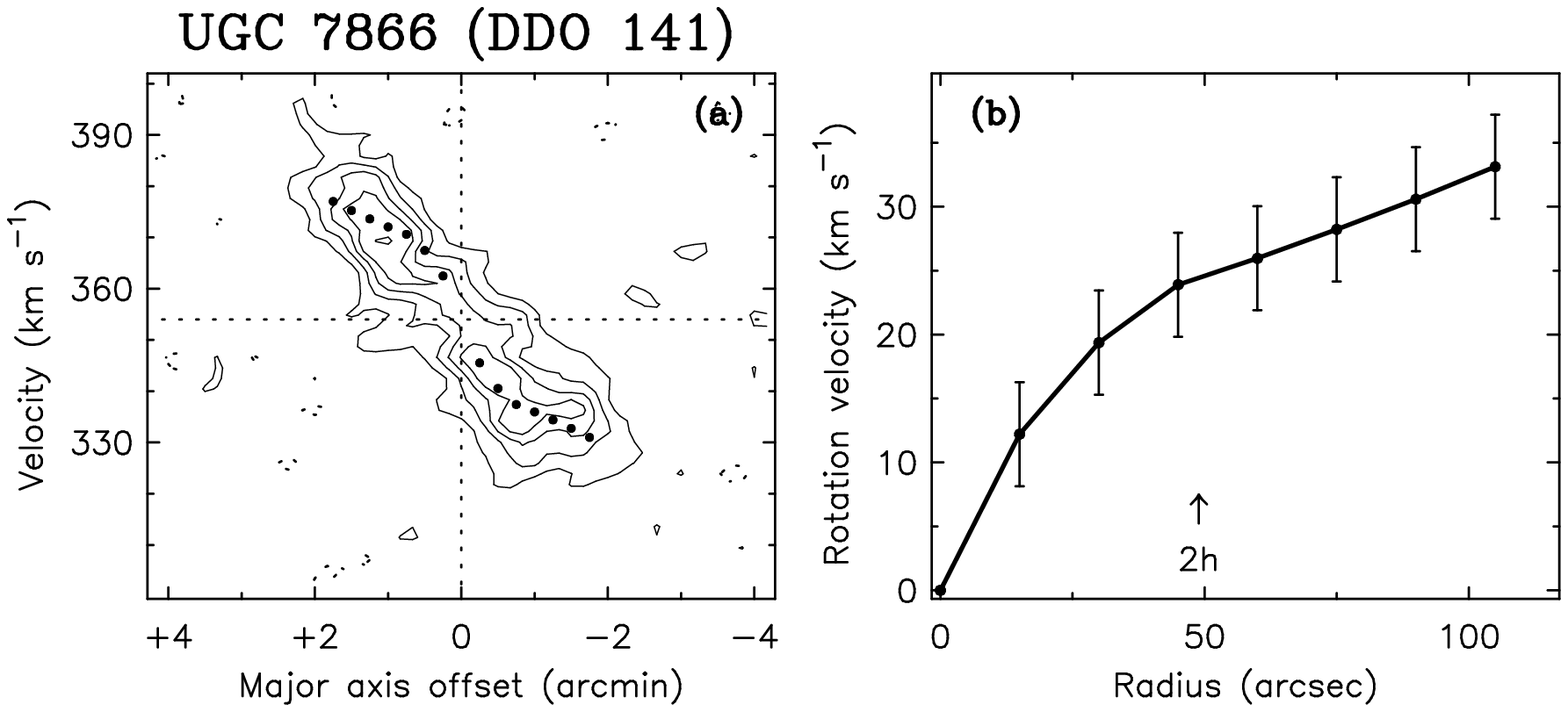}}
\kern0.5cm
\resizebox{0.45\hsize}{!}{\includegraphics{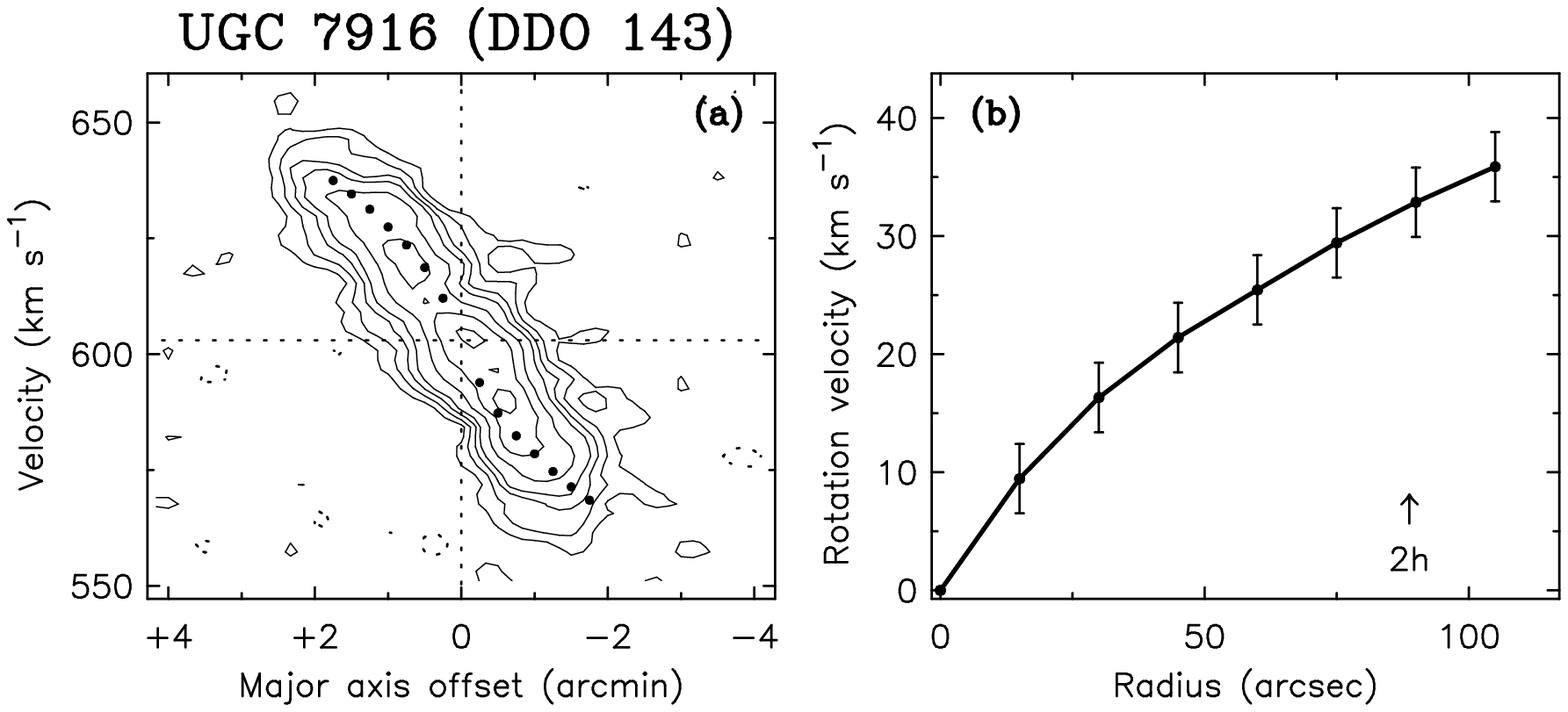}}
\newline
\vskip-0.15cm
\noindent
\resizebox{0.45\hsize}{!}{\includegraphics{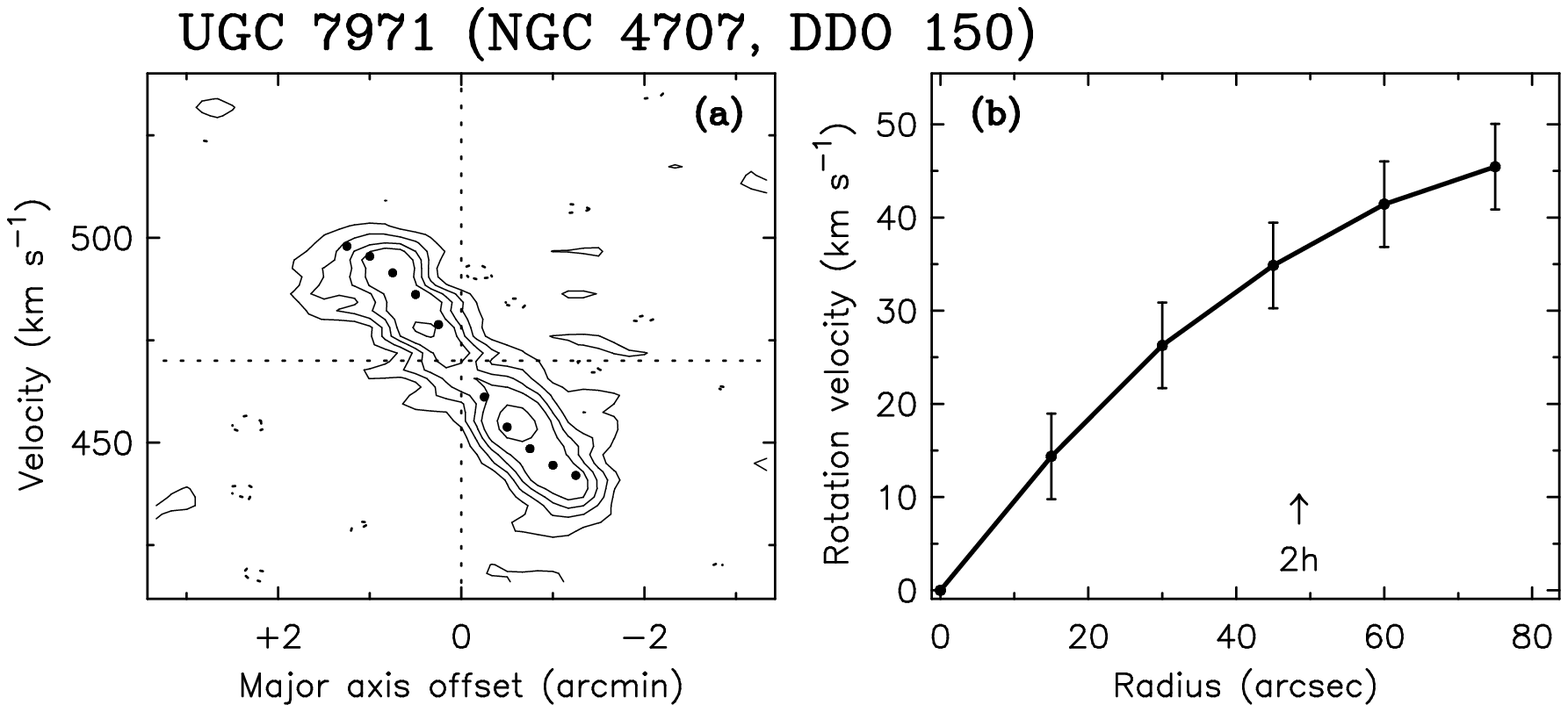}}
\kern0.5cm
\resizebox{0.45\hsize}{!}{\includegraphics{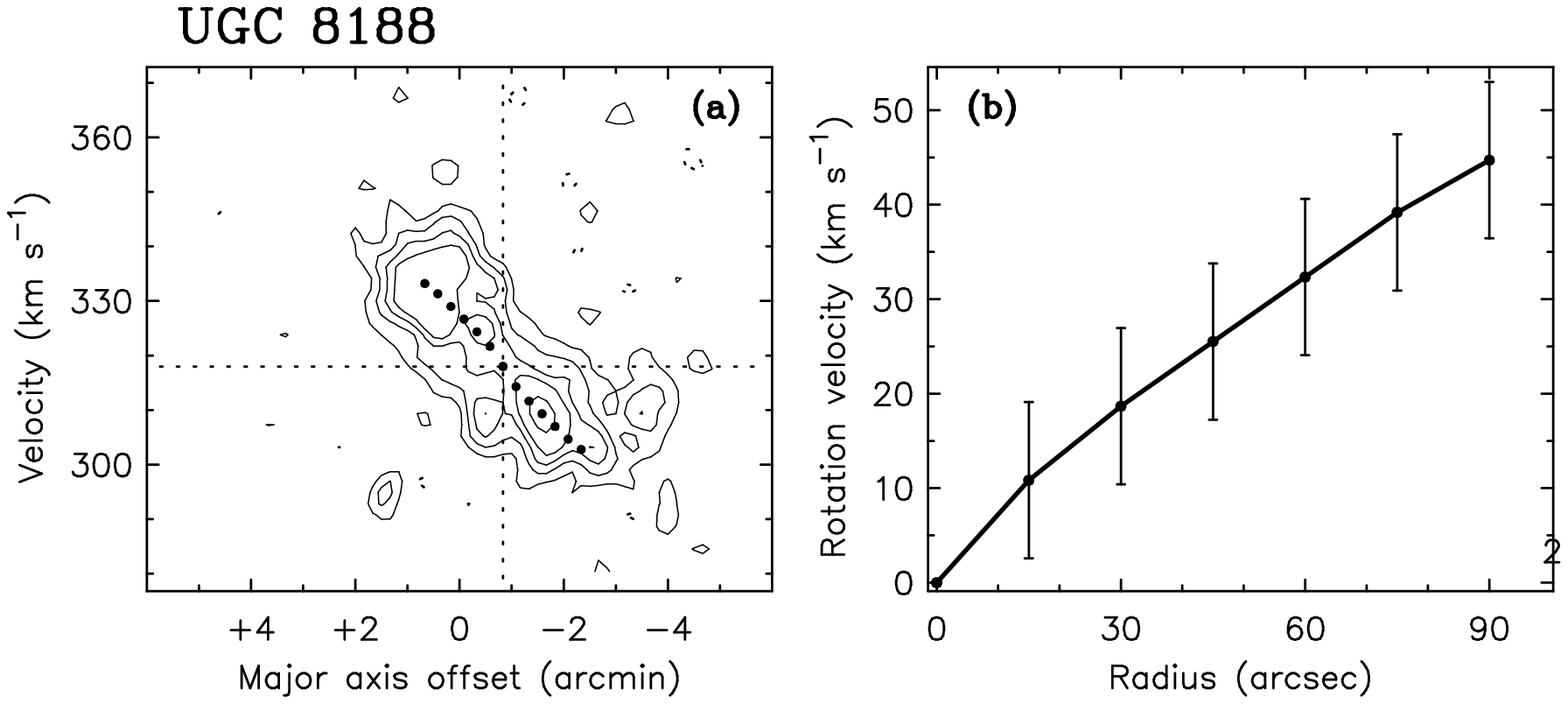}}
\newline
\vskip-0.15cm
\noindent
\resizebox{0.45\hsize}{!}{\includegraphics{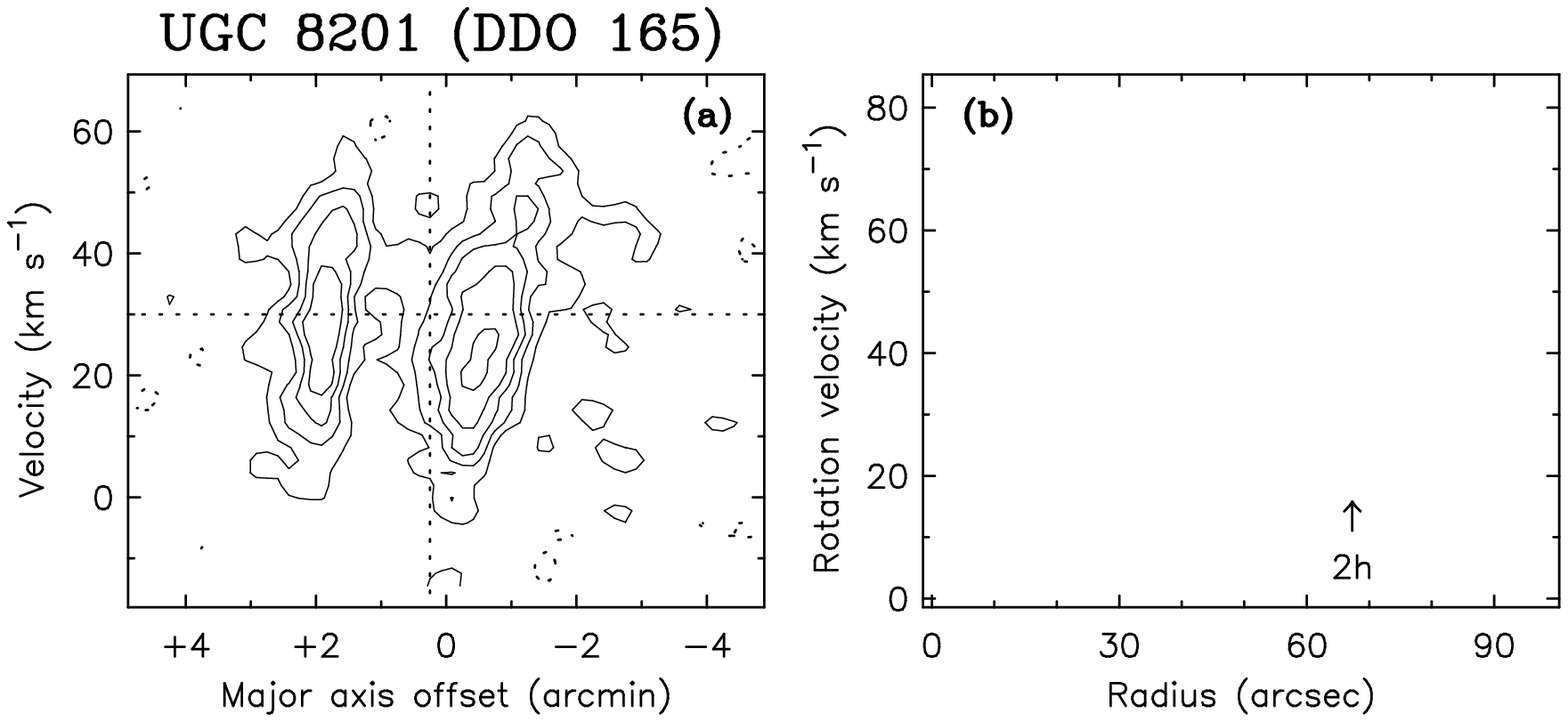}}
\kern0.5cm
\resizebox{0.45\hsize}{!}{\includegraphics{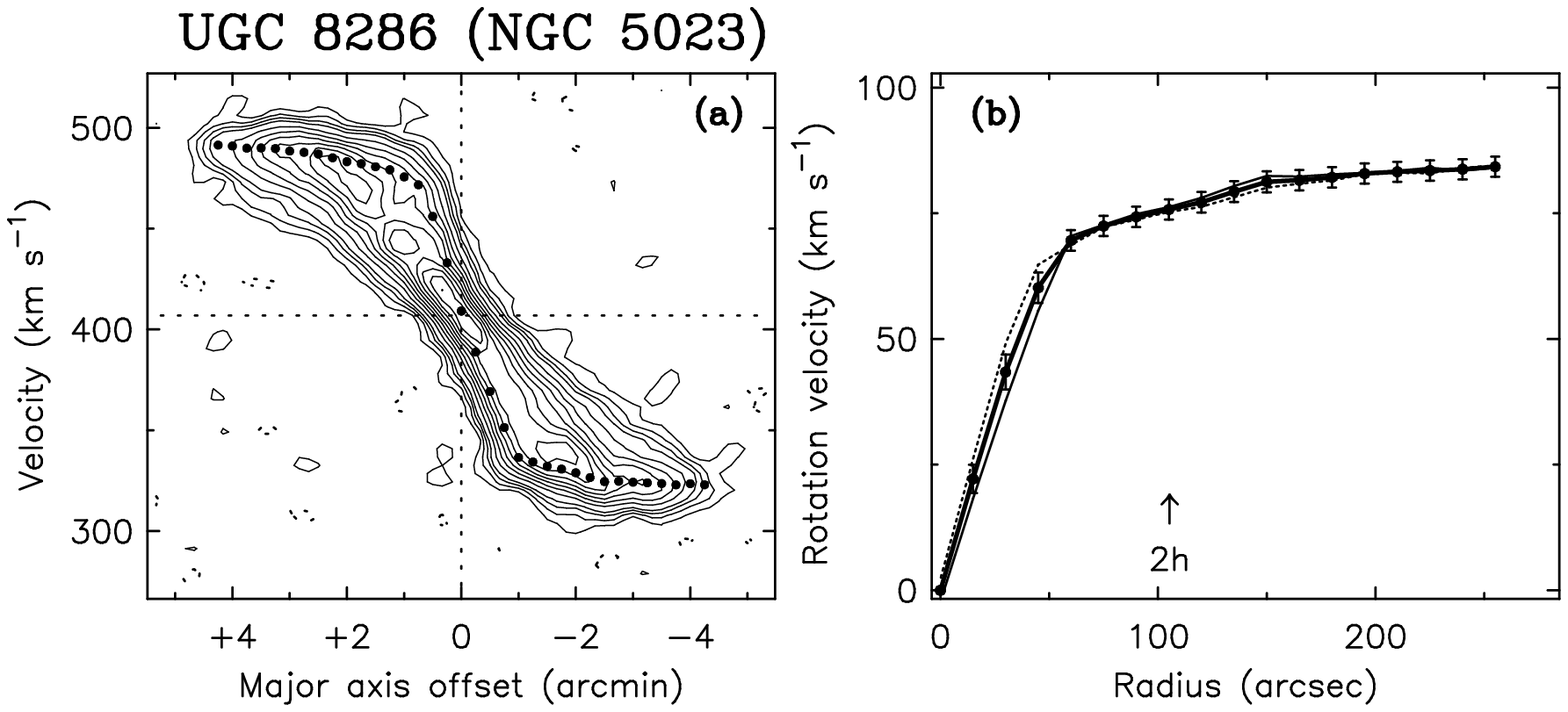}}
\newline
\vskip-0.15cm
\noindent
\resizebox{0.45\hsize}{!}{\includegraphics{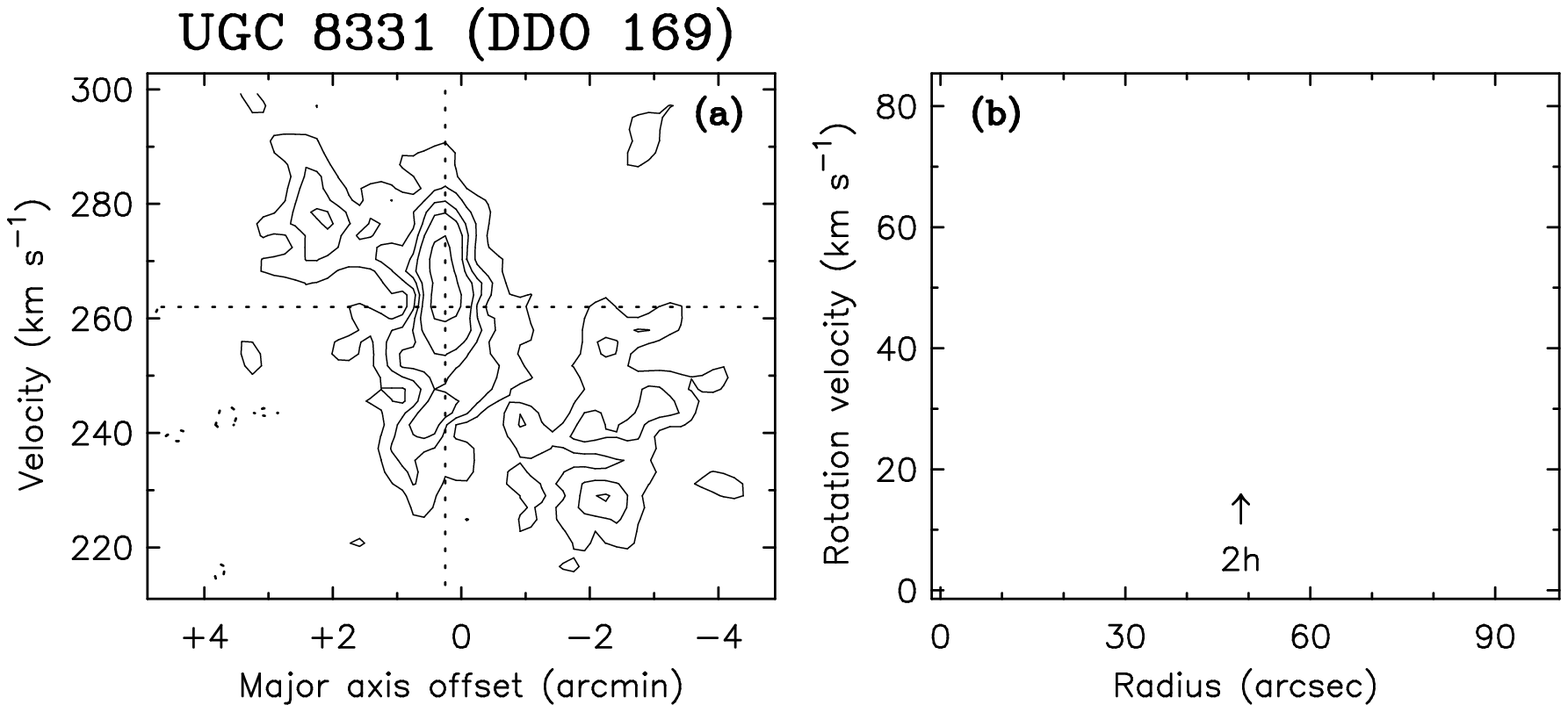}}
\kern0.5cm
\resizebox{0.45\hsize}{!}{\includegraphics{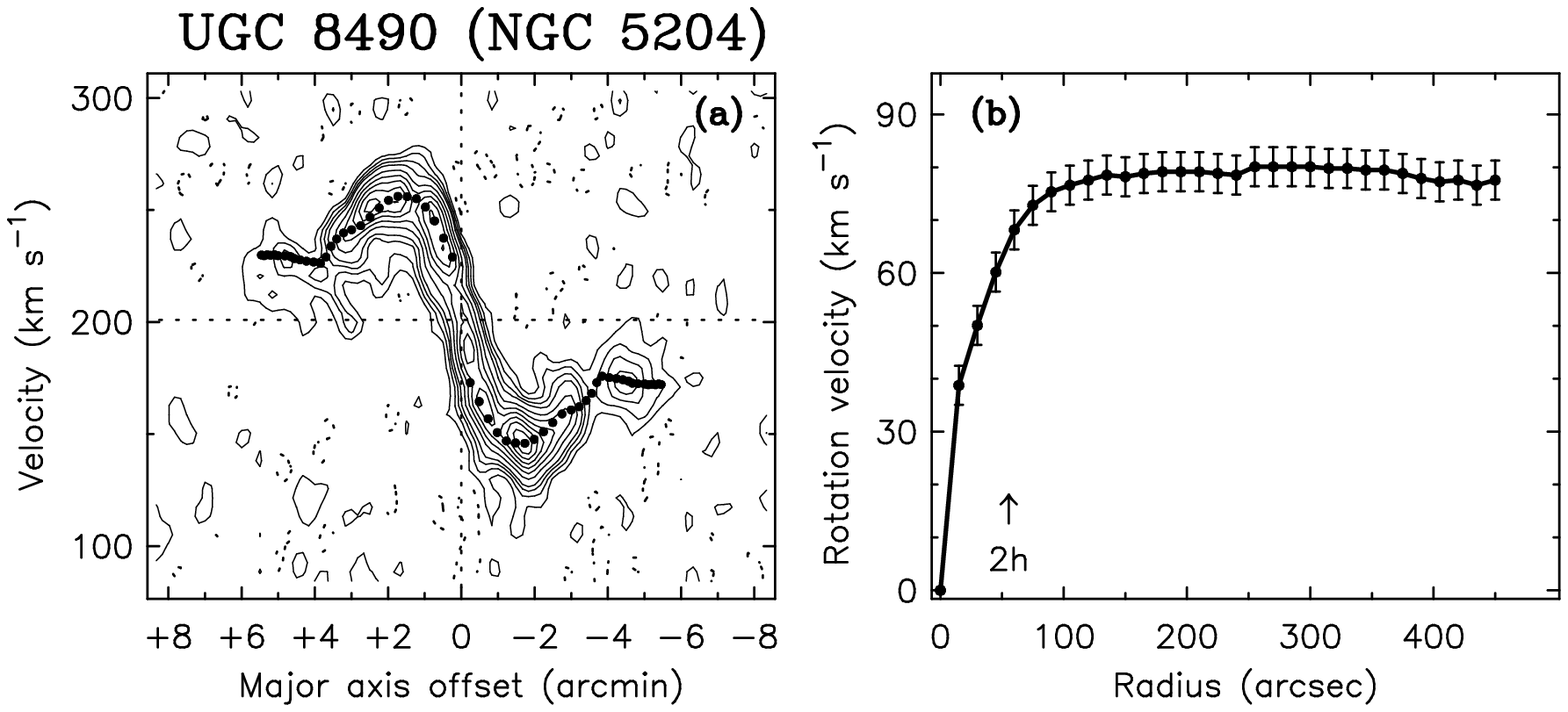}}
\newline
\vskip-0.15cm
\noindent
\resizebox{0.45\hsize}{!}{\includegraphics{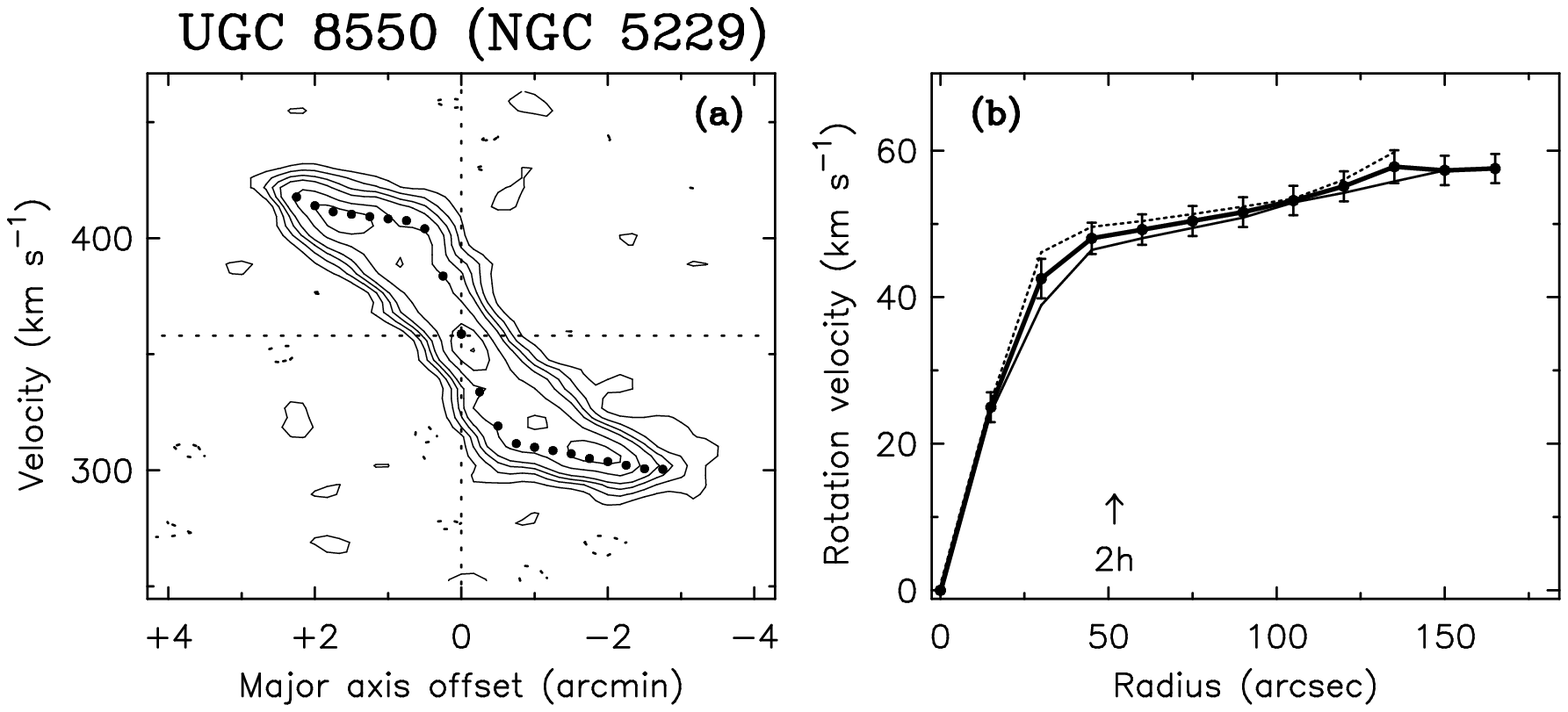}}
\kern0.5cm
\resizebox{0.45\hsize}{!}{\includegraphics{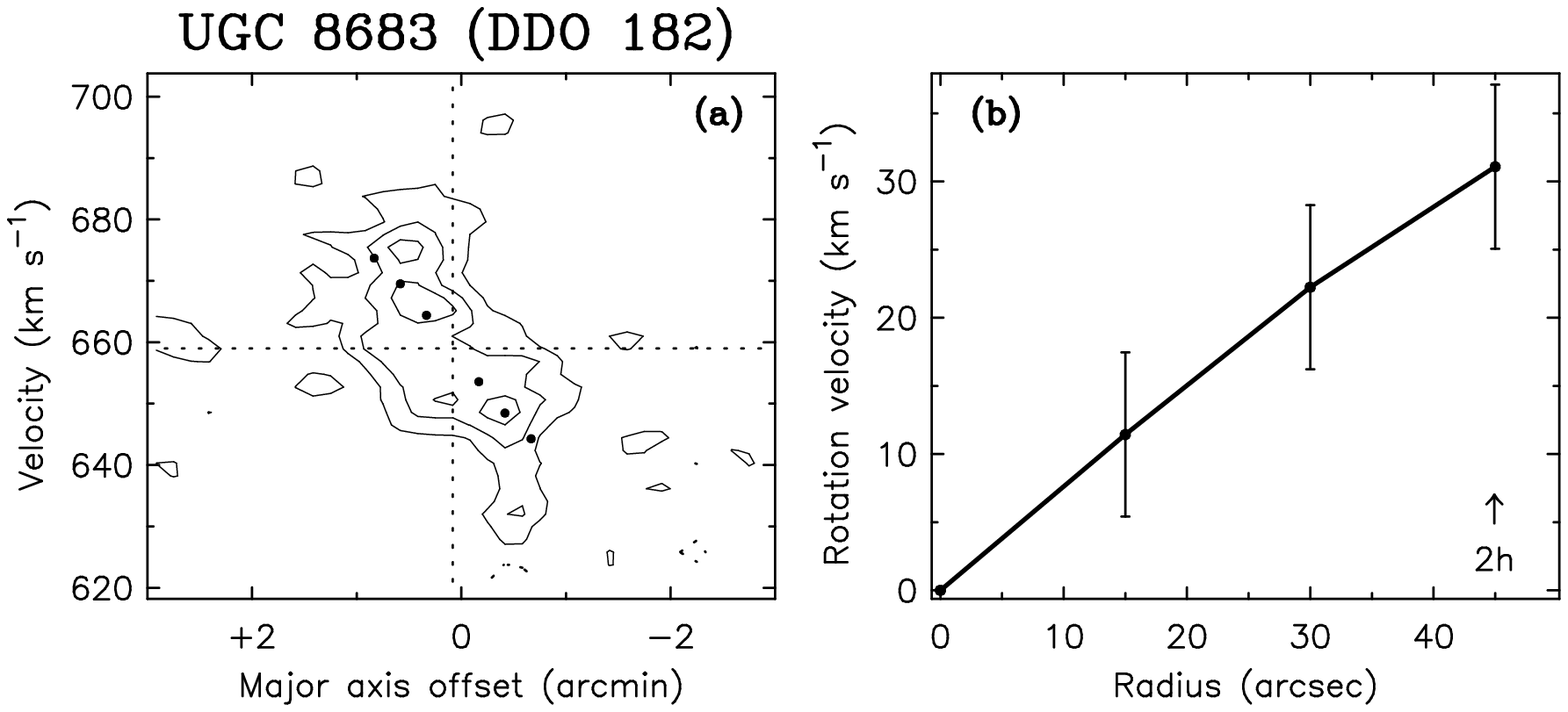}}
\newline
\vskip-0.15cm
\noindent
\resizebox{0.45\hsize}{!}{\includegraphics{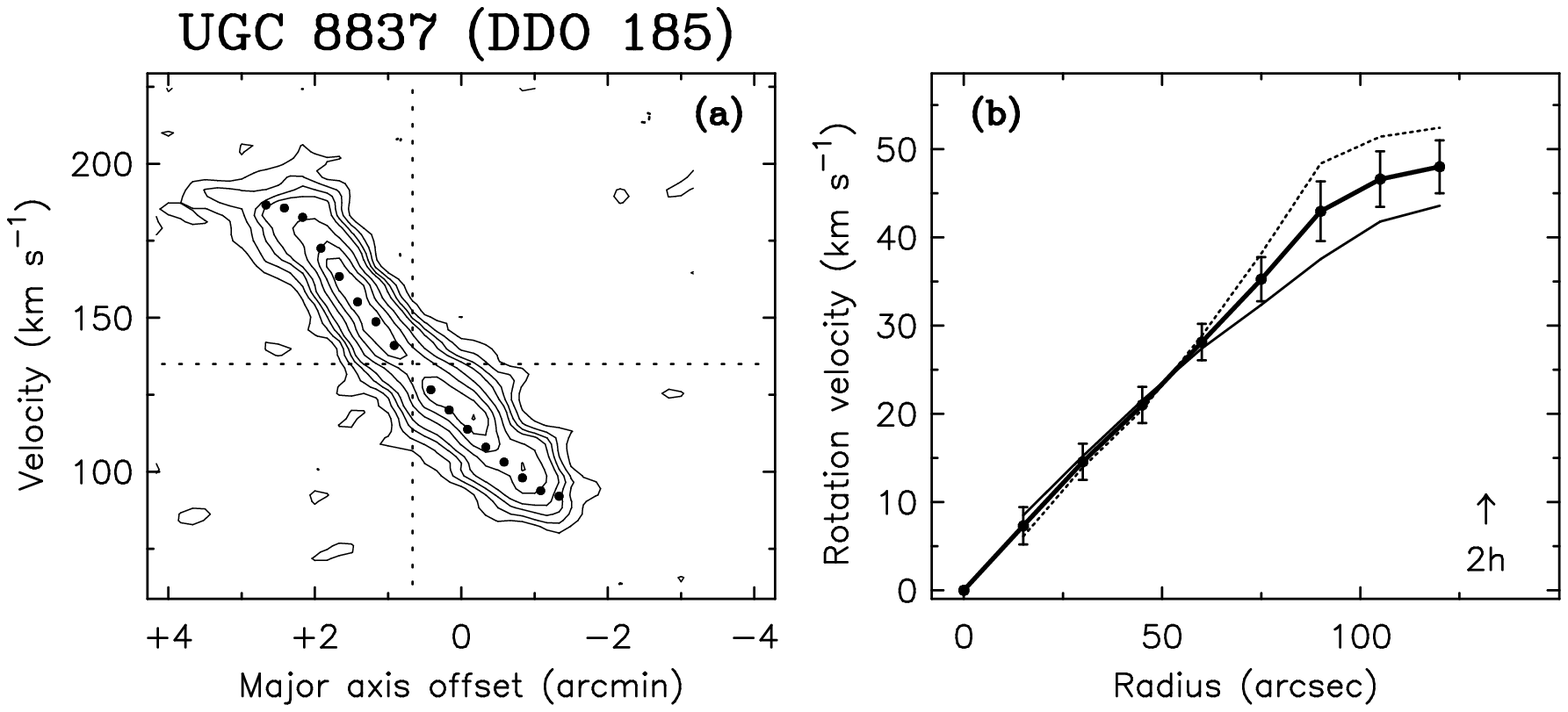}}
\kern0.5cm
\resizebox{0.45\hsize}{!}{\includegraphics{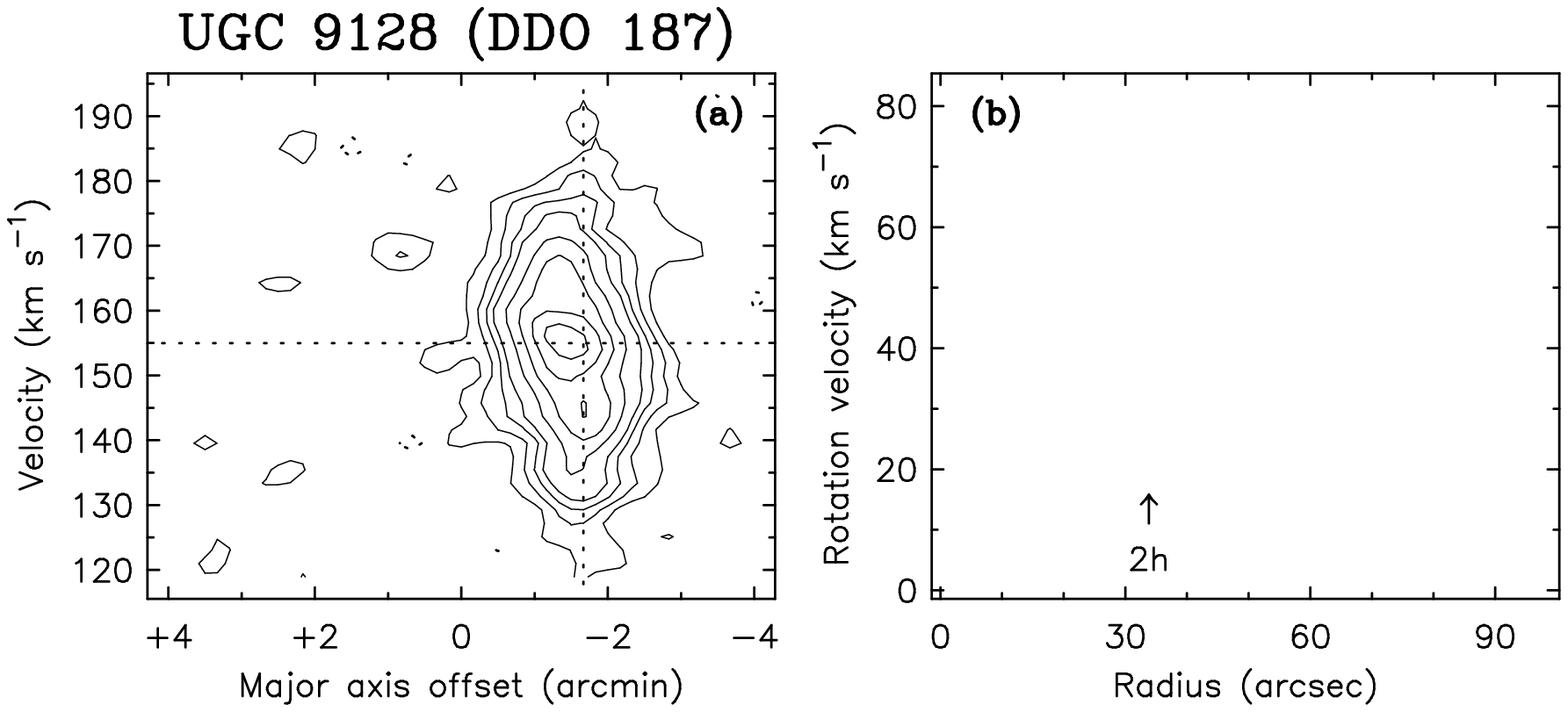}}
\clearpage
\vskip-0.15cm
\noindent
\resizebox{0.45\hsize}{!}{\includegraphics{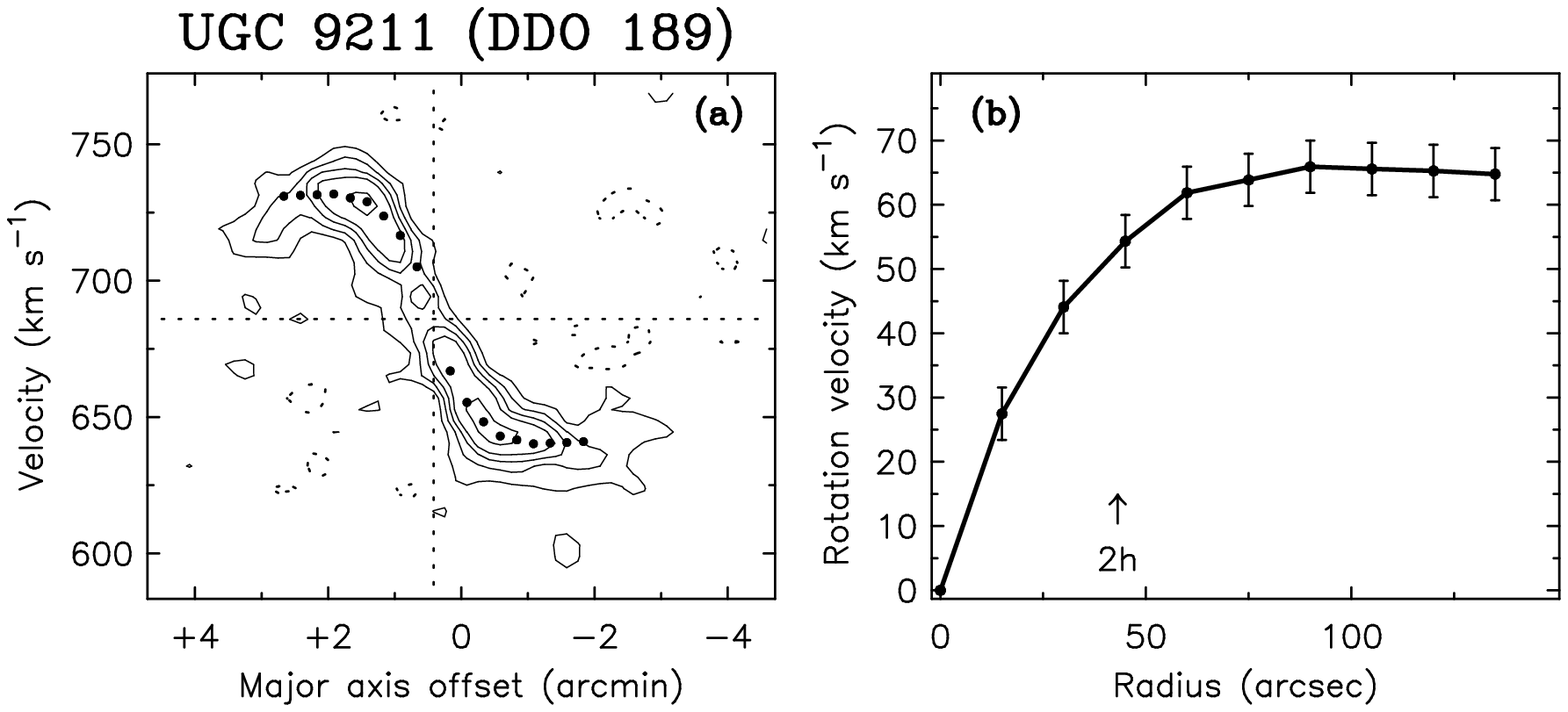}}
\kern0.5cm
\resizebox{0.45\hsize}{!}{\includegraphics{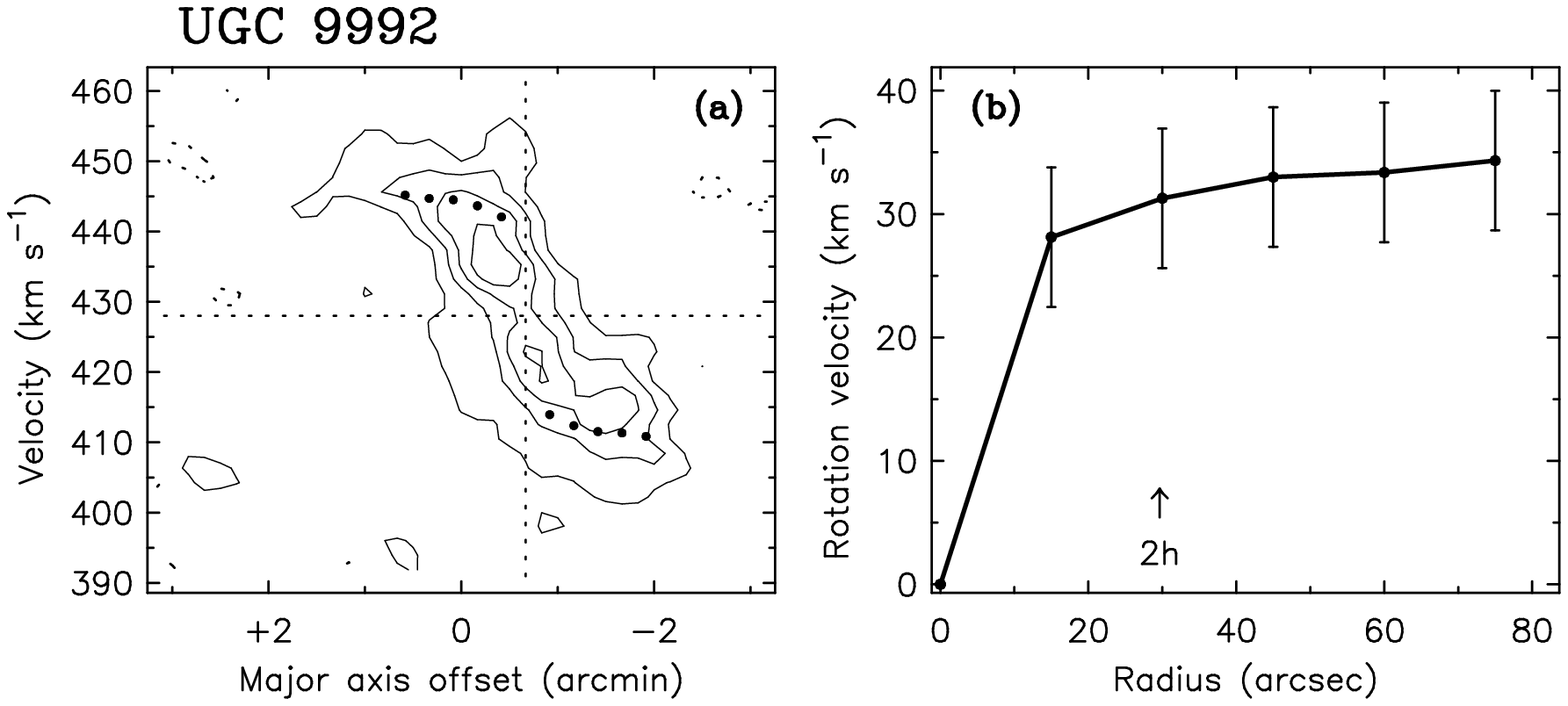}}
\newline
\vskip-0.15cm
\noindent
\resizebox{0.45\hsize}{!}{\includegraphics{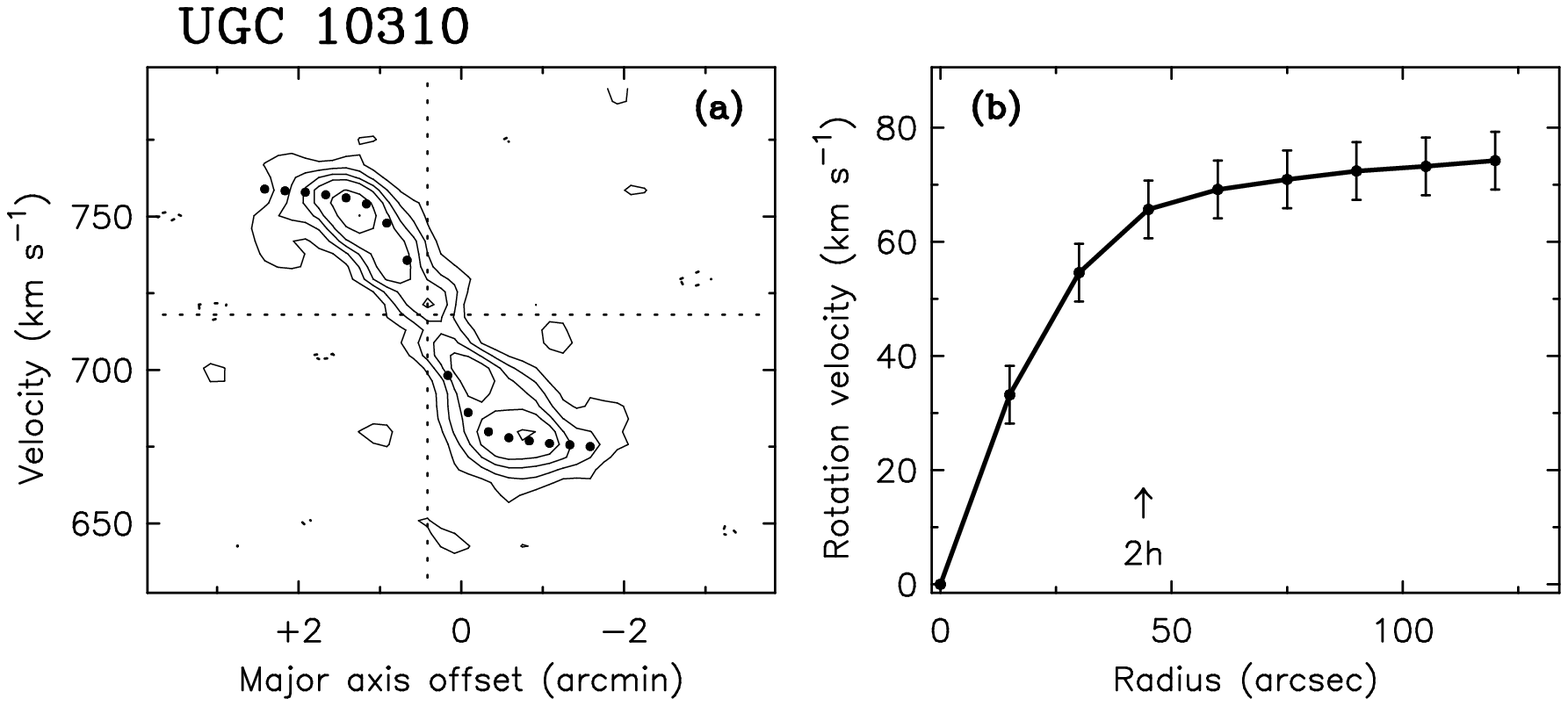}}
\kern0.5cm
\resizebox{0.45\hsize}{!}{\includegraphics{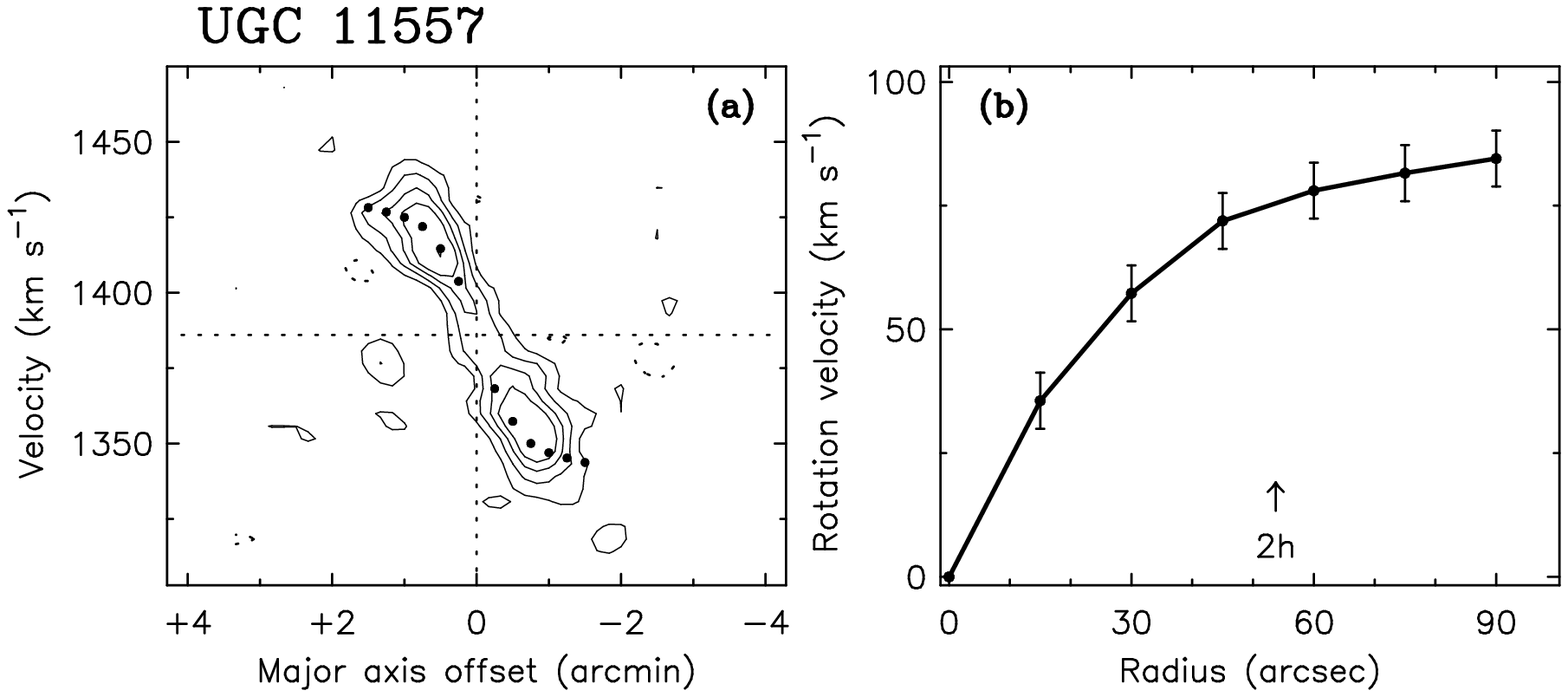}}
\newline
\vskip-0.15cm
\noindent
\resizebox{0.45\hsize}{!}{\includegraphics{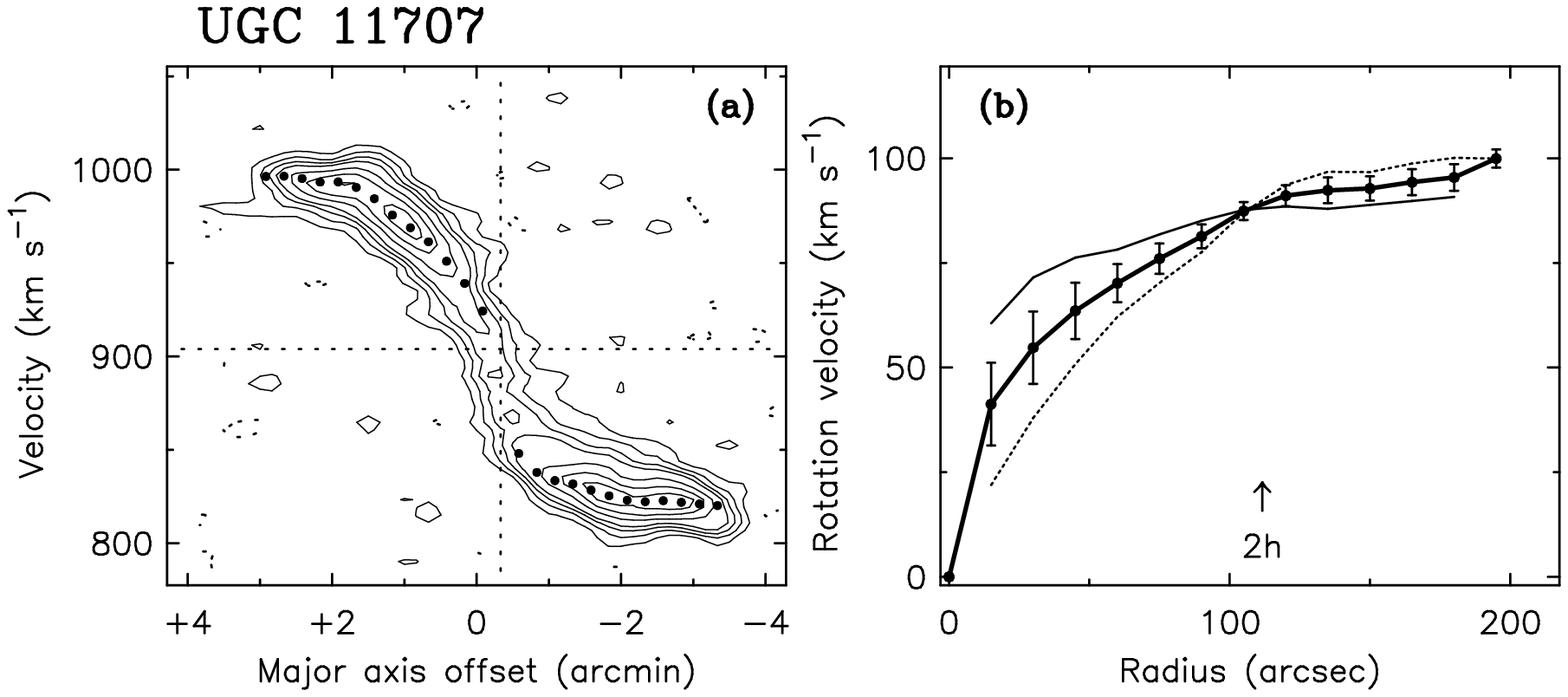}}
\kern0.5cm
\resizebox{0.45\hsize}{!}{\includegraphics{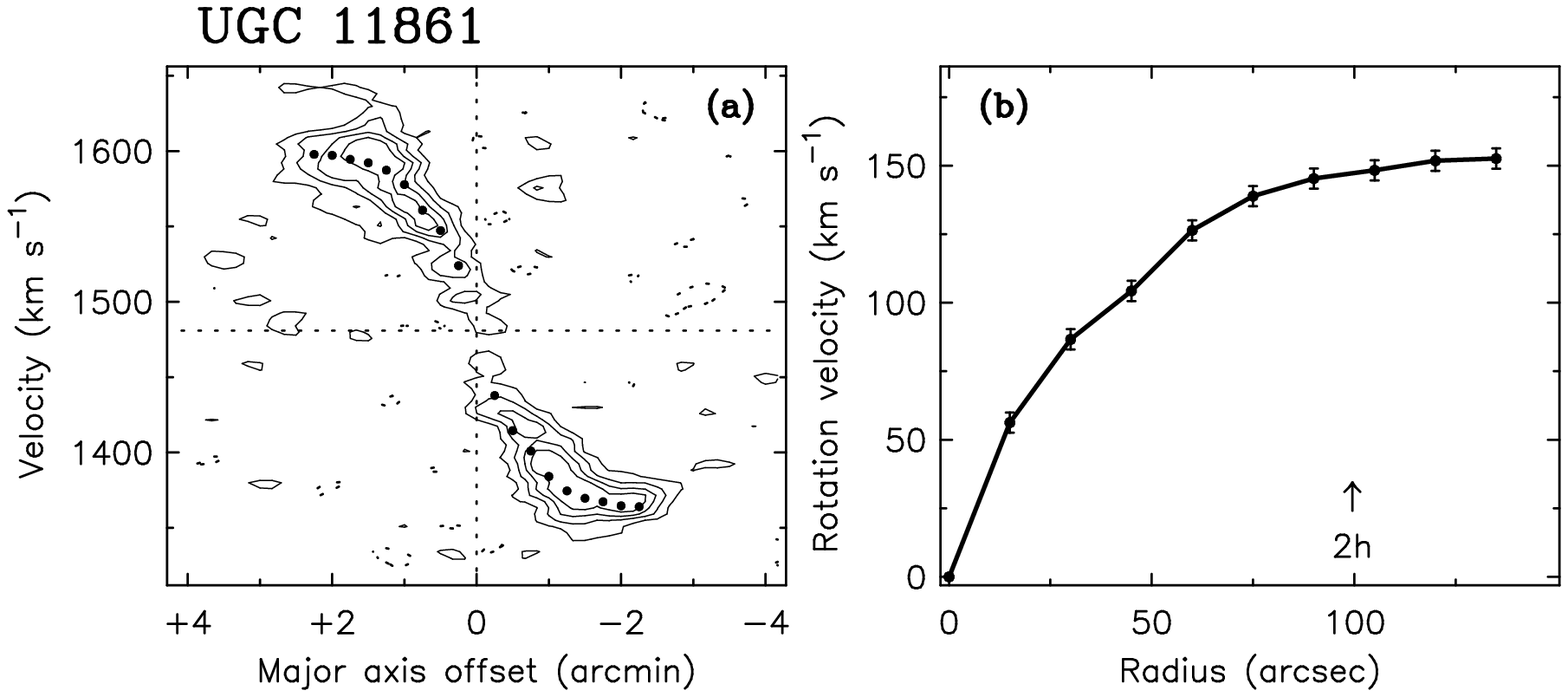}}
\newline
\vskip-0.15cm
\noindent
\resizebox{0.45\hsize}{!}{\includegraphics{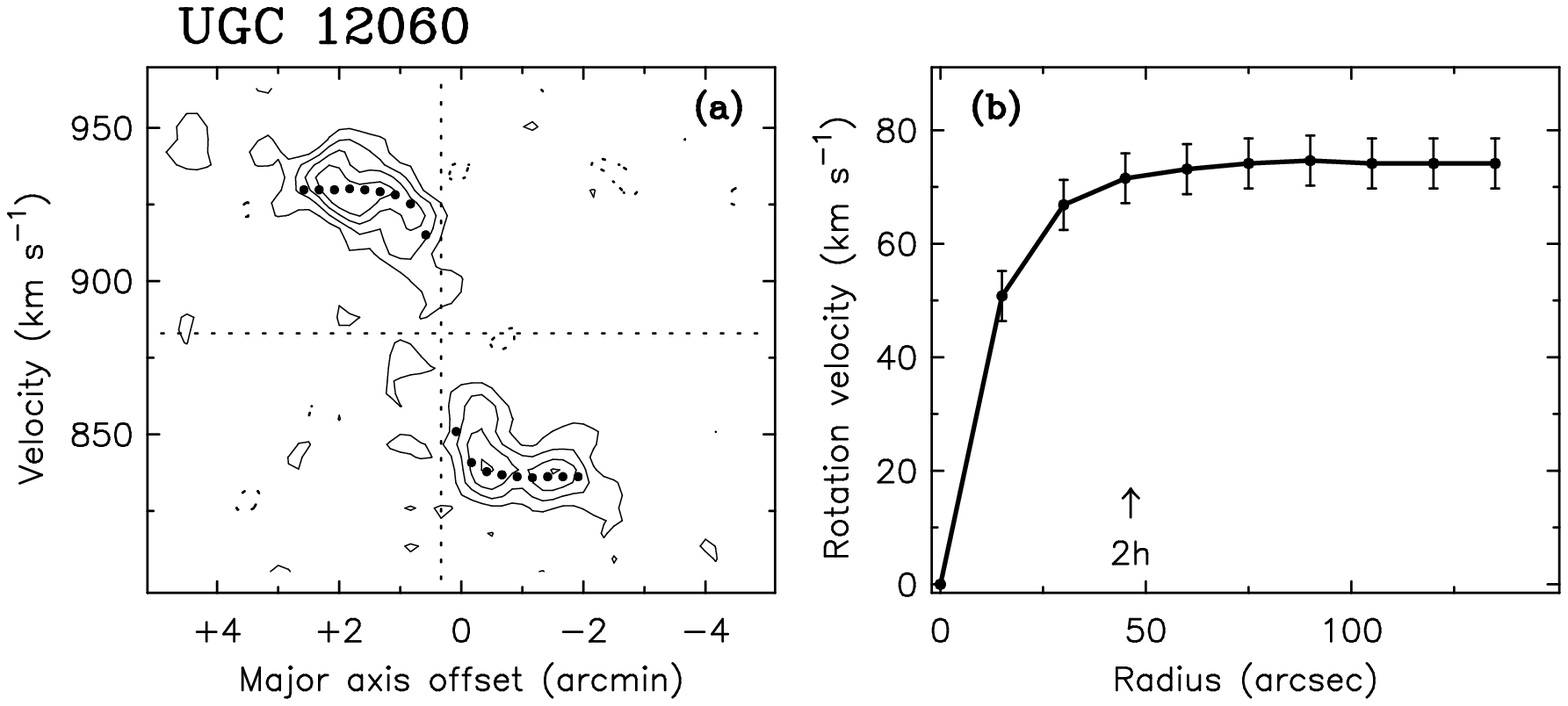}}
\kern0.5cm
\resizebox{0.45\hsize}{!}{\includegraphics{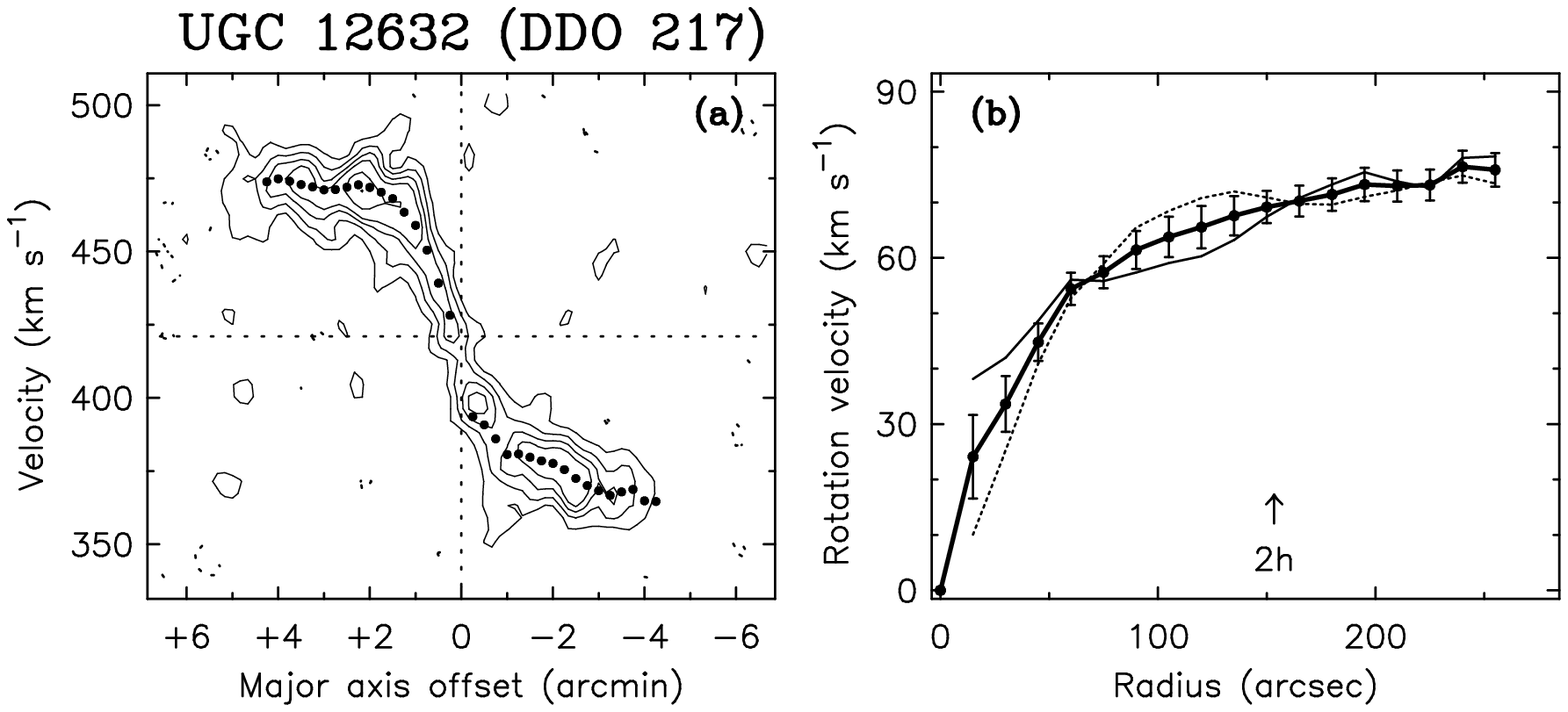}}
\newline
\vskip-0.15cm
\noindent
\resizebox{0.45\hsize}{!}{\includegraphics{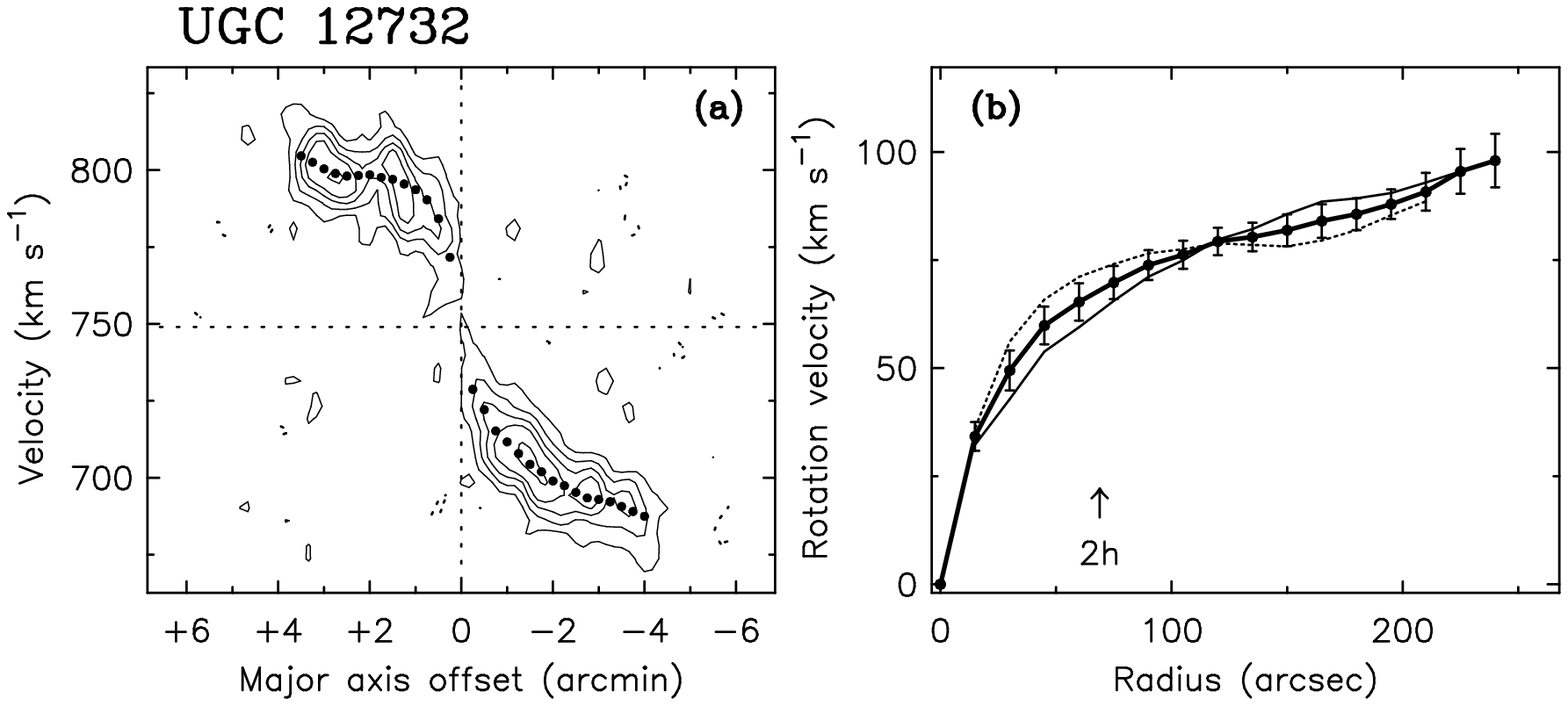}}
\kern0.5cm


\begin{thebibliography}{}

\bibitem[]{} Begeman, K., 1987, PhD thesis, Rijksuniversiteit Groningen

\bibitem[]{} Begeman, K.\ G.\ 1989, A\&A, 223, 47

\bibitem[]{} Begum, A., Chengalur, J.~N., Karachentsev, I.~D.,
  Sharina, M.~E., \& Kaisin, S.~S.\ 2008, MNRAS, 386, 1667

\bibitem[]{} Binney, J., Tremaine, S. 1987, Galactic Dynamics,
  Princeton University Press

\bibitem[]{} Blais-Ouellette, S., Amram, P., \& Carignan, C.\ 2001,
  AJ, 121, 1952

\bibitem[]{} Bosma, A. 1978, PhD thesis, Rijksuniversiteit Groningen

\bibitem[]{} Bosma, A. 1981a, AJ 86, 1791

\bibitem[]{} Bosma, A. 1981b, AJ 86, 1825

\bibitem[]{} Bottema, R., Shostak, G.~S., van der Kruit, P.~C. 1986, A\&A 167, 34

\bibitem[]{} Broeils, A.~H. 1992a, PhD thesis, Rijksuniversiteit Groningen

\bibitem[]{} Broeils, A.~H. 1992b, A\&A 256, 19

\bibitem[]{} Bullock, J.~S., Kolatt, T.~S., Sigad, Y., Somerville,
  R.~S., Kravtsov, A.~V., Klypin, A.~A., Primack, J.~R., \& Dekel,
  A.\ 2001, MNRAS, 321, 559

\bibitem[]{} Casertano, S., van Gorkom, J.~H. 1991, AJ 101, 1231 (CvG)

\bibitem[]{} Catinella, B., Giovanelli, R., \& Haynes, M.~P.\ 2006,
  ApJ, 640, 751

\bibitem[]{} Chengalur, J.~N., Begum, A., Karachentsev, I.~D., Sharina,
  M., Kaisin, S.~S. 2008, in Galaxies in the Local Volume,
  ed. B. Koribalski, H. Jerjen, in print (see also astro-ph 0711.2153)

\bibitem[]{} Corradi, R.~L.~M., Capaccioli, M. 1990, A\&A 237, 36

\bibitem[]{} C{\^o}t{\'e}, S., Carignan, C., \& Freeman, K.~C.\ 2000,
  AJ, 120, 3027

\bibitem[]{} de Blok, W.~J.~G., \& Bosma, A.\ 2002, A\&A, 385, 816
(dBB)

\bibitem[]{} de Blok, W.~J.~G., McGaugh, S.~S. 1997, MNRAS 290, 533

\bibitem[]{} de Blok, W.~J.~G., McGaugh, S.~S., van der Hulst, J.M.
  1996, MNRAS 283, 18

\bibitem[]{} de Blok, W.~J.~G., McGaugh, S.~S., \& Rubin, V.~C.\ 2001,
  \aj, 122, 2396

\bibitem[]{} de Vaucouleurs, G., de Vaucouleurs, A., Corwin, H.~G., et
  al. 1991, Third Reference Catalogue of Bright Galaxies (New
  York:Springer)

\bibitem[]{} Garc{\'{\i}}a-Ruiz, I., Sancisi, R., \& Kuijken, K.\
2002, A\&A, 394, 769

\bibitem[]{} Garrido, O., Marcelin, M., Amram, P., \& Boulesteix,
  J.\ 2002, A\&A, 387, 821

\bibitem[]{} Hunter, D.~A., Rubin, V.~C., Swaters, R.~A., Sparke,
  L.~S., \& Levine, S.~E.\ 2002, ApJ, 580, 194

\bibitem[]{} Kent, S.~M. 1986, AJ 91, 1301

\bibitem[]{} Kent, S.~M. 1987, AJ 93, 816

\bibitem[]{} Kregel, M., \& van der Kruit, P.~C.\ 2004, MNRAS, 352,
787

\bibitem[]{} Kuzio de Naray, R., McGaugh, S.~S., de Blok, W.~J.~G., \&
Bosma, A.\ 2006, ApJSS, 165, 461

\bibitem[]{} Marchesini, D., D'Onghia, E., Chincarini, G., Firmani,
  C., Conconi, P., Molinari, E., \& Zacchei, A.\ 2002, ApJ, 575, 801

\bibitem[]{} McGaugh, S.~S., Rubin, V.~C., \& de Blok, W.~J.~G.\ 2001,
  AJ, 122, 2381

\bibitem[]{} Navarro, J.~F., Frenk, C.~S., \& White, S.~D.~M.\ 1996,
  ApJ, 462, 563

\bibitem[]{} Navarro, J.~F., Frenk, C.~S., \& White, S.~D.~M.\ 1997,
  ApJ, 490, 493

\bibitem[]{} Nilson, P.  1973, Uppsala General Catalogue of Galaxies,
  Uppsala Astr.  Obs.  Ann., Vol.  6 (UGC)

\bibitem[]{} Noordermeer, E., van der Hulst, J.~M., Sancisi, R.,
  Swaters, R.~S., \& van Albada, T.~S.\ 2007, MNRAS, 376, 1513

\bibitem[]{} Persic, M., \& Salucci, P.\ 1988, MNRAS, 234, 131

\bibitem[]{} Persic, M., Salucci, P., Stel, F. 1996, MNRAS 281, 27

\bibitem[]{} Power, C., Navarro, J.~F., Jenkins, A., Frenk, C.~S.,
  White, S.~D.~M., Springel, V., Stadel, J., \& Quinn, T.\ 2003,
  MNRAS, 338, 14

\bibitem[]{} Puche, D., Westpfahl, D., Brinks, E., \& Roy,
  J.-R.\ 1992, \aj, 103, 1841

\bibitem[]{} Rhee, M.-Y. 1996, PhD thesis, Rijksuniversiteit Groningen

\bibitem[]{} Rhee, G., Valenzuela, O., Klypin, A., Holtzman, J., \&
  Moorthy, B.\ 2004, ApJ, 617, 1059

\bibitem[]{} Rubin, V.~C., Burstein, D., Ford, Jr., W.~K., Thonnard, N.
  1985, ApJ 289, 81

\bibitem[]{} Rubin, V.~C., Kenney, J.~D., Boss, A.~P., Ford, W.K. 1989,
  AJ 98, 1246

\bibitem[]{} Sancisi, R., Allen, R.~J. 1979, A\&A 74, 73

\bibitem[]{} Sicotte, V., Carignan, C. 1997, AJ 113, 609

\bibitem[]{} Sicotte, V., Carignan, C., Durand, D. 1996, AJ 112, 1423

\bibitem[]{} Simon, J.~D., Bolatto, A.~D., Leroy, A., Blitz, L., \&
  Gates, E.~L.\ 2005, ApJ, 621, 757

\bibitem[]{} Spekkens, K., \& Giovanelli, R.\ 2006, \aj, 132, 1426 

\bibitem[]{} Spekkens, K., Giovanelli, R., \& Haynes, M.~P.\ 2005,
  \aj, 129, 2119

\bibitem[]{} Stil, J.~M. 1999, PhD thesis, University of Leiden 

\bibitem[]{} Swaters, R.~A. 1999, PhD thesis, Rijksuniversiteit
  Groningen

\bibitem[]{} Swaters, R.~A., \& Balcells, M.\ 2002, A\&A, 390, 863
(Paper II)

\bibitem[]{} Swaters, R.~A., Sancisi, R., van der Hulst, J.M. 1997,
  ApJ 491, 140

\bibitem[]{} Swaters, R.~A., Schoenmakers, R.~H.~M., Sancisi, R., \&
van Albada, T.~S.\ 1999, MNRAS, 304, 330

\bibitem[]{} Swaters, R.~A., Madore, B.~F., \& Trewhella, M.\ 2000,
  ApJ, 531, L107

\bibitem[]{} Swaters, R.~A., van Albada, T.~S., van der Hulst, J.~M.,
  \& Sancisi, R.\ 2002, A\&A, 390, 829 (Paper~I)

\bibitem[]{} Swaters, R.~A., Madore, B.~F., van den Bosch, F.~C., \&
Balcells, M.\ 2003a, ApJ, 583, 732 (SMvdBB)

\bibitem[]{} Swaters, R.~A., Verheijen, M.~A.~W., Bershady, M.~A., \&
  Andersen, D.~R.\ 2003b, ApJ, 587, L19

\bibitem[]{} Tully, R.~B., Bottinelli, L., Fisher, J.~R., Gouguenheim,
  L., Sancisi, R., van Woerden, H. 1978 A\&A 63, 37

\bibitem[]{} van den Bosch, F.~C., \& Swaters, R.~A. 2001, MNRAS,
  325, 1017

\bibitem[]{} van den Bosch, F.~C., Robertson, B.~E., Dalcanton,
  J.~J., \& de Blok, W.~J.~G.\ 2000, \aj, 119, 1579

\bibitem[]{} Verheijen, M.~A.~W. 1997, PhD thesis, Rijksuniversiteit Groningen

\bibitem[]{} Verheijen, M.~A.~W., \& Sancisi, R.\ 2001, A\&A, 370, 765 

\end{thebibliography}
\end{document}